# The Complexity of Non-Monotone Markets

Xi Chen[*]     Dimitris Paparas[†]     Mihalis Yannakakis[‡]

**Abstract**

We introduce the notion of non-monotone utilities, which covers a wide variety of utility functions in economic theory. We then prove that it is PPAD-hard to compute an approximate Arrow-Debreu market equilibrium in markets with linear and non-monotone utilities. Building on this result, we settle the long-standing open problem regarding the computation of an approximate Arrow-Debreu market equilibrium in markets with CES utility functions, by proving that it is PPAD-complete when the Constant Elasticity of Substitution parameter $\rho$ is any constant less than $-1$.

---

[*]Columbia University. Email: xichen@cs.columbia.edu. Research supported by NSF CCF-1149257, a Sloan research fellowship, and start-up funds from Columbia University.

[†]Columbia University. Email: paparas@cs.columbia.edu. Research supported by NSF CCF-1017955.

[‡]Columbia University. Email: mihalis@cs.columbia.edu. Research supported by NSF CCF-1017955.

# 1  Introduction

General equilibrium theory [Deb59, Ell94] is regarded by many as the crown jewel of Mathematical Economics. It studies the interactions of price, demand and supply, and is established on the demand-equal-supply principle of Walras [Wal74]. A remarkable market model central to this field is the one of Arrow and Debreu [AD54], which has laid the foundation for competitive pricing mechanisms [AD54, Sca73].

In this model, traders exchange goods at a marketplace to maximize their utilities.[1] Formally, an Arrow-Debreu market $M$ consists of a set of traders and a set of goods, denoted by $\{G_1, \ldots, G_m\}$ for some $m \geq 1$. Each trader has an initial endowment $\mathbf{w} \in \mathbb{R}_+^m$, where $w_j$ denotes the amount of $G_j$ she brings to the market. Each trader also has a real-valued utility function $u$. Given a bundle $\mathbf{x} \in \mathbb{R}_+^m$ of goods, $u(\mathbf{x})$ is her utility if she obtains $\mathbf{x}$ after the exchange.

Now let $\boldsymbol{\pi} \in \mathbb{R}_+^m$ denote a price vector, where we use $\pi_j$ to denote the price of $G_j$. Each trader first sells her endowment $\mathbf{w}$ at $\boldsymbol{\pi}$ to obtain a budget of $\mathbf{w} \cdot \boldsymbol{\pi}$. She then spends it to purchase a bundle of goods $\mathbf{x}$ from the market to maximize her utility. We say $\boldsymbol{\pi}$ is a *market equilibrium* of $M$ if we can assign each trader an optimal bundle with respect to $\boldsymbol{\pi}$ such that the total demand equals the total supply and the market clears.

The celebrated theorem of Arrow and Debreu [AD54] asserts that, under mild conditions, every market has an equilibrium. Their proof, however, is based on Kakutani's fixed point theorem [Kak41] and is highly non-constructive and non-algorithmic, given that no efficient general fixed-point algorithm is known so far. Exponential lower bounds on the query complexity of a discrete fixed-point problem are proved for various query models in [HPV89, CD08, CT07, CST08].

The problem of finding a market equilibrium was first studied in the pioneering work of Scarf [Sca73]. During the past decade, starting with the work of Deng, Papadimitriou, and Safra [DPS03], the computation and approximation of equilibria have been studied intensively under various market models, and much progress has been made. This includes efficient algorithms for the market equilibrium problem [JMS03, DV03, CDSY04, DV04, GK04, GKV04, CMV05, CMPV05, CPV05, JVY05, JM05, JV06, CHT06, Jai07, Ye07, DK08, DPSV08, Ye08, Vaz10], many of which are based on the convex-programming approach of [EG59, NP83]. Several complexity-theoretic results have also been obtained for various market models [CSVY06, HT07, DD08, CDDT09, VY11, EY10, PW10, CT09, CT11].

**Markets with CES Utilities**

We study the complexity of approximating market equilibria in Arrow-Debreu markets with CES (constant elasticity of substitution) utilities [MCWG95]. A CES utility function takes the following form:

$$u(x_1, \ldots, x_m) = \left( \sum_{j=1}^m \alpha_j \cdot x_j^\rho \right)^{1/\rho}$$

where $\alpha_j \geq 0$ for all $j \in [m]$; and the parameter $\rho < 1$ and $\rho \neq 0$. The family of CES utility functions was first introduced in [Sol56, Dic54]. It was then used in [ACMS61] to model production functions and predict economic growth. It has been one of the most widely used families of utility functions in economics literature [SW92, dLG09] due to their versatility and flexibility in economic modeling. For example, the popular modeling language MPSGE [Rut99] for equilibrium analysis uses CES functions (and their generalization to nested CES functions) to model consumption and production. The parameter $\rho$ of a CES utility function is related to the elasticity of substitution $\sigma$, a measure on how easy it is to substitute different goods or resources [Hic32, Rob33] (namely $\rho = (\sigma - 1)/\sigma$). Selecting specific values for $\rho$ between 1 and $-\infty$ yields, as special cases, various basic utility functions and models different points in the substitutes-complements spectrum, ranging from the perfect substitutes case when $\rho = 1$, which corresponds to linear utilities, to the intermediate case when $\rho \to 0$, which corresponds to the Cobb-Douglas utilities, to the perfect complements case when $\rho \to -\infty$, which corresponds to Leontief utilities.

---

[1]The model of Arrow and Debreu also considers firms with production plans. Here we focus on the setting of exchange only.

Nenakov and Primak [NP83] gave a convex program that characterizes the set of equilibria when $\rho = 1$ i.e., all utility functions are linear. Jain [Jai07] discovered the same convex program independently and used the ellipsoid algorithm to give a polynomial-time exact algorithm. It turns out that this convex program can also be applied to characterize the set of equilibria in CES markets with $\rho > 0$. In [CMPV05], Codenotti, McCune, Penumatcha, and Varadarajan gave a different convex formulation for the set of equilibria in CES markets with $\rho : -1 \leq \rho < 0$. The range of $\rho < -1$ however has remained an intriguing open problem. For this range, it is known that the set of equilibria can be disconnected, and thus one cannot hope for a direct convex formulation. An example can be found in [Gje96] with three isolated market equilibria.

The failure of the convex-programming approach seems to suggest that the problem might be hard. On the other hand, as $\rho \to -\infty$, CES utilities converge to Leontief utilities for which finding an approximate market equilibrium is known to be PPAD-hard [CSVY06]. This argument, however, is less compelling due to the fact that a market with CES utilities converges to a Leontief market, as $\rho \to -\infty$, does not mean that the equilibria of the CES markets converge to an equilibrium of the Leontief market at the limit; actually it is easy to find examples where this is not the case, and in fact it is possible that the CES markets have equilibria that converge but the Leontief market at the limit does not even have any (approximate) equilibrium.

Moreover, with respect to the problem of determining whether a market equilibrium exists, CES utilities do not behave like the Leontief limit but rather like those tractable utilities. Typically, the tractability of the equilibrium existence problem conforms with that of the equilibrium computation problem (under standard sufficient conditions for existence). For example, the existence problem for linear utilities can be solved in polynomial time [Gal76] (as does the computation problem [Jai07]), and the same holds for Cobb-Douglas utilities [Eav85], whereas the existence problem is NP-hard for Leontief utilities [CSVY06] and for separable piecewise-linear utilities [VY11] (and their computation problem is PPAD-hard [CSVY06, CDDT09, VY11]). However, it is known that the existence problem for CES functions is polynomial-time solvable for all (finite) values of $\rho$ [CMPV05]. This suggests that the equilibrium computation problem for CES utilities might be also tractable.

The difficulty in resolving the complexity of the equilibrium computation problem for CES markets with a constant $\rho < -1$ is mainly due to the continuous nature of the problem. Most, if not all, of the problems shown to be PPAD-hard have a rich underlying combinatorial structure, whether it is to find an approximate Nash equilibrium in a normal-form game [DGP09, CDT09] or to compute an approximate equilibrium in a market with Leontief utilities [CSVY06] or with additively separable and concave piecewise-linear utilities [CDDT09, VY11]. In contrast, given a price vector $\boldsymbol{\pi}$, the optimal bundle $\mathbf{x}$ of a CES trader is a continuous function over $\boldsymbol{\pi}$, with an explicit algebraic form (see (1) in Section 2) which can be derived using the KKT conditions. The problem of finding a market equilibrium now boils down to solving a system of polynomial equations over variables $\boldsymbol{\pi}$, and it is not clear how to extract from it a useful combinatorial structure.

We settle the complexity of finding approximate equilibria in CES markets for all values of $\rho < -1$:

**Theorem 1.** *For any fixed rational number $\rho < -1$, the problem of finding an approximate market equilibrium in a CES market of parameter $\rho$ is* PPAD-complete.

It is worth pointing out that the notion of approximate market equilibria used in Theorem 1 is one-sided, i.e., $\boldsymbol{\pi}$ is an $\epsilon$-approximate market equilibrium if the excess demand of each good is bounded from above by $\epsilon$-fraction of the total supply. While the two-sided notion of approximate equilibria is more commonly used in the literature (which we will refer to as $\epsilon$-*tight* approximate market equilibria), i.e., the absolute value of excess demand is bounded, we present an unexpected CES market with $\rho < 0$ in Section 2.2 and prove that any of its $(1/2)$-tight approximate equilibria requires exponentially many bits to represent. By contrast, we show that for the one-sided notion there is always an $\epsilon$-approximate equilibrium with a polynomial number of bits and, furthermore, its computation is in PPAD. We show also that the problem of computing an actual equilibrium (to any desired precision) is in FIXP.

**PPAD-Hardness for Non-Monotone Families of Utilities**

The resolution of the complexity of CES markets with $\rho < -1$ inspired us to ask the following question:

> *Can we prove a complexity dichotomy for any given family of utility functions?*

Formulating it more precisely, we let $\mathcal{U}$ denote a generic family of utility functions that satisfy certain mild conditions (e.g., they should be continuous, quasi-concave). The question now becomes the following:

> *Does there exist a mathematically well-defined property on families of functions such that:*
> *For any $\mathcal{U}$ satisfying this property, the equilibrium problem it defines is in polynomial time;*
> *For any $\mathcal{U}$ that violates this property, the problem is hard, e.g., PPAD-hard or even FIXP-hard.*

For the algorithmic part of this question, a property that has played a critical role in the approximation of market equilibria is WGS (Weak Gross Substitutability). A family $\mathcal{U}$ of utilities satisfies WGS if for any market consisting of traders with utilities from $\mathcal{U}$, increasing the price of one good while keeping all other prices fixed cannot cause a decrease in the demand of any other good. WGS implies that the set of equilibria form a convex set. In [ABH59], Arrow, Block and Hurwicz showed that, given any market satisfying WGS, the continuous tatonnement process [Wal74, Sam47] converges. Recently in [CMV05], Codenotti, McCune and Varadarajan showed that a discrete tatonnement algorithm converges to an approximate equilibrium in polynomial time, if equipped with an excess demand oracle. Another general property that implies convexity of the set of equilibria is WARP (Weak Axiom of Revealed Preference, see [MCWG95] for its definition and background). While many families of utilities satisfy WGS or WARP, they do not seem to cover all the efficiently solvable market problems, e.g., the family of CES utilities with parameter $-1 \le \rho < 0$ does not satisfy WGS or WARP but has a convex formulation [CMPV05].

For the hardness part of this question, our knowledge is much more limited. Only for a few specific and isolated families of utilities mentioned earlier, the problem of finding an approximate equilibrium is shown to be hard. And the reduction techniques developed in these proofs are all different, each fine tuned for the family of utilities being considered.

Our second contribution is a PPAD-hardness result that is widely applicable to any generic family $\mathcal{U}$ of utility functions, as long as it satisfies the following condition:

> [*Informal*]: *There exists a market $M$ with utilities from $\mathcal{U}$, a special good $G$ in $M$, and a price vector $\boldsymbol{\pi}$ such that at $\boldsymbol{\pi}$, the excess demand of $G$ is nonnegative and raising the price of $G$, while keeping all other prices the same, strictly increases the demand of $G$.*

We call $M$ a *non-monotone* market. We also call $\mathcal{U}$ a *non-monotone* family if such a market $M$ exists.

Examples of simple non-monotone markets, constructed from various families of utilities, can be found in Section 2.3. We show that the existence of such a pair $(M, \boldsymbol{\pi})$ implies the following hardness result:

**Theorem 2** (Informal)**.** *If $\mathcal{U}$ is non-monotone, then the following problem is* PPAD-hard*: Given a market in which the utility of each trader is either linear or from $\mathcal{U}$, find an approximate market equilibrium.*

We remark that there is clearly a gap between WGS and non-monotonicity. It remains an open problem as whether we can further reduce the gap and whether we can remove the use of linear functions.

The reductions for both of our main results are quite involved, and start from the problem of computing a well-suported approximate equilibrium for a polymatrix game with 2 strategies per player, which we show is PPAD-hard (the problem of finding an exact equilibrium was shown previously to be hard in [DGP09]).

The rest of the paper is organized as follows. In Section 2, we give basic definitions and state formally our main results. We provide also a very brief outline of the proofs. Section 3 contains the PPAD-hardness proof for general non-monotone utilities (Theorem 2), and Section 4 contains the PPAD-hardness proof for CES utilities (Theorem 1). Section 5 shows that the problem of computing an equilibrium for CES markets is in FIXP, and Section 6 shows that computing an approximate equilibrium is in PPAD. Section 7 contains the hardness proof of the polymatrix problem that serves as the starting point in our reductions. Finally we conclude in Section 8.

## 2  Preliminaries and Main Results

**Notation.** We use $\mathbb{R}_+$ to denote the set of nonnegative real numbers and $\mathbb{Q}_+$ to denote the set of nonnegative rational numbers. Given a positive integer $n$, we use $[n]$ to denote the set $\{1, \ldots, n\}$. Given two integers $m$ and $n$, where $m \leq n$, we use $[m:n]$ to denote the set $\{m, m+1, \ldots, n\}$. Given a vector $\mathbf{y} \in \mathbb{R}^m$, we use $B(\mathbf{y}, c)$ to denote the set of $\mathbf{x}$ with $\|\mathbf{x} - \mathbf{y}\|_\infty \leq c$.

### 2.1  Arrow-Debreu Markets and Market Equilibria

An Arrow-Debreu exchange market $M$ consists of a finite set of traders, denoted by $\{T_1, \ldots, T_n\}$ for some $n \geq 1$, and a finite set of goods, denoted by $\{G_1, \ldots, G_m\}$ for some $m \geq 1$. Each trader $T_i$ owns an initial endowment $\mathbf{w}_i \in \mathbb{R}_+^m$, where $w_{i,j}$ denotes the amount of good $G_j$ she initially owns. Each trader $T_i$ also has a utility function $u_i : \mathbf{R}_+^m \to \mathbf{R}_+$, where $u_i(x_{i,1}, \ldots, x_{i,m})$ represents the utility she derives if the amount of $G_j$ she obtains by the end is $x_{i,j}$ for each $j \in [m]$. In the rest of the paper, we will refer to an Arrow-Debreu exchange market simply as a market for convenience.

Now let $\boldsymbol{\pi} = (\pi_1, \ldots, \pi_m) \neq \mathbf{0}$ denote a nonnegative price vector, with $\pi_j \geq 0$ being the price per unit of $G_j$. Each trader $T_i$ sells her initial endowment $\mathbf{w}_i$ at prices $\boldsymbol{\pi}$ and obtains a budget of $\sum_{j \in [m]} w_{i,j} \cdot \pi_j$. She then spends the budget to buy a bundle of goods $\mathbf{x}_i \in \mathbb{R}_+^m$ from the market to maximize her utility. We say $\boldsymbol{\pi}$ is a *market equilibrium* of $M$ if we can assign each trader $T_i$ an optimal bundle $\mathbf{x}_i$ with respect to $\boldsymbol{\pi}$, such that the total demand equals the total supply and the market clears. Formally, given $\boldsymbol{\pi}$ we use $\mathsf{opt}_i(\boldsymbol{\pi})$ to denote the set of optimal bundles of $T_i$ with respect to $\boldsymbol{\pi}$: $\mathbf{x} \in \mathbb{R}_+^m$ is in $\mathsf{opt}_i(\boldsymbol{\pi})$ if

$$\sum_{j \in [m]} x_j \cdot \pi_j \leq \sum_{j \in [m]} w_{i,j} \cdot \pi_j$$

and for any $\mathbf{x}' \in \mathbb{R}_+^m$ that satisfies the budget constraint above, we have $u_i(\mathbf{x}) \geq u_i(\mathbf{x}')$.

Next we define the (aggregate) excess demand of a good with respect to a given price vector $\boldsymbol{\pi}$:

**Definition 1** (Excess Demand). *Given $\boldsymbol{\pi}$, the* excess demand *$Z(\boldsymbol{\pi})$ consists of all vectors $\mathbf{z}$ of the form*

$$\mathbf{z} = \mathbf{x}_1 + \cdots + \mathbf{x}_m - (\mathbf{w}_1 + \cdots + \mathbf{w}_m)$$

*where $\mathbf{x}_i$ is an optimal bundle in $\mathsf{opt}_i(\boldsymbol{\pi})$ for each $i \in [n]$. For each good $G_j$ we also use $Z_j(\boldsymbol{\pi})$ to denote the projection of $Z(\boldsymbol{\pi})$ on the jth coordinate.*

In general, $Z(\boldsymbol{\pi})$ is a set and $Z$ is a correspondence. We usually refer to a subset of traders in a market as a submarket. Sometimes we are interested in the excess demand of a submarket, for which the sums of $\mathbf{x}_i$'s and $\mathbf{w}_i$'s are only taken over traders in the subset. Finally we define market equilibria:

**Definition 2** (Market Equilibria). *We say $\boldsymbol{\pi}$ is a* market equilibrium *of a market $M$ if $\mathbf{0} \in Z(\boldsymbol{\pi})$.*

Notice that if $z_j > 0$, then the traders request more than the total available amount of $G_j$ and if $z_j \leq 0$ then they request at most as much amount of it as is available in the market. As $\mathsf{opt}_i(\boldsymbol{\pi})$ is invariant under scaling of $\boldsymbol{\pi}$ (by a positive factor), it is easy to see that the set of market equilibria is closed under scaling.

We now define two versions of approximate market equilibria:

**Definition 3** ($\epsilon$-Approximate Market Equilibria). *We call $\boldsymbol{\pi}$ an $\epsilon$-approximate market equilibrium of $M$ for some $\epsilon > 0$, if there exists a vector $\mathbf{z} \in Z(\boldsymbol{\pi})$ such that $z_j \leq \epsilon \sum_{i \in [n]} w_{i,j}$ for all $j \in [m]$.*

**Definition 4** ($\epsilon$-Tight Approximate Market Equilibria). *We call $\boldsymbol{\pi}$ an $\epsilon$-tight approximate market equilibrium of $M$ for some $\epsilon > 0$, if there exists a vector $\mathbf{z} \in Z(\boldsymbol{\pi})$ such that $|z_j| \leq \epsilon \sum_{i \in [n]} w_{i,j}$ for all $j \in [m]$.*

Both notions of approximate equilibria have been used in the literature. Although the two-sided notion of tight approximate equilibria is more commonly used, we present an unexpected CES market with $\rho < 0$ in Section 2.2, and prove that any $(1/2)$-tight approximate equilibrium $\boldsymbol{\pi}$ must have one of the entries being doubly exponentially small, when $\sum_j \pi_j = 1$.

In general, a market equilibrium may not exist. The pioneering existence theorem of Arrow and Debreu [AD54] states that if all the utility functions are quasi-concave, then under certain mild conditions a market always has an equilibrium. In this paper, we use the weaker sufficient condition of Maxfield [Max97].

**Definition 5** (Local Non-Satiation). *We say $u : \mathbb{R}_+^m \to \mathbb{R}_+$ is* locally non-satiated *if for any vector $\mathbf{x} \in \mathbb{R}_+^m$ and any $\epsilon > 0$, there exists a $\mathbf{y} \in B(\mathbf{x}, \epsilon)$ such that $u(\mathbf{y}) > u(\mathbf{x})$. We say $u$ is* non-satiated *with respect to the kth good, if for any $\mathbf{x} \in \mathbb{R}_+^m$, there exists a $\mathbf{y} \in \mathbb{R}_+^m$ such that $u(\mathbf{y}) > u(\mathbf{x})$ and $y_j = x_j$ for all $j \neq k$.*

If the utility of a trader is locally non-satiated, then her optimal bundle must exhaust her budget. Therefore, if every trader in $M$ has a non-satiated utility, then Walras' law holds: $\mathbf{z} \cdot \boldsymbol{\pi} = 0$ for all $\mathbf{z} \in Z(\boldsymbol{\pi})$.

**Definition 6** (Economy Graphs). *Given a market $M$, we define a directed graph as follows. Each vertex of the graph corresponds to a good $G_j$ in $M$. For two goods $G_i$ and $G_j$ in $M$, we add an edge from $G_i$ to $G_j$, if there is a trader $T_k$ such that $w_{k,i} > 0$ and $u_k$ is non-satiated with respect to $G_j$, i.e., $T_k$ owns a positive amount of $G_i$ and is interested in $G_j$. We call this graph the* economy graph *of $M$ [Max97].[2]*

We then call a market $M$ strongly connected if its economy graph is strongly connected. Here is a simplified version of the existence theorem from Maxfield [Max97]:

---

[2] Maxfield defines this as a graph between the traders instead of the goods, but the sufficient condition of strong connectivity is equivalent between the two versions, as long as each trader owns some good and is non-satiated with respect to some good, and each good is owned by some trader and desired by some trader. Codenotti et al. [CMPV05] use in their analysis of CES markets also the trader-based version which they decompose into strongly connected components (scc's), but again it is not hard to show that there is a correspondence between the nontrivial scc's of the two graphs.

**Theorem 3** (Maxfield [Max97]). *If the following two conditions hold, then $M$ must have an equilibrium: 1). Every utility is continuous, quasi-concave and locally non-satiated; and 2). $M$ is strongly connected.*

## 2.2 CES Utility Functions

In this paper, we focus on the family of CES (Constant Elasticity of Substitution) utility functions:

**Definition 7.** *We call $u : \mathbb{R}_+^m \to \mathbb{R}_+$ a CES function with parameter $\rho < 1$, $\rho \neq 0$, if it is of the form*

$$u(x_1, \ldots, x_m) = \left( \sum_{j \in [m]} \alpha_j \cdot x_j^\rho \right)^{\frac{1}{\rho}}$$

*where the coefficients $\alpha_1, \ldots, \alpha_m \in \mathbb{R}_+$.*

Let $T$ denote a trader with a CES utility function $u$ in which $\alpha_j > 0$ if and only if $j \in S \subseteq [m]$. Let $\mathbf{w}$ denote the initial endowment of $T$ and let $\boldsymbol{\pi}$ denote a price vector with $\pi_j > 0$ for all $j \in [m]$. Then using the KKT conditions, one can show that the unique optimal bundle of $T$ consists of

$$x_j = \left( \frac{\alpha_j}{\pi_j} \right)^{1/(1-\rho)} \times \frac{\mathbf{w} \cdot \boldsymbol{\pi}}{\sum_{k \in S} \alpha_k^{1/(1-\rho)} \cdot \pi_k^{-\rho/(1-\rho)}} \tag{1}$$

units of $G_j$, $j \in S$. It is also clear that if $\pi_j = 0$ for some $j \in S$, then $T$ would demand an infinite amount of $G_j$. This implies that when a CES market is strongly connected, then $\pi_j$ must be positive for all $j \in [m]$ in any (exact or approximate) market equilibrium of $M$.

The problem of whether there exists an equilibrium in a CES market can be solved in polynomial time: a simple necessary and sufficient condition for the existence of an equilibrium in a CES market was shown in [CMPV05] based on the decomposition of the economy graph into strongly connected components. They furthermore proved that the computation of an equilibrium for the whole market (if the condition is satisfied) amounts to the computation of equilibria for the submarkets induced by the strongly connected components. Thus, we will focus on markets with a strongly connected economy graph.

We are interested in the problem of computing an equilibrium in a market with CES utilities. As such a market may not have a rational equilibrium in general, even when $\rho$ and all the coefficients are rational, we study the approximation of market equilibria. We define the following three problems:

1. **CES**: The input of the problem is a pair $(k, M)$, where $k$ is a positive integer encoded in unary ($k$ represents the desired number of bits of precision), and $M$ is a strongly connected market in which all utilities are CES, with the parameter $\rho_i < 1$ of each trader $T_i$ being rational and given in unary (because $\rho$ appears in the exponent in the utility and demand functions). The parameters $\rho_i$'s for different traders may be the same or different, and there may be a mixture of positive and negative parameters. The endowments $w_{i,j}$ and coefficients $\alpha_{i,j}$ are rational and encoded in binary. The goal is to find a price vector $\boldsymbol{\pi}$ that is within $1/2^k$ of some equilibrium in every coordinate, i.e., such that there exists an (exact) equilibrium $\boldsymbol{\pi}^*$ of $M$ with

$$\|\boldsymbol{\pi} - \boldsymbol{\pi}^*\|_\infty \leq 1/2^k$$

2. **CES-APPROX**: The input of the problem is the same as CES. The goal is to find an $\epsilon$-approximate market equilibrium of $M$, where $\epsilon = 1/2^k$.

3. For each (fixed) rational number $\rho < -1$, we also define the following problem $\rho$-**CES-APPROX**: The input is the same as **CES**, except that the utilities of all the traders have the same fixed parameter $\rho$, which is considered as a constant, not part of the input. The goal is to find an $\epsilon$-approximate market equilibrium of $M$, where $\epsilon = 1/k$.

The output of **CES** is usually referred to in the literature as a strongly approximate equilibrium. Besides, we can also define **CES** under a model of real computation and ask for an exact equilibrium.

Finally we present the following example to justify the use of $\epsilon$-*approximate market equilibria*, instead of $\epsilon$-*tight approximate market equilibria*, in both **CES-APPROX** and $\rho$-**CES-APPROX**.

**Example 2.1.** *Fix a $\rho < 0$ and let $r = |\rho| > 0$. Let $M$ denote the following CES market with parameter $\rho$. Here $M$ has $n$ goods $G_1, \ldots, G_n$ and $n$ traders $T_1, \ldots, T_n$. Each $T_i$, $i \in [n]$, has $2^{i(n+1)}$ units of good $G_i$ at the beginning. Each $T_i$, $i \in [n-1]$, is equally interested in $G_1$ and $G_{i+1}$ so in particular, $T_1$ is interested in only $G_1$ and $G_2$. $T_n$ is only interested in $G_1$. The economy graph of $M$ is clearly strongly connected.*

We prove the following lemma which implies that we need an exponential number of bits to represent any $(1/2)$-tight approximate equilibrium of this market.

**Lemma 1.** *If $\pi$ is a $(1/2)$-tight approximate market equilibrium of $M$, then we must have*

$$\frac{\max_j \pi_j}{\min_j \pi_j} > 2^{n(1+r)^{n-2}}$$

*Proof.* For each $i \in [n-1]$, since $T_i$ is the only trader interested in $G_{i+1}$, we must have from (1)

$$\text{the demand of } G_{i+1} \text{ from } T_i = \frac{2^{i(n+1)} \cdot \pi_i}{\pi_{i+1}^{1/(1+r)} \left(\pi_{i+1}^{r/(1+r)} + \pi_1^{r/(1+r)}\right)} \geq 2^{(i+1)(n+1)-1}$$

We denote $(\pi_i/\pi_1)^{1/(1+r)}$ by $t_i$ for each $i \in [n]$. Using $\pi_{i+1} > 0$, we have

$$2^n < \frac{\pi_i}{\pi_{i+1}^{1/(1+r)} \cdot \pi_1^{r/(1+r)}} = \left(\frac{\pi_i}{\pi_{i+1}}\right)^{1/(1+r)} \left(\frac{\pi_i}{\pi_1}\right)^{r/(1+r)} \;\Rightarrow\; t_{i+1} < 2^{-n} \cdot (t_i)^{1+r}$$

From $t_1 = 1$ and $t_2 < 2^{-n}$ we can use induction to show that $t_i < 2^{-n(1+r)^{i-2}}$ for all $i \in [2:n]$. □

We are now ready to state our main results for CES markets. First, in Section 5 and Section 6, we prove the membership of **CES** in FIXP [EY10] and membership of **CES-APPROX** in PPAD [Pap94], respectively:

**Theorem 4.** **CES** *is in* FIXP.

**Theorem 5.** **CES-APPROX** *is in* PPAD.

We show in Section 4 that CES markets are PPAD-hard to solve when $\rho < -1$:

**Theorem 6.** *For any rational number $\rho < -1$, the problem $\rho$-**CES-APPROX** is* PPAD-hard.

Combining Theorem 5 and Theorem 6, we have

**Corollary 1.** *For any rational number $\rho < -1$, the problem $\rho$-**CES-APPROX** is* PPAD-complete.

In the proof of Theorem 6 in Section 4, we present a polynomial-time reduction from a PPAD-hard problem (see Section 2.4) to $\rho$-**CES-APPROX**. The hard instances we construct are in fact very restricted in the sense that each trader is interested in one or two goods and applies one of the following utility functions:

$$u(x) = x, \quad u(x_1, x_2) = (x_1^\rho + x_2^\rho)^{1/\rho}, \quad \text{or} \quad u(x_1, x_2) = (\alpha \cdot x_1^\rho + x_2^\rho)^{1/\rho} \tag{2}$$

where $\alpha$ is a positive rational constant that depends on $\rho$ only.

## 2.3 Non-Monotone Markets and Families of Utilities

We use $\mathcal{U}$ to denote a generic family of *continuous, quasi-concave and locally non-satiated* functions, e.g., linear functions, piecewise-linear functions (see Example 2.4), CES functions for a specific parameter of $\rho$ e.g. $\rho = -3$, or even the finite set of three functions given in (2). Ideas behind the proof of Theorem 6 allow us to prove a PPAD-hardness result for the problem of computing an approximate equilibrium of a market in which the utility function of each trader is *either linear or from $\mathcal{U}$*, when the latter is "*non-monotone*" (to be defined shortly). For this purpose, we formally set up the problem as follows.

First we assume that $\mathcal{U}$ is countable, and each function $g \in \mathcal{U}$ corresponds to a unique binary string so that a trader can specify a function $g \in \mathcal{U}$ using a binary string. In a market with $m$ goods, we say a trader "*applies*" a function $g \in \mathcal{U}$ if her utility function $u$ is of the form

$$u(x_1, \ldots, x_m) = g\left(\frac{x_{\ell_1}}{b_1}, \ldots, \frac{x_{\ell_k}}{b_k}\right),$$

where $g \in \mathcal{U}$ has $k \leq m$ variables; $\ell_1, \ldots, \ell_k \in [m]$ are distinct indices; and $b_1, \ldots, b_k$ are positive rational numbers. In this way, each trader can be described by a finite binary string. We now use $\mathcal{M}_\mathcal{U}$ to denote the set of all markets in which every trader has a rational initial endowment and applies a utility function from $\mathcal{U}$. We also use $\mathcal{M}_\mathcal{U}^*$ to denote the set of markets in which every trader has a rational initial endowment and applies either a utility function from $\mathcal{U}$ or a linear utility with rational coefficients.

Second we assume that there exists a univariate function $g^* \in \mathcal{U}$ that is strictly monotone.

**Remark.** We always make these two assumptions on a family of utilities $\mathcal{U}$ throughout this paper. Both of them seem to be natural, and we only need them for technical reasons that will become clear later. When a trader applies a function from $\mathcal{U}$, she can always change units by scaling. The second assumption basically allows us to add single-minded traders who spend all their budget on one specific good.

We next define *non-monotone* markets as well as *non-monotone* families of utilities:

**Definition 8** (Non-monotone Markets and Families of Utilities). *Let $M$ be a market with $k \geq 2$ goods. We say $M$ is* non-monotone *at a price vector $\boldsymbol{\pi}$ if the following conditions hold: $\pi_j > 0$ for all $j \in [k]$ and*

> *For some $c > 0$, the excess demand $Z_1(y_1, \ldots, y_k)$ of $G_1$ is a continuous function (instead of a correspondence) over $\mathbf{y} \in B(\boldsymbol{\pi}, c)$, with $Z_1(\boldsymbol{\pi}) \geq 0$. The partial derivative of $Z_1$ with respect to $y_1$ exists over $B(\boldsymbol{\pi}, c)$, is continuous over $B(\boldsymbol{\pi}, c)$, and is (strictly) positive at $\boldsymbol{\pi}$.*

*We call $M$ a* non-monotone *market if there exists such a price vector $\boldsymbol{\pi}$. We call $\mathcal{U}$ a* non-monotone *family of utilities if there exists a non-monotone market in $\mathcal{M}_\mathcal{U}$.*

**Remark.** By the definition, $M$ being non-monotone at $\boldsymbol{\pi}$ means that, raising the price of $G_1$ while keeping the prices of all other goods the same, would actually increase the total demand of $G_1$. Also note that using the continuity of $Z_1$ as well as its partial derivative with respect to $y_1$, we can indeed require, without loss of generality, the price vector $\boldsymbol{\pi}$ to be rational in Definition 8: if $M$ is a non-monotone market at $\boldsymbol{\pi}$ but $\boldsymbol{\pi}$ is not rational, then a rational vector $\boldsymbol{\pi}^*$ close enough to $\boldsymbol{\pi}$ would have the same property. Therefore, whenever $\mathcal{U}$ is non-monotone, there is a market $M \in \mathcal{M}_\mathcal{U}$ that is non-monotone at a rational price vector $\boldsymbol{\pi}$. We would like to point out that $M$ is not necessarily strongly connected. Also the excess demand $Z_1(\boldsymbol{\pi})$ of $G_1$ as well as the partial derivative of $Z_1$ with respect to $y_1$ at $\boldsymbol{\pi}$ do not have to be rational.

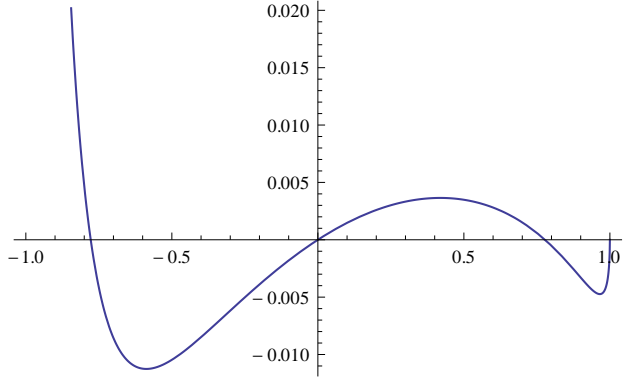

Figure 1: The excess demand function $Z_1(x)$ of Example 2.2.

Now we state our PPAD-hardness result for a non-monotone family $\mathcal{U}$ of functions. We use $\mathcal{U}$-**MARKET** to denote the following problem: the input is a pair $(k, M)$, where $k$ is a positive integer in unary and $M$ is a strongly connected market from $\mathcal{M}_\mathcal{U}^*$ encoded in binary. The goal is to output an $\epsilon$-approximate equilibrium of $M$ with $\epsilon = 1/k$. While our hardness result essentially states that $\mathcal{U}$-**MARKET** is PPAD-hard when $\mathcal{U}$ is non-monotone, we need the following definition to make a formal statement:

**Definition 9.** *We say a real number $\beta$ is* moderately computable *if there is an algorithm that, given $\gamma > 0$, outputs a $\gamma$-rational approximation $\beta'$ of $\beta$: $|\beta' - \beta| \leq \gamma$, in time polynomial in $1/\gamma$.*

**Theorem 7.** *Let $\mathcal{U}$ denote a non-monotone family of utility functions. If there exists a market $M \in \mathcal{M}_\mathcal{U}$ such that $M$ is non-monotone at a rational price vector $\boldsymbol{\pi}$ such that the excess demand $Z_1(\boldsymbol{\pi})$ of $G_1$ at $\boldsymbol{\pi}$ is moderately computable, then the problem $\mathcal{U}$-**MARKET** is PPAD-hard.*

**Remark.** From the definition, $\mathcal{U}$ being non-monotone implies the existence of $M$ and $\boldsymbol{\pi}$. The other assumption made in Theorem 7 only requires that there exists one such pair $(M, \boldsymbol{\pi})$ for which $Z_1(\boldsymbol{\pi})$ as a specific positive number is moderately computable. We also point out that when the assumptions of Theorem 7 hold such a pair $M$ and $\boldsymbol{\pi}$ is considered as a constant, which we later use in the proof of Theorem 7 as a gadget to give a polynomial-time reduction from a PPAD-hard problem (see Section 2.4) to $\mathcal{U}$-**MARKET**. As a result, all components of $M$, including the number of goods and traders, the endowments of traders, binary strings that specify their utility functions from $\mathcal{U}$, are all considered as constants and encoded by binary strings of constant length. This also includes the positive rational vector $\boldsymbol{\pi}$.

Now we present three examples of non-monotone markets, one with CES utilities of parameter $\rho < -1$, one with Leontief utilities, and one with additively separable and piecewise-linear utilities:

**Example 2.2** (A Non-Monotone Market with CES Utilities of $\rho < -1$)**.** *Consider the following market $M$ with two goods $G_1, G_2$ and two traders $T_1, T_2$. $T_1$ has 1 unit of $G_1$, $T_2$ has 1 unit of $G_2$ and the utilities are*

$$u_1(x_1, x_2) = (\alpha \cdot x_1^\rho + x_2^\rho)^{1/\rho} \quad \text{and} \quad u_2(x_1, x_2) = (x_1^\rho + \alpha \cdot x_2^\rho)^{1/\rho}$$

respectively. When $\rho < -1$ and $\alpha$ is large enough, [Gje96] proves that $M$ has $(1,1)$ as an equilibrium and is non-monotone at $(1,1)$. This implies that $M$ has multiple isolated equilibria, and the set of equilibria of a CES market with $\rho < -1$ is not convex (not even connected) in general. To see this, we let $Z_1(x)$ denote the excess demand function of $G_1$, when the price of $G_1$ is $1+x$ and the price of $G_2$ is $1-x$. We plot $Z_1$ in Figure 1. From the picture it is clear that the curve has three roots or equilibria. (When $x$ goes to 1, $Z_1(x)$ converges to 0 but is always negative.) We will formally prove properties of this curve later in Section 4.1 which play an important role in the proof of Theorem 6.

**Example 2.3** (A Non-Monotone Market with Leontief Utilities). *We say $u$ is a Leontief utility function if*

$$u(x_1,\ldots,x_k) = \min_{j \in S}\left\{\frac{x_j}{a_j}\right\}, \quad \text{where } a_j > 0 \text{ for all } j \in S \subseteq [k].$$

*Let $M$ denote the Leontief market consisting of the following two traders $T_1$ and $T_2$. $T_1$ has 1 unit of $G_1$, $T_2$ has 1 unit of $G_2$, and their utility functions are*

$$u_1(x_1,x_2) = \min\{x_1/2, x_2\} \quad \text{and} \quad u_2(x_1,x_2) = \min\{x_1, x_2/2\}$$

*respectively. It is easy to show that $M$ is non-monotone at $(1,1)$.*

**Example 2.4** (A Non-Monotone Market with Additively Separable and Piecewise-Linear Utilities). *We say a utility function is additively separable and piecewise-linear if it is of the form:*

$$u(x_1,\ldots,x_k) = f_1(x_1) + \cdots + f_k(x_k) \tag{3}$$

*where $f_1,\ldots,f_k$ are all piecewise-linear functions. Consider the following market $M$ with two goods $G_1, G_2$ and two traders $T_1, T_2$. $T_1$ has 1 unit of $G_1$, and $T_2$ has 1 unit of $G_2$. Their utility functions are*

$$u_1(x_1,x_2) = x_1 + f(x_2) \quad \text{and} \quad u_2(x_1,x_2) = f(x_1) + x_2 \quad \text{with} \quad f(x) = \begin{cases} 2x & \text{if } x \leq 1/3 \\ 2/3 & \text{if } x > 1/3 \end{cases}$$

*It can be shown that $M$ has $(1,1)$ as an equilibrium and is non-monotone at $(1,1)$. Note that in general, the excess demand of a market with such utilities is a correspondence instead of a map, and partial derivatives may not always exist. But in the definition of non-monotone markets, we only need these properties in a local neighborhood of $\pi$, like $(1,1)$ here.*

Since linear functions are special cases of additively separable and piecewise-linear functions, we get a corollary from Theorem 7 and Example 2.4, that finding an approximate equilibrium in a market with additively separable and concave piecewise-linear utilities is PPAD-hard, shown earlier in [CDDT09]. Combining it with the membership of PPAD proved in [VY11], we have

**Corollary 2.1.** *The problem of computing an approximate market equilibrium in a market with additively separable and concave piecewise-linear utilities is* PPAD-complete, *even when each univariate function $f_j$ in (3) is either linear or has the form of $f$ in Example 2.4, i.e., linear function with a threshold.*

## 2.4 Polymatrix Games and Nash Equilibria

To prove Theorem 6 and 7, we give a polynomial-time reduction from the problem of computing an approximate Nash equilibrium in a polymatrix game [Jan68] with two pure strategies for each player. Such a game with $n$ players can be described by a $2n \times 2n$ rational matrix $\mathbf{P}$, with all entries between 0 and 1.[3]

An $\epsilon$-well-supported Nash equilibrium is a vector $\mathbf{x} \in \mathbb{R}_+^{2n}$ such that $x_{2i-1} + x_{2i} = 1$ and

$$\mathbf{x}^T \cdot \mathbf{P}_{2i-1} > \mathbf{x}^T \cdot \mathbf{P}_{2i} + \epsilon \Rightarrow x_{2i} = 0$$
$$\mathbf{x}^T \cdot \mathbf{P}_{2i} > \mathbf{x}^T \cdot \mathbf{P}_{2i-1} + \epsilon \Rightarrow x_{2i-1} = 0$$

for all $i \in [n]$, where $\mathbf{P}_{2i-1}$ and $\mathbf{P}_{2i}$ denote the $(2i-1)$th and $(2i)$th column vectors of $\mathbf{P}$, respectively.

Let **POLYMATRIX** denote the following problem:

Given a polymatrix game $\mathbf{P}$, compute an $\epsilon$-well-supported Nash equilibrium with $\epsilon = 1/n$.

It was shown in [DGP09] that finding an exact Nash equilibrium of a polymatrix game with two strategies for each player is PPAD-hard (it is not stated explicitly there, but it follows from the proof of Lemma 6.3). We prove in Section 7 that **POLYMATRIX** is also PPAD-hard. The proof follows techniques developed in previous work on Nash equilibria [DGP09, CDT09]. While its PPAD-hardness is used here as a bridge to establish Theorem 6 and Theorem 7, we think the result on **POLYMATRIX** is interesting for its own right.

**Theorem 8.** **POLYMATRIX** *is* PPAD-complete.

## 2.5 Proof Sketch of the Hardness Reductions

We give a rough overview of the constructions for the main results.

For Theorem 7, given any $2n \times 2n$ polymatrix game $\mathbf{P}$ we construct a market $M_{\mathbf{P}}$ in which the utility of each trader is either linear or from $\mathcal{U}$. We then show that given any $\epsilon$-approximate equilibrium $\boldsymbol{\pi}$ of $M_{\mathbf{P}}$ for some polynomially small $\epsilon$, we can recover a $(1/n)$-well-supported Nash equilibrium in polynomial time.

A building block of our construction is the *linear price-regulating markets* [CDDT09, VY11]. We let $\tau$ and $\alpha$ denote two positive parameters. Such a market consists of two traders $T_1, T_2$ and two goods $G_1, G_2$. $T_i$ owns $\tau$ units of $G_i$, $i \in \{1, 2\}$. The utility of $T_1$ is $(1+\alpha)x_1 + x_2$ and the utility of $T_2$ is $x_1 + (1+\alpha)x_2$. Let $\pi_i$ denote the price of $G_i$, then we have the following useful property: Even if we add more traders to the market, as long as their total endowment of $G_1$ and $G_2$ is negligible compared to that of $T_1$ and $T_2$, the ratio of $\pi_1$ and $\pi_2$ must lie between $(1-\alpha)/(1+\alpha)$ and $(1+\alpha)/(1-\alpha)$ at an approximate equilibrium.

Our construction starts with the following blueprint of encoding a vector $\mathbf{x}$ of $2n$ variables and the $2n$ linear forms $\mathbf{x}^T \cdot \mathbf{P}_j$, $j \in [2n]$, in $M_{\mathbf{P}}$. Let $G_1, \ldots, G_{2n}$ and $H_1, \ldots, H_{2n}$ denote $4n$ goods. Let $\tau$ denote a large enough polynomial in $n$. Let $\alpha$ and $\beta$ denote two polynomially small parameters with $\alpha \ll \beta$. For each $i \in [n]$, we first create a price-regulating market over $G_{2i-1}$ and $G_{2i}$ with parameters $\tau$ and $\alpha$, and a price-regulating market over $H_{2i-1}$ and $H_{2i}$ with parameters $\tau$ and $\beta$. Then for each $i \in [2n]$ and $j \in [2n]$, we add a trader, denoted by $T_{i,j}$, who owns $P_{i,j}$ units of $H_i$ and is only interested in $G_j$.

At this moment, the property of price-regulating markets mentioned above implies that at any approximate equilibrium $\boldsymbol{\pi}$, the ratio of $\pi(H_{2i-1})$ and $\pi(H_{2i})$ is between $(1-\beta)/(1+\beta)$ and $(1+\beta)/(1-\beta)$ and the ratio of $\pi(G_{2i-1})$ and $\pi(G_{2i})$ is between $(1-\alpha)/(1+\alpha)$ and $(1+\alpha)/(1-\alpha)$, where we use $\pi(G)$ to denote the price of a good $G$ in $\boldsymbol{\pi}$. Here is some wishful thinking: If for every $i \in [n]$,

$$\pi(H_{2i-1}) + \pi(H_{2i}) = \pi(G_{2i-1}) + \pi(G_{2i}) = 2$$

---

[3]Usually in a polymatrix game the $2 \times 2$ block diagonal matrices are set to 0: $P_{2i-1,2i-1} = P_{2i-1,2i} = P_{2i,2i-1} = P_{2i,2i} = 0$ for all $i \in [n]$. We do not impose such a requirement to simplify the reduction from polymatrix games to markets later.

then we can extract a vector $\mathbf{x}$ from $\boldsymbol{\pi}$ as follows: For each $i \in [2n]$, let

$$x_i = \frac{\pi(H_i) - (1-\beta)}{2\beta}$$

It is clear that $\mathbf{x}$ is nonnegative; and $x_{2i-1} + x_{2i} = 1$ for all $i \in [n]$. The linear forms $\mathbf{x} \cdot \mathbf{P}_j$ appear in the market $M_\mathbf{P}$ as follows: The total money that traders $T_{i,j}$, $i \in [2n]$, spend on $G_j$ is

$$\sum_{i \in [2n]} P_{i,j} \cdot \pi(H_i) = \sum_{i \in [2n]} P_{i,j} \cdot \big(2\beta x_i + (1-\beta)\big) = 2\beta \cdot \mathbf{x}^T \cdot \mathbf{P}_j + (1-\beta) \sum_{i \in [2n]} P_{i,j}$$

Again with some wishful thinking, we assume that the sums $\sum_{i \in [2n]} P_{i,j}$ are the same for all $j$. As $\beta \gg \gamma$, $\mathbf{x}^T \cdot \mathbf{P}_{2j-1} > \mathbf{x}^T \cdot \mathbf{P}_{2j} + 1/n$ would imply that the total demand for $G_{2j-1}$ from traders $T_{i,2j-1}$, $i \in [2n]$ must be strictly larger than that of $G_{2j}$ from traders $T_{i,2j}$, $i \in [2n]$. To achieve an approximate equilibrium, the price-regulating market over $G_{2j-1}$ and $G_{2j}$ must demand strictly more $G_{2j}$ than $G_{2j-1}$, to balance the deficit. But this can only happen when $\pi(G_{2j-1})$ and $\pi(G_{2j})$ are $1 + \alpha$ and $1 - \alpha$, respectively.

However, what we really need to finish the construction is $\pi(H_{2j-1}) = 1 + \beta$ and $\pi(H_{2j-1}) = 1 - \beta$, so that $x_{2j-1} = 1$, $x_{2j} = 0$ and the Nash constraint is met. The big missing piece of the puzzle is then how to enforce at any approximate equilibrium the following *ratio amplification*:

$$\frac{\pi(G_{2j-1})}{\pi(G_{2j})} = \frac{1+\alpha}{1-\alpha} \quad \Rightarrow \quad \frac{\pi(H_{2j-1})}{\pi(H_{2j})} = \frac{1+\beta}{1-\beta}$$

It turns out that our goal can be achieved by adding carefully a long chain of copies of a non-monotone market $M$ as well as price-regulating markets and traders who transfer money between them (like the $T_{i,j}$'s above). For each $j \in [n]$ we add such a chain that starts from $G_{2j-1}, G_{2j}$ and ends at $H_{2j-1}, H_{2j}$. The non monotone markets together with price-regulating markets, can step-by-step amplify the ratio of two goods, either from $(1+\alpha)/(1-\alpha)$ to $(1+\beta)/(1-\beta)$ or from $(1-\alpha)/(1+\alpha)$ to $(1-\beta)/(1+\beta)$.

The tricky part of the construction is that all the actions happen in the local neighborhood of $M$, where the phenomenon of non-monotonicity appears. Once the chains are added to $M_\mathbf{P}$, we show that the wishful thinking assumed earlier actually holds, approximately though, and we get a polynomial-time reduction.

For Theorem 6 the major challenge is that we can no longer use the linear price-regulating markets, but only CES utilities with a fixed $\rho < -1$. Note that we used the following two properties of price-regulating markets: The price ratio is bounded between $(1-\alpha)/(1+\alpha)$ and $(1+\alpha)/(1-\alpha)$; and must be equal to one of them if the demand of $G_1$ from the price-regulating market is different from that of $G_2$. The continuous nature of CES utilities however, makes it difficult, if not impossible, to construct a CES market that behaves similarly.

Instead we use the simple two-good two-trader market $M$ from [Gje96] which is itself a non-monotone market with three isolated market equilibria. The high-level picture of the construction is similar to that of Theorem 7, in which we add to $M_\mathbf{P}$ a long chain of copies of the non-monotone market $M$ for each $j \in [n]$ starting from $G_{2j-1}, G_{2j}$ and ending at $H_{2j-1}, H_{2j}$. The proof of correctness, however, is more challenging for which we need to first prove a few global properties of $M$. With these properties we show that whenever the ratio of $\pi(G_{2j-1})$ and $\pi(G_{2j})$ deviates from 1 by a non-negligible amount, the chain would amplify the ratio step by step. By the end of the chain at $H_{2j-1}$ and $H_{2j}$, the ratio converges to one of two constants that correspond to the two nontrivial equilibria of $M$. Correctness of the reduction then follows.

# 3 From Polymatrix to Markets with Non-Monotone and Linear Utilities

We prove Theorem 7 in this section. Let $\mathcal{U}$ denote a non-monotone family of utilities and $M \in \mathcal{M}_\mathcal{U}$ denote a market that is non-monotone at a rational price vector $\boldsymbol{\pi}$. We use $k \geq 2$ to denote the number of goods in $M$. We also assume that the excess demand $Z_1(\boldsymbol{\pi})$ of $G_1$ at $\boldsymbol{\pi}$ is moderately computable.

## 3.1 Normalized Polymatrix Games

To prove Theorem 7, we give a polynomial-time reduction from **POLYMATRIX** to strongly connected markets in $\mathcal{M}_\mathcal{U}^*$. Let $\mathbf{P}$ be a rational $2n \times 2n$ matrix that has entries between 0 and 1.

We can normalize $\mathbf{P}$ to get a $2n \times 2n$ matrix $\mathbf{P}'$ as follows: For all $i \in [2n]$ and $j \in [n]$, set

$$P'_{i,2j-1} = 1/2 + (P_{i,2j-1} - P_{i,2j})/2 \quad \text{and} \quad P'_{i,2j} = 1/2 - (P_{i,2j-1} - P_{i,2j})/2$$

It is clear that $\mathbf{P}'$ is also a rational matrix with entries between 0 and 1. In addition, $\mathbf{P}'$ satisfies

$$P'_{i,2j-1} + P'_{i,2j} = 1, \quad \text{for all } i \in [2n] \text{ and } j \in [n]. \tag{4}$$

From the definition of $\epsilon$-well-supported Nash equilibria, it is easy to show that

**Lemma 2.** *For any $\epsilon \geq 0$, $\mathbf{P}$ and $\mathbf{P}'$ have the same set of $\epsilon$-well-supported Nash equilibria.*

From now on we assume, without loss of generality, that $\mathbf{P}$ is normalized, meaning that $\mathbf{P}$ satisfies (4).

## 3.2 Normalized Non-Monotone Markets

Note that in Example 2.2, 2.3, and 2.4, the market we construct not only is non-monotone at $\mathbf{1} = (1, 1)$ but also has $Z_1(\mathbf{1}) = 0$. (Indeed $\mathbf{1}$ is an equilibrium in all three examples.) The lemma below shows that this is not really a coincidence. Recall that $M \in \mathcal{M}_\mathcal{U}$ is a market that is non-monotone at a rational vector $\boldsymbol{\pi}$, with $k \geq 2$ goods, such that $Z_1(\boldsymbol{\pi})$ is moderately computable. We use $M$ and $\boldsymbol{\pi}$ to prove the following lemma:

**Lemma 3.1** (Normalized Non-Monotone Markets). *There exist two (not necessarily rational) positive constants $c$ and $d$ with the following property. Given any $\gamma > 0$, one can build a market $M_\gamma \in \mathcal{M}_\mathcal{U}$ with $k \geq 2$ goods $G_1, \ldots, G_k$, in time polynomial in $1/\gamma$, such that*

> *Let $f_\gamma(x)$ denote the excess demand function of $G_1$ when the price of $G_1$ is $1 + x$ and the prices of all other $(k-1)$ goods are $1 - x$. Then $f_\gamma$ is well defined over $[-c, c]$ with $|f_\gamma(0)| \leq \gamma$ and its derivative $f'_\gamma(0) = d > 0$. For any $x \in [-c, c]$, $f_\gamma(x)$ also satisfies*
>
> $$\bigl|f_\gamma(x) - f_\gamma(0) - dx\bigr| \leq |x/D|, \quad \text{where } D = \max\{20, 20/d\}.$$

*Proof.* First, we construct $M'$ from $M$ by scaling: For each trader with utility function $u$ and initial endowment vector $\mathbf{w} \in \mathbb{Q}_+^k$, replace them by $w'_j = w_j \cdot \pi_j$ for every $j \in [k]$ and

$$u'(x_1, \ldots, x_k) = u\left(\frac{x_1}{\pi_1}, \ldots, \frac{x_k}{\pi_k}\right)$$

Since $\boldsymbol{\pi}$ is rational and positive, we have $M' \in \mathcal{M}_\mathcal{U}$. It is also easy to verify that $M'$ now is non-monotone at $\mathbf{1}$. Let $g(x)$ denote the excess demand function of $G_1$ when the price of $G_1$ is $1 + x$ and the prices of all other goods are $1 - x$, then by the definition of non-monotone markets, there exist two positive constants $c$ and $d$ such that $g$ is well defined over $[-c, c]$, $g(0) \geq 0$ and $g'(0) = d > 0$. The latter follows from the fact that the excess demand at $(1 + x, 1 - x, \ldots, 1 - x)$ is the same as that at $((1 + x)/(1 - x), 1, \ldots, 1)$. As $d$ (and thus, $D$) is a constant, it follows from $g'(0) = d$ that by setting $c$ to be a small enough constant:

$$\big|g(x) - g(0) - dx\big| \leq |x/D|, \quad \text{for all } x \in [-c, c].$$

Next, let $Z'_1 = g(0)$ denote the excess demand of $G_1$ in $M'$ at $\mathbf{1}$. Then $Z'_1 = \pi_1 \cdot Z_1(\boldsymbol{\pi})$ and thus, $Z'_1$ is also moderately computable. Given any $\gamma > 0$, we compute a $\gamma$-rational approximation $z$ of $Z'_1$. We assume without loss of generality, that $z$ is nonnegative; otherwise, simply set $z = 0$. Finally, we construct $M_\gamma$ from $M'$ by adding a trader with $z$ units of $G_1$ who is only interested in $G_k$.

Let $f_\gamma(x)$ denote the excess demand function of $G_1$ in $M_\gamma$, when the price of $G_1$ is $1 + x$ and all other goods have price $1 - x$. The construction of $M_\gamma$ then implies that $f_\gamma(x) = g(x) - z$ and thus, $|f_\gamma(0)| \leq \gamma$. It follows that $M_\gamma$ and $f_\gamma$ satisfy all the desired properties with respect to constants $c$ and $d$ above. □

### 3.3 Our Construction

Given a normalized $2n \times 2n$ polymatrix game $\mathbf{P}$, we construct a market $M_{\mathbf{P}} \in \mathcal{M}^*_{\mathcal{U}}$ in polynomial time (in the input size of $\mathbf{P}$) as follows. First of all, the two main building blocks of $M_{\mathbf{P}}$ are

**Normalized Non-Monotone Market**: We use the following notation. Given two positive rational numbers $\mu$ and $\gamma$, we use $\mathbf{NM}\,(\mu, \gamma, G_1, \ldots, G_k)$ to denote the creation of the following set of traders in $M_{\mathbf{P}}$. First, we make a new copy of $M_\gamma$ in which the $k$ goods that they are interested in are $G_1, \ldots, G_k$. Then for each trader in $M_\gamma$ with utility function $u(x_1, \ldots, x_k)$ and endowment $\mathbf{w} = (w_1, \ldots, w_k)$, where we let $x_j$ denote the amount of $G_j$ she buys and let $w_j$ denote the amount of $G_j$ she owns, replace $\mathbf{w}$ by $\mu \mathbf{w}$ and $u$ by

$$u'(x_1, \ldots, x_k) = u\left(\frac{x_1}{\mu}, \ldots, \frac{x_k}{\mu}\right)$$

It is clear that when $\mu$ is polynomially bounded and $\gamma$ is polynomially small in $n$, it takes time polynomial in $n$ to create these traders. Let $f_{\mu,\gamma}(x)$ denote the excess demand of $G_1$ when the price of $G_1$ is $1 + x$ and the prices of all other goods are $1 - x$, then we have $f_{\mu,\gamma}(x) = \mu \cdot f_\gamma(x)$. From the properties of $f_\gamma$ stated in Lemma 3.1, $f_{\mu,\gamma}$ is well defined over $[-c, c]$, satisfies $|f_{\mu,\gamma}(0)| \leq \mu\gamma$, and

$$\big|f_{\mu,\gamma}(x) - f_{\mu,\gamma}(0) - \mu dx\big| \leq |\mu x/D|, \quad \text{for all } x \in [-c, c], \text{ with } D = \max\{20, 20/d\}. \tag{5}$$

Recall here $c$ and $d$ are positive constants from Lemma 3.1, which do not depend on $\gamma$ or $\mu$.

**Price-Regulating Market**: Let $G_1, \ldots, G_\ell$ denote $\ell \geq 2$ goods in $M_{\mathbf{P}}$ and let $\lambda$ and $\alpha$ denote two positive rational numbers, where $\alpha < 1$. We use $\mathbf{PR}\,(\lambda, \alpha, G_1, \ldots, G_\ell)$ below to denote the creation of the following two traders $T_1, T_2$, and refer to the submarket they form as a *price-regulating market* [CDDT09, VY11].

The endowment of $T_1$ is $(\ell - 1)\lambda$ units of $G_1$, and the endowment of $T_2$ is $\lambda$ units of $G_2, \ldots, G_\ell$ each. Let $u_1$ and $u_2$ denote their utility functions, then both are linear functions and we have

$$u_1(x_1, \ldots, x_\ell) = (1 + \alpha)x_1 + \sum_{2 \leq j \leq \ell}(1 - \alpha)x_j \quad \text{and} \quad u_2(x_1, \ldots, x_\ell) = (1 - \alpha)x_1 + \sum_{2 \leq j \leq \ell}(1 + \alpha)x_j$$

where in both $u_1$ and $u_2$ we used $x_j$ to denote the amount of $G_j$ bought.

We will see that, when $\lambda$ is large enough and certain conditions are satisfied, a price-regulating market basically requires the prices of $G_2, \ldots, G_\ell$ to be the same when $\ell > 2$; and the ratio of prices of $G_1$ and $G_2$ to be between $(1-\alpha)/(1+\alpha)$ and $(1+\alpha)/(1-\alpha)$, in any approximate market equilibrium.

Other than these two building blocks, all other traders in $M_\mathbf{P}$ are indeed *single-minded*: Each of them is only interested in one specific good and spends all her budget on it. We use the following notation. First we say a trader is a $(\tau, G_1 : G_2)$-trader if her endowment consists of $\tau$ units of $G_1$ and she is only interested in $G_2$. Second we say a trader is a $(\tau, G_1, G_2 : G_3)$-trader if her endowment consists of $\tau$ units of $G_1$ and $G_2$ each, and she is only interested in $G_3$.

Now we describe the construction of $M_\mathbf{P}$. Without loss of generality, we always assume that $n = 2^t$ for some integer $t$. Then the market $M_\mathbf{P}$ consists of the following $O(ntk) = O(n \log n)$ goods:

$$\text{AUX}_i, \ G_{2i-1,j}, \ G_{2i,j}, \ \text{and} \ S_{i,\ell,r}, \quad \text{for } i \in [n], j \in [0:4t], \ell \in [4t] \text{ and } r \in [3:k].$$

Note that when $k = 2$, we do not have any of the goods $S_{i,\ell,r}$ in $M_\mathbf{P}$.

We also divide the goods, except the $\text{AUX}_i$'s, into the following $n(4t+1)$ groups $\{\mathcal{R}_{i,j}\}$, where $i \in [n]$ and $j \in [0:4t]$. For each $i \in [n]$ and $j \in [4t]$, we use $\mathcal{R}_{i,j}$ to denote the following group of $k$ goods:

$$\mathcal{R}_{i,j} = \{G_{2i-1,j}, G_{2i,j}, S_{i,j,3}, \ldots, S_{i,j,k}\}$$

and for each $i \in [n]$, we use $\mathcal{R}_{i,0}$ to denote the following group of two goods $\mathcal{R}_{i,0} : \{G_{2i-1,0}, G_{2i,0}\}$.

Next we list all the parameters used in the construction. We use $\alpha_i$ to denote $2^i/n^5$ for each $i \in [0:4t]$ so $\alpha_0 = 1/n^5$ and $\alpha_{4t} = 1/n$. Recall the positive constant $d$ from Lemma 3.1. We let $d^*$ denote a positive rational number (a constant) that satisfies

$$1 - 1/D \le d^* d \le 1, \quad \text{where } D = \max\{20, 20/d\}.$$

We list the parameters: $\beta = 1/n$, $\mu = d^*n$, $\tau = n^2$, $\gamma = 1/n^6$, $\xi = \epsilon nt$, $\delta = \epsilon t$ and $\epsilon = 1/n^8$.

**Construction of $M_\mathbf{P}$.** First we use **NM** and **PR** to build a *closed* economy over each group $\mathcal{R}_{i,j}$. Here by a closed economy over a group of goods, we mean a set of traders whose endowments consist of goods from this group only and they are interested in goods from this group only.

1. For each group $\mathcal{R}_{i,j}$, where $i \in [n]$ and $j \in [4t]$, we add a price-regulating market

    $$\mathbf{PR}\left(\tau, \alpha_j, G_{2i-1,j}, G_{2i,j}, S_{i,j,3}, \ldots, S_{i,j,k}\right)$$

    We also add a non-monotone market

    $$\mathbf{NM}\left(\mu, \gamma, G_{2i-1,j}, G_{2i,j}, S_{i,j,3}, \ldots, S_{i,j,k}\right)$$

    We will refer to them simply as *the* **PR** market and *the* **NM** market over $\mathcal{R}_{i,j}$, respectively.

2. For each group $\mathcal{R}_{i,0}$ of $\{G_{2i-1,0}, G_{2i,0}\}$, where $i \in [n]$, we add a price-regulating market

    $$\mathbf{PR}\left(\tau, \alpha_0, G_{2i-1,0}, G_{2i,0}\right)$$

    We will refer to it as *the* **PR** market over $\mathcal{R}_{i,0}$.

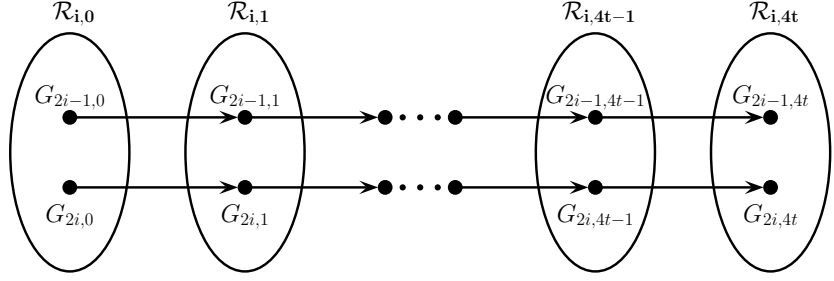

Figure 2: A chain of markets over groups $\mathcal{R}_{i,j}$ of goods, where $j \in [0 : 4t]$.

Next we add a number of single-minded traders who trade between different groups. The initial endowment of each such trader consists of $G_{2i-1,j}$ and $G_{2i,j}$ of a group $\mathcal{R}_{i,j}$ (one of them or both) and she is only interested in either $G_{2i'-1,j'}$ or $G_{2i',j'}$ of another group $\mathcal{R}_{i',j'}$, where $(i,j) \ne (i',j')$. We will refer to her as a trader who trades from $\mathcal{R}_{i,j}$ to $\mathcal{R}_{i',j'}$.

At the same time we construct a weighted directed graph $\mathcal{G} = (V, E)$ which will be used in the proof of correctness only. Here each group of goods $\mathcal{R}_{i,j}$ corresponds to a vertex in the graph $\mathcal{G}$ so $|V| = n(4t+1)$. Now given two groups $\mathcal{R}_{i,j}$ and $\mathcal{R}_{i',j'}$, we add an edge from $\mathcal{R}_{i,j}$ to $\mathcal{R}_{i',j'}$ in $\mathcal{G}$ whenever we create a set of traders who trade from $\mathcal{R}_{i,j}$ to $\mathcal{R}_{i',j'}$. Our construction below always makes sure that, whenever we create a set of traders who trade from $\mathcal{R}_{i,j}$ to $\mathcal{R}_{i',j'}$, the total initial endowment of these traders must consist of the same amount, say $w > 0$, of $G_{2i-1,j}$ and $G_{2i,j}$. We then set $w$ as the weight of this edge. We will prove, by the end of the construction that $\mathcal{G}$ is a strongly connected graph and for each group $\mathcal{R}_{i,j}$, its total in-weight is the same as its total out-weight.

Here is the construction:

1. For each $i \in [2n]$, we use $G_i$ to denote $G_{i,0}$ and $H_i$ to denote $G_{i,4t}$ for convenience. For each pair $i, j \in [n]$, we add to $M_{\mathbf{P}}$ the following four traders who trade from group $\mathcal{R}_{i,4t}$ to group $\mathcal{R}_{j,0}$: one $(P_{2i-1,2j-1}, H_{2i-1} : G_{2j-1})$-trader, one $(P_{2i-1,2j}, H_{2i-1} : G_{2j})$-trader, one $(P_{2i,2j-1}, H_{2i} : G_{2j-1})$ trader, and one $(P_{2i,2j}, H_{2i} : G_{2j})$-trader. Since $\mathbf{P}$ is normalized, we have

$$P_{2i-1,2j-1} + P_{2i-1,2j} = P_{2i,2j-1} + P_{2i,2j} = 1$$

    Thus, the total endowment of these four traders consists of one unit of $H_{2i-1}$ and $H_{2i}$ each, so we add an edge in $\mathcal{G}$ from $\mathcal{R}_{i,4t}$ to $\mathcal{R}_{j,0}$ with weight 1. At this moment, the total out-weight of each $\mathcal{R}_{i,4t}$ in $\mathcal{G}$ (a complete bipartite graph) is $n$, and the total in-weight of each $\mathcal{R}_{i,0}$ in $\mathcal{G}$ is $n$.

2. For each $i \in [n]$ and $j \in [0 : 4t-1]$, we add two traders who trade from group $\mathcal{R}_{i,j}$ to $\mathcal{R}_{i,j+1}$: one $(n, G_{2i-1,j} : G_{2i-1,j+1})$-trader and one $(n, G_{2i,j} : G_{2i,j+1})$-trader. We also add an edge in graph $\mathcal{G}$ from $\mathcal{R}_{i,j}$ to $\mathcal{R}_{i,j+1}$ with weight $n$.

This finishes the construction of $\mathcal{G}$. It is also easy to verify that $\mathcal{G}$ is a strongly connected graph, and every vertex (group) has both its total in-weight and out-weight equal to $n$.

Finally, we add traders between $\text{AUX}_j$ and $\mathcal{R}_{j,0}$ for each $j \in [n]$. Let $r_{2j-1}$ and $r_{2j}$ denote

$$r_{2j-1} = 2n - \sum_{i \in [2n]} P_{i,2j-1} > 0 \quad \text{and} \quad r_{2j} = 2n - \sum_{i \in [2n]} P_{i,2j} > 0 \tag{6}$$

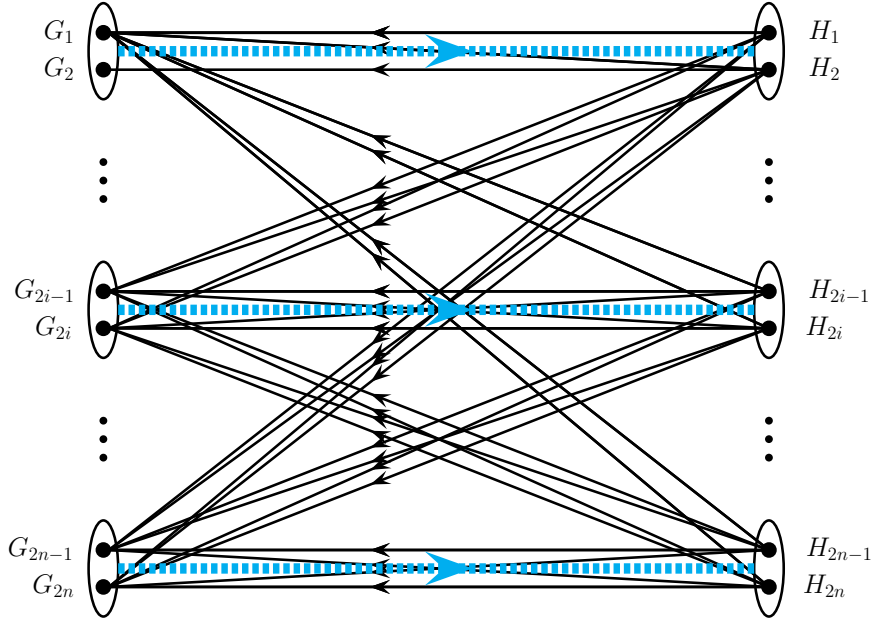

Figure 3: Market $M_{\mathbf{P}}$: Black arrows correspond to single minded-traders from $H_i$'s to $G_j$'s. Dashed arrows correspond to the chains of markets over $\mathcal{R}_{i,j}$'s pictured in Figure 2.

Because the polymatrix game $\mathbf{P}$ is normalized, note that

$$r_{2j-1} + r_{2j} = 2n, \quad \text{for any } j \in [n].$$

Recall $\beta = \alpha_{4t} = 1/n$. We add to $M_{\mathbf{P}}$ the following three traders: one $((1-\beta)r_{2j-1}, \text{AUX}_j : G_{2j-1})$-trader, one $((1-\beta)r_{2j}, \text{AUX}_j : G_{2j})$-trader, and one $((1-\beta)n, G_{2j-1}, G_{2j} : \text{AUX}_j)$-trader.

This finishes the construction of $M_{\mathbf{P}}$. It follows immediately from the strong connectivity of $\mathcal{G}$ that the economy graph of $M_{\mathbf{P}}$ is strongly connected and thus, $M_{\mathbf{P}}$ is a valid input of problem $\mathcal{U}$-**MARKET** and can be constructed from $\mathbf{P}$ in polynomial time. We also record the following properties of $M_{\mathbf{P}}$:

**Lemma 3.** *For each $i \in [n]$, the total supply of $\text{AUX}_i$ is $2(1-\beta)n$;*
*For each $i \in [2n]$, the total supply of $G_{i,0}$ is $n^2 + O(n)$;*
*For each $i \in [n]$ and $j \in [4t]$, the total supply of $G_{2i-1,j}$ is $(k-1)n^2 + O(n)$; and*
*For each $i \in [n]$, $j \in [4t]$ and $\ell \in [3:k]$, the total supply of $G_{2i,j}$ and $S_{i,j,\ell}$ is $n^2 + O(n)$.*

### 3.4 Proof of Correctness

First we introduce additively approximate market equilibria to simplify the presentation:

**Definition 10.** *We say $\boldsymbol{\pi}$ is an $\epsilon$-additively approximate market equilibrium of a market $M$, for some $\epsilon \geq 0$, if there exists a vector $\mathbf{z} \in Z(\boldsymbol{\pi})$ such that $z_j \leq \epsilon$ for all $j$.*

Let $\boldsymbol{\pi}$ denote a $(1/(kn^{10}))$-approximate equilibrium of $M_{\mathbf{P}}$, then by Lemma 3 it must be an $\epsilon$-additively approximate equilibrium of $M_{\mathbf{P}}$ as well, where $\epsilon = 1/n^8$. We prove in the rest of this section that given an $\epsilon$-additively approximate equilibrium $\boldsymbol{\pi}$ of $M_{\mathbf{P}}$, we can compute a $(1/n)$-well-supported Nash equilibrium of $\mathbf{P}$ in polynomial time. Theorem 7 then follows. In the proof below we use $\pi(G)$ to denote the price of a good $G$ in the price vector $\boldsymbol{\pi}$. We use $a = b \pm c$, where $c > 0$, to denote the inequality $b - c \leq a \leq b + c$.

First, from the **PR** markets in $M_\mathbf{P}$, we prove the following lemma:

**Lemma 4.** *Let $\pi$ denote an $\epsilon$-additively approximate equilibrium of $M_\mathbf{P}$ with $\epsilon = 1/n^8$, then we have*

$$\frac{1-\alpha_j}{1+\alpha_j} \leq \frac{\pi(G_{2i-1,j})}{\pi(G_{2i,j})} \leq \frac{1+\alpha_j}{1-\alpha_j}, \quad \text{for all } i \in [n] \text{ and } j \in [0:4t].$$

*Furthermore, we also have $\pi(G_{2i,j}) = \pi(S_{i,j,3}) = \cdots = \pi(S_{i,j,k})$ for all $i \in [n]$ and $j \in [4t]$.*

*Proof.* We consider the case when $i \in [n]$ and $j \in [4t]$ since the case of $j = 0$ is simpler.

Denote the two traders in the **PR** market over $\mathcal{R}_{i,j}$ by $T_1$ and $T_2$. We let

$$p_{\min} = \min\left\{\pi(G_{2i,j}), \pi(S_{i,j,3}), \ldots, \pi(S_{i,j,k})\right\} \quad \text{and} \quad p_{\max} = \max\left\{\pi(G_{2i,j}), \pi(S_{i,j,3}), \ldots, \pi(S_{i,j,k})\right\}$$

First, assume for contradiction that

$$\frac{1+\alpha_j}{\pi(G_{2i-1,j})} < \frac{1-\alpha_j}{p_{\min}}$$

It follows that neither $T_1$ nor $T_2$ is interested in $G_{2i-1,j}$ and they only buy goods from $\mathcal{R}_{i,j}$ that are priced at $p_{\min}$. Let $F_{\min} \subset \mathcal{R}_{i,j}$ denote the set of such goods then we have $G_{2i-1,j} \notin F_{\min}$. On the other hand, by the definition of $p_{\min}$, the budget of both $T_1$ and $T_2$ is at least $(k-1)\tau p_{\min}$. It follows that the total demand for goods in $F_{\min}$ is at least $2(k-1)\tau$. However, the total supply of goods in $F_{\min}$ is at most $(k-1)\tau + O(n)$, contradicting with the assumption that $\pi$ is an $\epsilon$-additively approximate equilibrium.

Next, assume for contradiction that

$$\frac{1-\alpha_j}{\pi(G_{2i-1,j})} > \frac{1+\alpha_j}{p_{\max}}$$

and we let $F_{\max} \subset \mathcal{R}_{i,j}$ denote the set of goods priced at $p_{\max}$. Then neither $T_1$ nor $T_2$ is interested in goods from $F_{\max}$ and they only buy goods from $\mathcal{R}_{i,j} - F_{\max}$. In particular, $T_2$ spends the part of budget she earns from selling $F_{\max}$ on goods in $\mathcal{R}_{i,j} - F_{\max}$ as well. As goods in $F_{\max}$ are the most expensive among $\mathcal{R}_{i,j}$, the demand for one of the goods in $\mathcal{R}_{i,j} - F_{\max}$ must be larger than the supply by $\Omega(\tau)$, contradicting with the assumption that $\pi$ is an $\epsilon$-additively approximate equilibrium.

Combining these two steps, we immediately get

$$\frac{1-\alpha_j}{1+\alpha_j} \leq \frac{\pi(G_{2i-1,j})}{p_{\max}} \leq \frac{\pi(G_{2i-1,j})}{\pi(G_{2i,j})} \leq \frac{\pi(G_{2i-1,j})}{p_{\min}} \leq \frac{1+\alpha_j}{1-\alpha_j} \tag{7}$$

In the rest of the proof, we show that $\pi(G_{2i,j}) = \pi(S_{i,j,3}) = \cdots = \pi(S_{i,j,k})$.

Assume for contradiction that this is not the case. Then $p_{\max} > p_{\min}$ which implies that neither $T_1$ nor $T_2$ is interested in $F_{\max}$. This leads us to the same contradiction, following the argument of the second step. The only difference is that $\pi(G_{2i-1,j})$ now might be larger than $p_{\max}$ but can be bounded using (7). □

From now on, for each group $\mathcal{R}_{i,j}$, $i \in [n]$ and $j \in [0:4t]$, we let $\pi_{i,j} = \pi(G_{2i-1,j}) + \pi(G_{2i,j})$.

Next note that only one trader is interested in $\text{AUX}_j$ and her budget is $(1-\beta)n\pi_{j,0}$. From this we have

**Lemma 5.** *Let $\pi$ denote an $\epsilon$-additively approximate market equilibrium of $M_\mathbf{P}$ with $\epsilon = 1/n^8$. If we scale $\pi$ so that $\pi_{j,0} = 2$ for some $j \in [n]$, then we have $\pi(\text{AUX}_j) \geq 1 - O(\epsilon/n)$*

*Proof.* As the total supply of $\text{AUX}_j$ is $2n(1-\beta)$, we have $2n(1-\beta) \leq (2n(1-\beta) + \epsilon)\pi(\text{AUX}_j)$. □

Now by using the strong connectivity of the graph $\mathcal{G}$ and the property that each vertex in $\mathcal{G}$ has the same total in-weight and out-weight, we prove the following lemma:

**Lemma 6.** *Let $\boldsymbol{\pi}$ denote an $\epsilon$-additively approximate equilibrium of $M_\mathbf{P}$. Let $\pi_{\max}$ and $\pi_{\min}$ denote*

$$\pi_{\max} = \max_{i,j} \pi_{i,j} \quad \text{and} \quad \pi_{\min} = \min_{i,j} \pi_{i,j}$$

*where the $\max$ and $\min$ are both taken over all $i \in [n]$ and $j \in [0:4t]$. If we scale $\boldsymbol{\pi}$ so that $\pi_{\min} = 2$ then we must have $\pi_{\max} = 2 + O(\epsilon t)$.*

*Proof.* For convenience, we use $u, v$ to denote vertices (groups) in $\mathcal{G}$. For each $u$ in $\mathcal{G}$, we use $\pi_u$ to denote $\pi_{i,j}$ if $u$ corresponds to $\mathcal{R}_{i,j}$. An edge from $u$ to $v$ of weight $w$ means traders from $u$ to $v$ spend $w\pi_u$ on $v$.

Now fix a vertex $v$ and let $\mathcal{R}$ denote its corresponding group of goods. By Lemma 4 we know the prices of all goods in $\mathcal{R}$ are close to each other. As $\boldsymbol{\pi}$ is an $\epsilon$-approximate equilibrium, we must have

$$\text{total money spent on goods in } \mathcal{R} - \text{total worth of goods in } \mathcal{R} \leq O(\epsilon k \pi_v) = O(\epsilon \pi_v) \tag{8}$$

For those traders in the closed economy over $\mathcal{R}$, by Walras' law, the money they spend on $\mathcal{R}$ is equal to the total worth of their initial endowments of $\mathcal{R}$, so they cancel each other in (8). We next list all other traders in $M_\mathbf{P}$ who either own goods in $\mathcal{R}$ at the beginning or are interested in goods in $\mathcal{R}$:

1. Let $N^-(v)$ denote the set of predecessors of $v$, then for each $u \in N^-(v)$, the amount of money that traders from $u$ to $v$ spend on $\mathcal{R}$ is $w_{u,v} \cdot \pi_u$, where $w_{u,v}$ denotes the weight of edge $(u,v)$.

2. Let $N^+(v)$ denote the set of successors of $v$, then for each $u \in N^+(v)$, the total worth of goods in $\mathcal{R}$ owned by traders from $v$ to $u$ at the beginning is $w_{v,u} \cdot \pi_v$.

3. If $\mathcal{R} = \mathcal{R}_{j,0}$ for some $j \in [n]$, then there are three more traders: one $((1-\beta)r_{2j-1}, \text{AUX}_j : G_{2j-1})$ trader, one $((1-\beta)r_{2j}, \text{AUX}_j : G_{2j})$-trader, and one $((1-\beta)n, G_{2j-1}, G_{2j} : \text{AUX}_j)$-trader.

Since these are all the traders in $M_\mathbf{P}$ relevant to goods in $\mathcal{R}$, from Lemma 5 and (8), we have

$$\sum_{u \in N^-(v)} w_{u,v} \cdot \pi_u - \sum_{u \in N^+(v)} w_{v,u} \cdot \pi_v \leq O(\epsilon \pi_v), \quad \text{for each } v \in V. \tag{9}$$

Now we use (9) to prove the lemma:

1. First of all, each group $\mathcal{R}_{i,j}$, where $i \in [n]$ and $j \in [4t-1]$, has exactly one predecessor $\mathcal{R}_{i,j-1}$ and one successor $\mathcal{R}_{i,j+1}$, both with weight $n$. From (9), we have

$$\pi_{i,j-1} - \pi_{i,j} \leq O(\epsilon \pi_{i,j}/n), \quad \text{for all } i \in [n] \text{ and } j \in [4t-1]. \tag{10}$$

2. Next, each group $\mathcal{R}_{i,4t}$, where $i \in [n]$, has only one predecessor $\mathcal{R}_{i,4t-1}$ with weight $n$, and $n$ successors each with weight 1. From (9), we have

$$\pi_{i,4t-1} - \pi_{i,4t} \leq O(\epsilon \pi_{i,4t}/n), \quad \text{for all } i \in [n]. \tag{11}$$

3. Finally, each group $\mathcal{R}_{i,0}$, where $i \in [n]$, has $n$ predecessors $\{\mathcal{R}_{\ell,4t}\}_{\ell \in [n]}$, all of weight 1, and has one successor $\mathcal{R}_{i,1}$ with weight $n$. From (9), we have

$$\sum_{\ell \in [n]} \pi_{\ell,4t} - n\pi_{i,0} \leq O(\epsilon \pi_{i,0}), \quad \text{for all } i \in [n]. \tag{12}$$

Let $\pi_{i,j} = \pi_{\min} = 2$ after scaling and $\pi_{x,y} = \pi_{\max}$. Using (10) and (11), we have

$$\pi_{i,0} \leq \big(1 + O(\epsilon/n)\big)^{4t} \cdot \pi_{i,j} = 2\big(1 + O(\epsilon t/n)\big) = 2 + O(\epsilon t/n)$$

where we used the fact that $\epsilon t/n \ll 1$. Similarly, we also have

$$\pi_{x,4t} \geq \big(1 + O(\epsilon/n)\big)^{-4t} \cdot \pi_{\max} \geq \big(1 - O(\epsilon t/n)\big) \pi_{\max}$$

Combining these two bounds with (12), we get

$$(n + O(\epsilon))(2 + O(\epsilon t/n)) \geq (n + O(\epsilon))\pi_{i,0} \geq \sum_{\ell \in [n]} \pi_{\ell,4t} \geq 2(n-1) + \big(1 - O(\epsilon t/n)\big)\pi_{\max}$$

Solving it for $\pi_{\max}$ gives us $\pi_{\max} \leq 2 + O(\epsilon t)$, and the lemma is proven. □

Using Lemma 6, we can also prove the following upper bound for $\pi(\text{AUX}_j)$:

**Lemma 7.** *Let $\boldsymbol{\pi}$ denote an $\epsilon$-additively approximate market equilibrium of $M_{\mathbf{P}}$ with $\epsilon = 1/n^8$. If we scale $\boldsymbol{\pi}$ so that $\pi_{j,0} = 2$ for some $j \in [n]$, then we have $\pi(\text{AUX}_j) \leq 1 + O(\epsilon t)$.*

*Proof.* We revisit (8). Let $v$ denote the vertex that corresponds to $\mathcal{R}_{j,0}$.

Plugging in (8) the list of traders enumerated in the proof of Lemma 6, we have

$$\sum_{\ell \in [n]} \pi_{\ell,4t} + 2n(1 - \beta) \cdot \pi(\text{AUX}_j) - n\pi_{j,0} - (1 - \beta)n\pi_{j,0} \leq O(\epsilon \pi_{j,0})$$

The lemma then follows directly from Lemma 6. □

From now on, we use $\boldsymbol{\pi}$ to denote the scaled price vector with $\pi_{\min} = 2$. By Lemma 5, 6 and 7,

$$2 \leq \pi_{i,j} = \pi(G_{2i-1,j}) + \pi(G_{2i,j}) \leq 2 + O(\epsilon t) \quad \text{and} \quad \pi(\text{AUX}_i) = 1 \pm O(\epsilon t) \tag{13}$$

for all $i \in [n]$ and $j \in [0:4t]$. For convenience, we let $\delta = \epsilon t$.

Recall that we use $H_i$ to denote the good $G_{i,4t}$. For each $i \in [n]$, we let

$$\theta_i = \frac{\pi(H_{2i-1}) + \pi(H_{2i})}{2}$$

From (13) we get the following corollary:

**Corollary 2.** *For every $i \in [n]$, we have $1 \leq \theta_i \leq 1 + O(\delta)$.*

Next we use Walras' law to show that the excess demand of each good is close to $0$ from both sides:

**Lemma 8.** *If $\boldsymbol{\pi}$ is an $\epsilon$-additively approximate equilibrium of $M_{\mathbf{P}}$, then there exists a $\mathbf{z} \in Z(\boldsymbol{\pi})$ such that*

$$|\mathbf{z}|_\infty \leq O(\epsilon nt)$$

*Proof.* Given a vector $\mathbf{z} \in Z(\boldsymbol{\pi})$ and a good $G$ in $M_{\mathbf{P}}$, we let $z(G)$ denote the excess demand of $G$ in $\mathbf{z}$. By definition, we know there exists a vector $\mathbf{z} \in Z(\boldsymbol{\pi})$ such that $z(G) \le \epsilon$ for all $G$, thus $|z(G)| \le \epsilon$ for goods $G$ with positive excess demand. By Walras' law, we also have $\mathbf{z} \cdot \boldsymbol{\pi} = 0$. By Lemma 4, 5, 6 and 7 we know that all prices are close to each other. Since the total number of goods in $M_{\mathbf{P}}$ is $O(nt)$ and $z(G) \le \epsilon$ for all $G$, it follows from Walras' law that $|z(G)| \le O(\epsilon nt)$ for all $G$ with negative excess demand. $\square$

From now on, we let $\xi = \epsilon nt = \log n/n^7$.

Now we are ready to recover a $(1/n)$-well-supported Nash equilibrium of the polymatrix game $\mathbf{P}$ from the price vector $\boldsymbol{\pi}$. Set $\mathbf{x}$ to be the following $2n$-dimensional nonnegative vector:

$$x_{2i-1} = \frac{\pi(H_{2i-1}) - (1-\beta)\theta_i}{2\beta\theta_i} \quad \text{and} \quad x_{2i} = \frac{\pi(H_{2i}) - (1-\beta)\theta_i}{2\beta\theta_i} \tag{14}$$

Recall that $\beta = \alpha_{4t} = 1/n$. It is easy to verify that $x_{2i-1} + x_{2i} = 1$ for each $i \in [n]$. Here $x_i \ge 0$ follows directly from Lemma 4. To finish the proof, we prove the following theorem:

**Theorem 9.** *When $n$ is sufficiently large, $\mathbf{x}$ built above is a $(1/n)$-well-supported Nash equilibrium of $\mathbf{P}$.*

To prove the main theorem, we need the following key lemma. Recall that $G_i$ denotes the good $G_{i,0}$.

**Lemma 9.** *For every $i \in [2n]$, we have*

$$\frac{1+\alpha_0}{\pi(G_{2i-1})} = \frac{1-\alpha_0}{\pi(G_{2i})} \Rightarrow \frac{1+\beta}{\pi(H_{2i-1})} = \frac{1-\beta}{\pi(H_{2i})} \quad \text{and}$$

$$\frac{1-\alpha_0}{\pi(G_{2i-1})} = \frac{1+\alpha_0}{\pi(G_{2i})} \Rightarrow \frac{1-\beta}{\pi(H_{2i-1})} = \frac{1+\beta}{\pi(H_{2i})}$$

Before proving Lemma 9, we use it to prove Theorem 9:

*Proof of Theorem 9.* Assume for contradiction that the vector $\mathbf{x}$ we construct from $\boldsymbol{\pi}$ in (14) is not a $(1/n)$ well-supported Nash equilibrium of $\mathbf{P}$. Without loss of generality, we assume that

$$\mathbf{x}^T \cdot \mathbf{P}_1 > \mathbf{x}^T \cdot \mathbf{P}_2 + 1/n \tag{15}$$

where $\mathbf{P}_1$ and $\mathbf{P}_2$ denote the first and second columns of $\mathbf{P}$, respectively, but $x_2 > 0$. To reach a contradiction, by Lemma 9, it suffices to show that (15) implies that

$$\frac{1+\alpha_0}{\pi(G_1)} = \frac{1-\alpha_0}{\pi(G_2)} \tag{16}$$

because it then implies that $(1+\beta)/\pi(H_1) = (1-\beta)/\pi(H_2)$ and thus, $x_2 = 0$ by (14).

To this end, we compare the total money spent by all traders in $M_{\mathbf{P}}$ on $G_1$ and $G_2$ of group $\mathcal{R}_{1,0}$, except the two traders in the **PR** market over $\mathcal{R}_{1,0}$. Here is the list of such traders:

1. For each $i \in [2n]$, there is a $(P_{i,1}, H_i : G_1)$-trader. The total money these traders spend on $G_1$ is

$$\sum_{i \in [2n]} P_{i,1} \cdot \pi(H_i) = \sum_{i \in [2n]} P_{i,1} \cdot (1 - \beta + 2\beta \cdot x_i) \cdot \theta_{\lceil i/2 \rceil}$$

2. For each $i \in [2n]$, there is a $(P_{i,2}, H_i : G_2)$-trader. The total money these traders spend on $G_2$ is

$$\sum_{i \in [2n]} P_{i,2} \cdot \pi(H_i) = \sum_{i \in [2n]} P_{i,2} \cdot (1 - \beta + 2\beta \cdot x_i) \cdot \theta_{\lceil i/2 \rceil}$$

3. Recall the definition of $r_{2j-1}$ and $r_{2j}$ in (6).
   There is one $((1-\beta)r_1, \text{AUX}_1 : G_1)$-trader, who spends her budget $(1-\beta)r_1 \cdot \pi(\text{AUX}_1)$ on $G_1$.
   There is one $((1-\beta)r_2, \text{AUX}_1 : G_2)$-trader, who spends her budget $(1-\beta)r_2 \cdot \pi(\text{AUX}_1)$ on $G_2$.

We denote by $M_1$ (or $M_2$) the total money these traders spend on $G_1$ (or $G_2$, respectively). Then

$$M_1 = \sum_{i \in [2n]} P_{i,1} \cdot (1 - \beta + 2\beta \cdot x_i) \cdot \theta_{\lceil i/2 \rceil} + (1-\beta)r_1 \cdot \pi(\text{AUX}_1)$$

Plugging in $\theta_{\lceil i/2 \rceil} \geq 1$, $\pi(\text{AUX}_1) \geq 1 - O(\delta)$ and the definition of $r_1$, we get

$$M_1 \geq 2n(1-\beta) + 2\beta \cdot \mathbf{x}^T \cdot \mathbf{P}_1 - O(n\delta)$$

Similarly, we also have the total money spent on $G_2$ is

$$M_2 = \sum_{i \in [2n]} P_{i,2} \cdot (1 - \beta + 2\beta \cdot x_i) \cdot \theta_{\lceil i/2 \rceil} + (1-\beta)r_2 \cdot \pi(\text{AUX}_1)$$

Plugging in $\theta_{\lceil i/2 \rceil} \leq 1 + O(\delta)$, $\pi(\text{AUX}_2) \leq 1 + O(\delta)$ and the definition of $r_2$, we get

$$M_2 \leq 2n(1-\beta) + 2\beta \cdot \mathbf{x}^T \cdot \mathbf{P}_2 + O(n\delta)$$

Combining these two bounds with (15), we get

$$M_1 \geq M_2 + 2\beta \cdot (1/n) - O(n\delta) = M_2 + \Theta(\beta/n)$$

since $\beta/n = 1/n^2 \gg n\delta$. So the difference between the demands for $G_1$ and $G_2$ from these traders is:

$$\frac{M_1}{\pi(G_1)} - \frac{M_2}{\pi(G_2)} \geq \frac{M_2 + \Theta(\beta/n)}{\pi(G_1)} - \frac{M_2(1+\alpha_0)}{\pi(G_1)(1-\alpha_0)} = \frac{\Theta(\beta/n)}{\pi(G_1)} - \frac{M_2}{\pi(G_1)} \cdot \frac{2\alpha_0}{1-\alpha_0} = \omega(\xi)$$

where the last inequality used the fact that $M_2 = O(n)$, $\alpha_0 = 1/n^5$, $\beta = 1/n$, and $\xi = \log n/n^7$.

The only other traders interested in $G_1, G_2$ are the two traders in the price-regulating market over $\mathcal{R}_{1,0}$ denoted by $T_1$ and $T_2$. Also from the construction of $M_\mathbf{P}$, the total supply of $G_1$ is exactly the same as that of $G_2$. By Lemma 8, we know that the total demand of $G_1$ from $T_1$ and $T_2$ must be strictly smaller than the total demand of $G_2$ from them, which in turn implies that the total demand of $G_1$ from $T_1$ and $T_2$ must be strictly smaller than the total supply of $G_1$ from $T_1$ and $T_2$ by Walras' law.

Assume (16) does not hold, then by Lemma 4 we must have

$$\frac{1+\alpha_0}{\pi(G_1)} > \frac{1-\alpha_0}{\pi(G_2)}$$

This implies that the (unique) optimal bundle of $T_1$ is to buy back her initial endowment of $G_1$ and thus, the total demand of $T_1$ and $T_2$ for $G_1$ is at least as much as the total supply of $G_1$ from $T_1$ and $T_2$, contradicting with Lemma 8. The theorem then follows. □

Finally, we prove Lemma 9. Using induction, it suffices to prove the following lemma:

**Lemma 10.** *For every $i \in [n]$ and $j \in [4t]$, we have*

$$\frac{1+\alpha_{j-1}}{\pi(G_{2i-1,j-1})} = \frac{1-\alpha_{j-1}}{\pi(G_{2i,j-1})} \Rightarrow \frac{1+\alpha_j}{\pi(G_{2i-1,j})} = \frac{1-\alpha_j}{\pi(G_{2i,j})} \quad \text{and}$$

$$\frac{1-\alpha_{j-1}}{\pi(G_{2i-1,j-1})} = \frac{1+\alpha_{j-1}}{\pi(G_{2i,j-1})} \Rightarrow \frac{1-\alpha_j}{\pi(G_{2i-1,j})} = \frac{1+\alpha_j}{\pi(G_{2i,j})}$$

To this end, we examine a group $\mathcal{R}_{i,j}$, $i \in [n]$ and $j \in [4t]$, more closely. For convenience, we scale the price vector $\boldsymbol{\pi}$ again so that $\pi_{i,j} = \pi(G_{2i-1,j}) + \pi(G_{2i,j}) = 2$. Note that what we need to prove in Lemma 10 remains the same after scaling. We are interested in the total demand of $G_{2i-1,j}$ from all traders in $M_\mathbf{P}$ except those two traders in the price-regulating market **PR** over $\mathcal{R}_{i,j}$.

First of all, for the **NM** market over $\mathcal{R}_{i,j}$, we let $f(x)$ denote the excess demand (within the **NM** market only) for $G_{2i-1,j}$, when the price of $G_{2i-1,j}$ is $1 + x$ and the prices of $G_{2i,j}, S_{i,j,3}, \ldots, S_{i,j,k}$ are $1 - x$. We let $\mu = d^*n = O(n)$ and $\gamma = 1/n^6$, then $f$ is exactly $f_{\mu,\gamma}$ in (5) and satisfies

$$|f(0)| = O(\mu\gamma) \quad \text{and} \quad |f(x) - f(0) - \mu dx| \le |\mu x/D|, \quad \text{for all } x \in [-c, c] \tag{17}$$

where $D = \max\{20, 20/d\}$ and $c > 0$ are both constants independent of $n$. So when $n$ is sufficiently large, we have $\beta = \alpha_{4t} = 1/n \ll c$. Next we use $h(x, y)$ to denote the following function:

$$h(x, y) = \text{excess demand of } G_{2i-1,j} \text{ from all traders except those two in the } \mathbf{PR} \text{ over } \mathcal{R}_{i,j}$$

when the price of $G_{2i-1,j-1}$ is $1 + y$, the price of $G_{2i-1,j}$ is $1 + x$, and the prices of $G_{2i,j}, S_{i,j,3}, \ldots, S_{i,j,k}$ are $1 - x$. By Lemma 4 and 6, we are only interested in $x, y$ satisfying $|x| \le \alpha_j$ and $|y| \le \alpha_{j-1} + O(\delta)$.

Using $f$, we obtain the following more explicit form of $h$ since other than the **NM** and **PR** markets over $\mathcal{R}_{i,j}$, there are $n$ units of $G_{2i-1,j}$ and only one $(n, G_{2i-1,j-1} : G_{2i-1,j})$-trader interested in $G_{2i-1,j}$:

$$h(x, y) = f(x) + \frac{n(1+y)}{1+x} - n = f(x) - \frac{nx}{1+x} + \frac{ny}{1+x}$$

We now use (17) to prove the following useful lemma about $h(x, y)$:

**Lemma 11.** *For all $x$ and $y$ with $|x| \le 3|y|$ and $|y| = \alpha_{j-1} \pm O(\delta)$, we have*

$$h(x, y) > ny/2 \text{ if } y > 0 \quad \text{and} \quad h(x, y) < ny/2 \text{ if } y < 0$$

*Proof.* For $x/(1+x)$, we can approximate it by $x$ when $|x|$ is small:

$$|x/(1+x) - x| = x^2/(1+x) \le 2x^2$$

For $f(x)$, by (17) we can approximate it by $\mu dx$:

$$|f(x) - \mu dx| \le |f(0)| + |\mu x/D| = O(\mu\gamma) + |nx/20|$$

where we used $D = \max\{20, 20/d\}$ and the assumption that $1 - 1/D \le d^*d \le 1$.

As a result, we can approximate $f(x) - nx/(1+x)$ using $(\mu d - n)x$ where the absolute value of error is bounded by $2nx^2 + O(\mu\gamma) + |nx/20|$. On the other hand, by the definition of $d^*$ we have

$$-nx/20 \le -nx/D \le (\mu d - n)x \le 0$$

Therefore, we can bound the absolute value $|f(x) - nx/(1+x)|$ by

$$2nx^2 + O(\mu\gamma) + |nx/10|$$

From $\mu = O(n), \gamma = 1/n^6, |x| \le 3|y|$ and $|y| = \alpha_{j-1} \pm O(\delta)$, this can be trivially bounded from above by $|ny/3|$. The lemma then follows since $|ny/(1+x)| > |5ny/6|$. $\square$

We are now ready to prove Lemma 10:

*Proof of Lemma 10.* We start by scaling $\pi$ so that $\pi(G_{2i-1,j}) + \pi(G_{2i,j}) = 2$. Depending on whether

$$\frac{1+\alpha_{j-1}}{\pi(G_{2i-1,j-1})} = \frac{1-\alpha_{j-1}}{\pi(G_{2i,j-1})} \quad \text{or} \quad \frac{1-\alpha_{j-1}}{\pi(G_{2i-1,j-1})} = \frac{1+\alpha_{j-1}}{\pi(G_{2i,j-1})}$$

we have either $y = \alpha_{j-1} \pm O(\delta)$ or $-\alpha_{j-1} \pm O(\delta)$ by Lemma 4 and Lemma 6. Moreover, from Lemma 4 we have $|x| \le \alpha_j$ and thus, it always holds that $|x| \le 3|y|$ since $\alpha_j = 2\alpha_{j-1} = \omega(\delta)$.

Therefore we can conclude from Lemma 11 that either

$$h(x,y) > ny/2 \quad \text{or} \quad h(x,y) < ny/2$$

respectively. Because $n\alpha_{j-1} \ge n\alpha_{4t} \gg \xi$, Lemma 8 implies that the excess demand of $G_{2i-1,j}$, within the price-regulating market **PR** over $\mathcal{R}_{i,j}$, must be either strictly negative or strictly positive, respectively.

When it is strictly negative, we know that the first trader $T_1$ of the price-regulating market do not spend all her budget on $G_{2i-1,j}$. This combined with Lemma 4 implies that

$$\frac{1+\alpha_j}{\pi(G_{2i-1,j})} = \frac{1-\alpha_j}{\pi(G_{2i,j})}$$

Similarly when it is strictly positive, we know that the second trader $T_2$ must be interested in $G_{2i-1,j}$ as well. This combined with Lemma 4 implies that

$$\frac{1-\alpha_j}{\pi(G_{2i-1,j})} = \frac{1+\alpha_j}{\pi(G_{2i,j})}$$

The lemma then follows. $\square$

## 4  From Polymatrix to Markets with CES Utilities

We prove Theorem 6 in this section. Let $\rho < -1$ denote a fixed rational number. Let $\mathbf{P}$ denote a normalized $2n \times 2n$ polymatrix game. First, we examine more closely the non-monotone market described in Example 2.2 with two goods and two traders. We then describe the construction of a strongly connected market $M_\mathbf{P}$ in which every trader uses a CES utility function of parameter $\rho$. Finally, we show that given any approximate equilibrium $\pi$ of $M_\mathbf{P}$, one can recover a well-supported Nash equilibrium of $\mathbf{P}$ efficiently. As we will see, the construction of $M_\mathbf{P}$ is similar to that of Section 3, but the proof of correctness is more involved.

## 4.1 Properties of the Excess Spending Function of Example 2.2

We need the following notion of *excess spending*. Let $S$ denote a subset of traders. Given $\boldsymbol{\pi}$ and a good $G$, the excess spending on $G$ from traders in $S$ is the product of $\pi(G)$ and the excess demand of $G$ from $S$:

$$\big(\text{total demand of } G \text{ from } S - \text{total supply of } G \text{ from } S\big) \times \pi(G)$$

For convenience we always use $r > 1$ to denote $-\rho$.

We use $M$ to denote the following market described earlier in Example 2.2 with two goods $G_1, G_2$ and two traders $T_1, T_2$. Here $T_1$ has 1 unit of $G_1$, $T_2$ has 1 unit of $G_2$, and their utility functions are

$$u_1(x_1, x_2) = \big(\alpha \cdot x_1^\rho + x_2^\rho\big)^{1/\rho} \quad \text{and} \quad u_2(x_1, x_2) = \big(x_1^\rho + \alpha \cdot x_2^\rho\big)^{1/\rho}$$

for some rational number $\alpha > 0$. By using the KKT conditions, one can prove that, give any positive prices $\pi_1$ and $\pi_2$, the optimal bundles $(x_{1,1}, x_{1,2})$ and $(x_{2,1}, x_{2,2})$ are unique and must satisfy

$$\frac{x_{1,1}}{x_{1,2}} = \left(\alpha \cdot \frac{\pi_2}{\pi_1}\right)^{1/(1+r)} \quad \text{and} \quad \frac{x_{2,1}}{x_{2,2}} = \left(\frac{1}{\alpha} \cdot \frac{\pi_2}{\pi_1}\right)^{1/(1+r)} \tag{18}$$

It is clear that $(1, 1)$ is a market equilibrium of $M$.

From now on, we always assume that $\alpha$ is a positive rational number such that $a = \alpha^{1/(r+1)}$ is rational as well. We are interested in the excess spending $f(x)$ on $G_1$ from $T_1$ and $T_2$, when the prices $\pi_1 = 1 + x$ and $\pi_2 = 1 - x$ with $x \in (-1, 1)$. Let $m_{i,j}$ denote the amount of money $T_i$ spends on $G_j$, then

$$\frac{m_{1,1}}{m_{1,2}} = a \left(\frac{\pi_1}{\pi_2}\right)^{r/(1+r)} \quad \text{and} \quad \frac{m_{2,1}}{m_{2,2}} = \frac{1}{a}\left(\frac{\pi_1}{\pi_2}\right)^{r/(1+r)}$$

We also have $m_{1,1} + m_{1,2} = \pi_1$. This gives us the following explicit form of $m_{1,1}$, as a function of $x$:

$$m_{1,1}(x) = \frac{\pi_1}{1 + \frac{1}{a}\left(\frac{\pi_2}{\pi_1}\right)^{\frac{r}{1+r}}} = \frac{1 + x}{1 + \frac{1}{a}\left(\frac{1-x}{1+x}\right)^{\frac{r}{1+r}}}$$

Similarly, we have the following explicit form of $m_{2,1}$, as a function of $x$:

$$m_{2,1}(x) = \frac{\pi_2}{1 + a\left(\frac{\pi_2}{\pi_1}\right)^{\frac{r}{1+r}}} = \frac{1 - x}{1 + a\left(\frac{1-x}{1+x}\right)^{\frac{r}{1+r}}}$$

The excess spending function $f(x)$ on $G_1$ from $T_1$ and $T_2$ is then

$$f(x) = m_{1,1}(x) + m_{2,1}(x) - (1 + x), \quad \text{for } x \in (-1, 1).$$

It is easy to show that $f(0) = 0$ and $f(x) = -f(-x)$ for any $x \in (-1, 1)$. From its symmetry, we have

$$f(x) = -f(-x) \;\Rightarrow\; f'(x) = f'(-x) \;\Rightarrow\; f''(x) = -f''(-x) \;\Rightarrow\; f''(0) = 0$$

Our first goal is to prove the following properties about the excess spending function $f$:

**Lemma 12.** *When $a > (r+1)/(r-1)$ is rational, $f'(0) > 0$ is rational and $f$ has three roots in $(-1, 1)$. Let $\{-\theta, 0, \theta\}$ denote the three roots of $f$ with $\theta > 0$, then we have $f'(\theta) < 0$.*

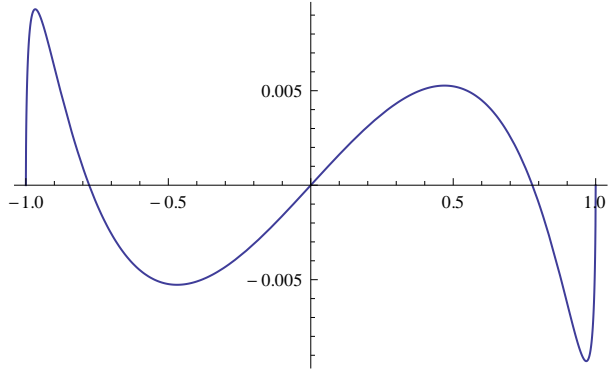

Figure 4: The excess spending function $f$

*Proof.* First we replace $x$ by the following variable $y$. Let

$$y^{1+r} = \frac{1-x}{1+x} \quad \text{and} \quad x = \frac{1-y^{1+r}}{1+y^{1+r}} \tag{19}$$

It suffices to show when $a > (r+1)/(r-1)$, the following function $p$ has three roots over $(0, +\infty)$:

$$p(y) = \frac{2}{(1+y^{1+r})(1+y^r/a)} + \frac{2y^{1+r}}{(1+y^{1+r})(1+ay^r)} - \frac{2}{1+y^{1+r}}$$

Let $q(y) = (1+y^{1+r})(1+y^r/a)(1+ay^r) \cdot p(y)$. Then it suffices to show that $q(y)$ has three roots:

$$q(y) = 2(1+ay^r) + 2y^{1+r}(1+y^r/a) - 2(1+y^r/a)(1+ay^r) = \frac{2}{a} y^r \cdot (y^{1+r} - ay^r + ay - 1)$$

Taking the derivative of $h(y) = y^{1+r} - ay^r + ay - 1$, we get

$$h'(y) = (r+1)y^r - ary^{r-1} + a$$

It is easy to see that $h(0) = -1 < 0$, $h(1) = 0$, and $h(y) \to +\infty$ when $y \to +\infty$. Moreover, we have

$$h'(1) = (r+1) - ar + a = (r+1) - a(r-1) < 0$$

when $a > (r+1)/(r-1)$. This immediately implies that $h$ has at least three roots in $(0, +\infty)$ and thus, $f$ has at least three roots in $(-1, 1)$. Next we show that $h$ has at most three roots. To see this, we have

$$h''(y) = r(r+1)y^{r-1} - ar(r-1)y^{r-2} = ry^{r-2}((r+1)y - a(r-1))$$

Therefore, there is a threshold $b = a(r-1)/(r+1) > 0$ such that $h''(y) > 0$ when $y > b$; and $h''(y) < 0$ when $y < b$. This implies that $h'(b)$ is the minimum of $h'$ over $[0, +\infty)$. It follows from $h'(b) \le h'(1) < 0$ that $h'$ has exactly one root in $(0, b)$ and exactly one root in $(b, +\infty)$. This implies that $h$ has at most three roots in $(0, +\infty)$ and thus, $f$ has at most three roots in $(-1, 1)$. As a result, $f$ has exactly three roots.

Let $\{-\theta, 0, \theta\}$ denote the three roots of $f$ with $\theta > 0$, then $\{y(-\theta), 1, y(\theta)\}$ are exactly the three roots

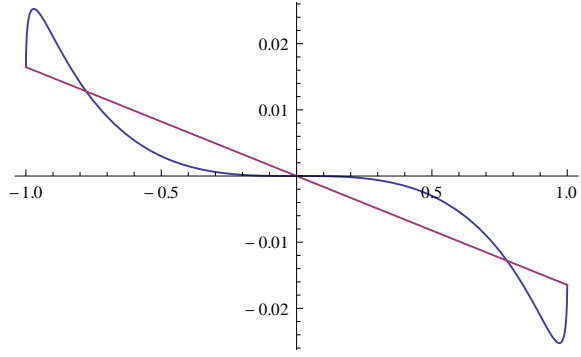

Figure 5: The function $g$ and the line $-\lambda x$

of $h$. From the proof above we also have $f'(0) > 0$ and $f'(\theta) < 0$. To see this, we have

$$f(x) = p(y(x)) \Rightarrow f'(x) = p'(y) \cdot \left(\frac{1}{1+r}\right) \cdot \left(\frac{1-x}{1+x}\right)^{\frac{-r}{1+r}} \cdot \frac{-2}{(1+x)^2}$$

This implies that $f'(0) = -2p'(1)/(1+r)$. Taking the derivative of

$$(1 + y^{1+r})(1 + y^r/a)(1 + ay^r) \cdot p(y) = 2y^r h(y)/a$$

and plugging in $h(1) = p(1) = 0$ we get $p'(1) = h'(1)/(1+a)^2 < 0$ and thus, $f'(0) > 0$ is rational. Next by using $h(y(\theta)) = 0$ and $h'(y(\theta)) > 0$, we can similarly show that $f'(\theta) < 0$. The lemma follows. □

From now on, we always assume that $a > (r+1)/(r-1)$ and use $\{-\theta, 0, \theta\}$ to denote the three roots of $f$ over $(-1, 1)$ with $\theta > 0$. Let $\lambda = f'(0)$, then we know that $\lambda$ is rational and positive. Let

$$g(x) = f(x) - \lambda x, \quad \text{for } x \in (-1, 1).$$

From the definition of $g(x)$, we have $g(0) = 0$, $g'(0) = 0$, and $g''(0) = 0$.

Next we show that when $a$ is chosen carefully, $g$ satisfies the following property:

**Lemma 13.** *Given any rational number $r > 1$, there is a rational number $a$ such that $a > (r+1)/(r-1)$ $\alpha = a^{1+r}$ is rational, and $g(x) < 0$ for all $x \in (0, 1)$. By the symmetry of $g$, $g(x) > 0$ for all $x \in (-1, 0)$.*

*Proof.* Assume for contradiction that there is an $x^* \in (0, 1)$ such that $g(x^*) \geq 0$, and $g(-x^*) \leq 0$. Similar to the proof of Lemma 12, we use $y$ in (19) to replace $x$. We are interested in $p(y)$ over $y \in (0, +\infty)$:

$$p(y) = \frac{2}{(1+y^{1+r})(1+y^r/a)} + \frac{2y^{1+r}}{(1+y^{1+r})(1+ay^r)} - \frac{2}{1+y^{1+r}} - \lambda \cdot \frac{1-y^{1+r}}{1+y^{1+r}}$$

By the definition of $p(y)$, we have

$$g(x) = p(y(x)) \Rightarrow p(1) = 0,\ p'(1) = 0 \text{ and } p''(1) = 0 \tag{20}$$

using the chain rule and the fact that $y'(x)$ is nonzero at $x = 0$. Let $y_1 = y(x^*)$ and $y_2 = y(-x^*)$, then we have $0 < y_1 < 1 < y_2$, $p(y_1) \geq 0$, and $p(y_2) \leq 0$. Next we use $q(y)$ to denote the following function:

$$q(y) = (1 + y^{1+r})(1 + y^r/a)(1 + ay^r) \cdot p(y)/2$$

then we have

$$q(y) = (1 + ay^r) + y^{1+r}(1 + y^r/a) - (1 + y^r/a)(1 + ay^r) - (\lambda/2)(1 - y^{1+r})(1 + y^r/a)(1 + ay^r)$$

By the definition of $q(y)$, we have $q(y_1) \geq 0$ and $q(y_2) \leq 0$. We use $u, v, w > 0$ to denote

$$u = \frac{\lambda}{2}, \quad v = \frac{a\lambda}{2} + \frac{\lambda}{2a} + \frac{1}{a} \quad \text{and} \quad w = 1 + \frac{\lambda}{2}$$

then we can rewrite $q(y)$ as follows:

$$q(y) = u \cdot y^{1+3r} + v \cdot y^{1+2r} - w \cdot y^{2r} + w \cdot y^{1+r} - v \cdot y^r - u$$

Taking its derivative, we get

$$q'(y) = u(1 + 3r) \cdot y^{3r} + v(1 + 2r) \cdot y^{2r} - 2wr \cdot y^{2r-1} + w(1 + r) \cdot y^r - vr \cdot y^{r-1}$$

Let $q'(y) = y^{r-1} \cdot s(y)$, then we have

$$s(y) = u(1 + 3r) \cdot y^{1+2r} + v(1 + 2r) \cdot y^{1+r} - 2wr \cdot y^r + w(1 + r) \cdot y - vr$$

Taking its derivative, we get

$$s'(y) = u(1 + 3r)(1 + 2r) \cdot y^{2r} + v(1 + 2r)(1 + r) \cdot y^r - 2wr^2 \cdot y^{r-1} + w(1 + r) \tag{21}$$

and its second-order derivative

$$s''(y) = 2ur(1 + 3r)(1 + 2r) \cdot y^{2r-1} + vr(1 + 2r)(1 + r) \cdot y^{r-1} - 2wr^2(r - 1) \cdot y^{r-2}$$

Let $s''(y) = y^{r-2} \cdot t(y)$, then we have

$$t(y) = 2ur(1 + 3r)(1 + 2r) \cdot y^{r+1} + vr(1 + 2r)(1 + r) \cdot y - 2wr^2(r - 1) \tag{22}$$

We prove properties about these functions. First we show that $s''(1)$ is indeed positive when $a$ is close enough to $(r + 1)/(r - 1)$. By (21), we have

$$s''(1) = 2ur(1 + 3r)(1 + 2r) + vr(1 + 2r)(1 + r) - 2wr^2(r - 1)$$

Let $c = a + 1/a$. Plugging in $v = cu + 1/a$ and $w = 1 + u$, we have

$$s''(1) = 2ur(1 + 3r)(1 + 2r) + (cu + 1/a)(1 + 2r)(1 + r)r - 2(1 + u)r^2(r - 1)$$

The trouble here is that $\lambda$ (and $u$) depends on the choice of $a$. But note that the coefficient of $u$ in $s''(1)$ is

$$2r(1 + 3r)(1 + 2r) + cr(1 + 2r)(1 + r) - 2r^2(r - 1) > 0$$

and $u$ is positive when $a > (r + 1)/(r - 1)$. The rest of $s''(1)$ is the following:

$$(1/a)(1 + 2r)(1 + r)r - 2r^2(r - 1)$$

Let $a = (1+\epsilon)(r+1)/(r-1)$. When $\epsilon$ goes to 0, the expression above converges to

$$r(r-1)(1+2r) - 2r^2(r-1) = r(r-1)(1+2r-2r) = r(r-1) > 0$$

Therefore there exists a positive rational number $a > (r+1)/(r-1)$ such that $s''(1) > 0$ and $\alpha = a^{1+r}$ is rational (note we do not care about the number of bits needed to encode it). We use such an $a$ from now on. From the definition of $q$ and $s$ from $p$ as well as the chain rule, one can show that $p(1) = p'(1) = p''(1) = 0$ (equation 20) implies that $q(1) = q'(1) = q''(1) = 0$ and $s(1) = s'(1) = 0$; Furthermore, since $s''(1) > 0$ we have $p'''(1) > 0$. Together with (20), we know there is a small enough $\epsilon > 0$ that satisfies:

$$p(1+\epsilon) > 0, \quad p(1-\epsilon) < 0 \quad \text{and} \quad y_1 < 1 - \epsilon < 1 + \epsilon < y_2$$

Recall that $p(y_1) \geq 0$ and $p(y_2) \leq 0$. By the definition of $q(y)$ from $p(y)$, we have

$$q(y_1) \geq 0, \quad q(1-\epsilon) < 0, \quad q(1+\epsilon) > 0 \quad \text{and} \quad q(y_2) \leq 0 \tag{23}$$

In the rest of the proof we show that this cannot happen.

First it is easy to see that $t(0) < 0$; $t(y) > 0$ when $y \to +\infty$; and $t'(y) > 0$ for any $y > 0$. This shows that there is a unique $b \in (0, \infty)$ such that $t(y) < 0$ for any $y < b$, $t(b) = 0$, and $t(y) > 0$ for any $y > b$.

Next using $s''(y) = y^{r-2} \cdot t(y)$, the same statement above also holds for $s''(y)$.

Now we examine $s'(y)$. Notice that $s'(0) > 0$ and $s'(y) > 0$ when $y \to \infty$. It follows from the property of $s''(y)$ that going from $y = 0$ to $+\infty$, the sign of $s'(y)$ can change at most twice from positive to negative and then back to positive.

Finally, regarding $s(y)$, we have $s(0) < 0$; and $s(y) > 0$ when $y \to +\infty$. From the property of $s'(y)$ we know $s(y)$ can have at most three roots in $(0, +\infty)$. From $q'(y) = y^{r-1} \cdot s(y)$, the same statement holds for $q'(y)$ as well. However, this contradicts with (23) because:

1. From $q(0) < 0$ and $q(y_1) \geq 0$, there exists a $y \in (0, y_1)$ such that $q'(y) > 0$;
2. From $q(y_1) \geq 0$ and $q(1-\epsilon) < 0$, there exists a $y \in (y_1, 1-\epsilon)$ such that $q'(y) < 0$;
3. From $q(1-\epsilon) < 0$ and $q(1+\epsilon) > 0$, there exists a $y \in (1-\epsilon, 1+\epsilon)$ such that $q'(y) > 0$;
4. From $q(1+\epsilon) > 0$ and $q(y_2) \leq 0$, there exists a $y \in (1+\epsilon, y_2)$ such that $q'(y) < 0$;
5. From $q(y_2) \leq 0$ and $q(y) > 0$ when $y \to +\infty$, there exists a $y \in (y_2, +\infty)$ such that $q'(y) > 0$.

It follows that $q'(y)$ has at least four roots in $(0, +\infty)$ and we get a contradiction. The lemma follows. $\square$

From now on we always assume that $a$ is positive and rational such that $\alpha = a^{1+r}$ is rational, $f$ satisfies conditions of Lemma 12 and $g$ satisfies conditions of Lemma 13. We also use $\lambda$ to denote $f'(0)$, a positive rational number, and use $\theta$ to denote the positive root of $f$. While $\theta$ is not rational in general, we can use $f$ (and $h$ in the proof of Lemma 12) to compute a $\gamma$-rational approximation $\theta^*$ of $\theta$, i.e. $|\theta^* - \theta| \leq \gamma$, in time polynomial in $1/\gamma$. We let $\sigma$ denote $f'(\theta) < 0$. The following corollaries follow from Lemma 12 and 13.

**Corollary 3.** *We have $g(x) < -\lambda x < -\lambda\theta$ for any $x \in (\theta, 1)$; $g(x) > -\lambda x > -\lambda\theta$ for any $x \in (0, \theta)$.*

*Proof.* By Lemma 12 we have $f(x) < 0$ for any $x \in (\theta, 1)$ and thus, $g(x) = f(x) - \lambda x < -\lambda x$.
By Lemma 12 we have $f(x) > 0$ for any $x \in (0, \theta)$ and thus, $g(x) = f(x) - \lambda x > -\lambda x$. $\square$

**Corollary 4.** $g(\theta) = -\lambda\theta$ and $g'(\theta) = \sigma - \lambda < -\lambda$.

**Corollary 5.** *There exists a positive constant $c$ such that for any $x \in [-c, c]$, we have*

$$\left|f(x) - \lambda x\right| \leq |\lambda x/2| \quad \text{and} \quad \left|f(\theta + x) - \sigma x\right| \leq |\sigma x/2|$$

Given a sufficiently large positive integer $N$, we will be interested in the behavior of $f$ and $g$ over:

$$A_N = [-\delta, \delta], \quad B_N = [\delta, \theta - \delta], \quad C_N = [\theta - \delta, \theta + \delta] \tag{24}$$
$$B'_N = [-\theta + \delta, -\delta], \quad C'_N = [-\theta - \delta, -\theta + \delta] \quad \text{and} \quad S_N = [-\theta - \delta, \theta + \delta]$$

where $\delta = 1/N$. We use Lemma 12 and Lemma 13 to prove the following lemmas that will be useful later.

**Lemma 14.** *When $N$ is sufficiently large, we have $|g(x)| \leq |\lambda x/2|$ for any $x \in A_N$.*

*Proof.* The lemma follows directly from the first part of Corollary 5. □

**Lemma 15.** *When $N$ is sufficiently large, we have $f(x) \geq \min(\lambda, |\sigma|)\delta/2$ for all $x \in B_N$.*

*Proof.* Assume for contradiction this is not the case, meaning that there is an infinite sequence of $N$ and $x_N$ such that $x_N \in B_N$ but $f(x_N) < \min(\lambda, |\sigma|)\delta/2$. As $x_N \in [0, \theta]$ is compact, there is a subsequence of $x_N$ that converges to a root $x^*$ of $f$ in $[0, \theta]$. As $0$ and $\theta$ are the only nonnegative roots of $f$, $x^* = 0$ or $\theta$. But no matter which case it is, the derivative of $f$ at $x^*$ is smaller than we expect and we get a contradiction. □

Using Lemma 15, we prove the following lemma:

**Lemma 16.** *Assume that $N$ is sufficiently large. If*

$$g(x) = -\lambda\theta \pm \Delta$$

*where $\Delta = \delta(\lambda - \sigma/2)$, then we must have that $x \in C_N$.*

*Proof.* First, $g(x) < 0$ when $N$ is sufficiently large. By Lemma 13 we have $x > 0$. Replacing $x$ by $\theta + y$:

$$f(\theta + y) - \lambda(\theta + y) = -\lambda\theta \pm \Delta \quad \Rightarrow \quad f(\theta + y) = \lambda y \pm \Delta$$

As $f(\theta + y) < 0$ when $y > 0$, and $f(\theta + y) > 0$ when $y < 0$ (and $x = \theta + y > 0$), we have $|y| < \Delta/\lambda$ and thus, Corollary 5 applies when $N$ is sufficiently large: If $y > 0$, we have

$$3\sigma y/2 \leq \lambda y \pm \Delta = f(\theta + y) \leq \sigma y/2$$

which implies that $0 < y \leq \Delta/(\lambda - \sigma/2) = \delta$. The case when $y < 0$ can be proved similarly. □

We also note that by the symmetry of $f$ and $g$, similar lemmas can be proved for $B'_N$ and $C'_N$.

## 4.2 Our Construction

Let $\rho < -1$ be a fixed rational number, with $r = |\rho|$. Given a normalized $2n \times 2n$ polymatrix game **P**, we construct a market $M_\mathbf{P}$ in which every trader uses a CES utility function of parameter $\rho$. The main building block in the construction of $M_\mathbf{P}$ is the following:

**Non-Monotone CES Markets:** We use $M$ to denote the non-monotone CES market discussed in Example 2.2 and Section 4.1, with rational constants $\alpha$ and $a$ satisfying all conditions of Lemma 12 and Lemma 13. We use the following notation. Given a positive rational number $\mu$, we use $\textbf{CES}(\mu, G_1, G_2)$ to denote the creation of the following two traders $T_1$ and $T_2$ in $M_\textbf{P}$. $T_1$ and $T_2$ are only interested in $G_1$ and $G_2$ and have the same utility functions as those of the two traders in $M$. $T_1$ has $\mu$ units of $G_1$ and $T_2$ has $\mu$ units of $G_2$. We use $f_\mu(x)$ to denote the excess spending function on $G_1$ from these two traders when the price of $G_1$ is $1+x$ and the price of $G_2$ is $1-x$; then we have $f_\mu(x) = \mu \cdot f(x)$.

Recall $\lambda = f'(0)$ is positive and rational, $\theta$ is the positive root of $f$, and $\sigma = f'(\theta) < 0$. Let $m = n^7$.

**Construction of $M_\textbf{P}$.** The market $M_\textbf{P}$ consists of the following $O(nm) = O(n^8)$ goods:

$$\text{AUX}_i, \ G_{2i-1,j} \text{ and } G_{2i,j}, \quad \text{for } i \in [n] \text{ and } j \in [0:m].$$

We also divide the goods into $n(m+1)$ groups: $\mathcal{R}_{i,j} = \{G_{2i-1,j}, G_{2i,j}\}$, for each $i \in [n]$ and $j \in [0:m]$.

First for each $i \in [n]$, we add a trader with $\tau = n^4$ units of $G_{2i-1,0}$ and $G_{2i,0}$ each, and set her utility

$$u(x_1, x_2) = (x_1^\rho + x_2^\rho)^{1/\rho}$$

where $x_1$ (or $x_2$) denotes the amount of $G_{2i-1,0}$ (or $G_{2i,0}$, respectively) she obtains.

Next for each $\mathcal{R}_{i,j}$, $i \in [n]$ and $j \in [m]$, we create a market $\textbf{CES}(\mu, G_{2i-1,j}, G_{2i,j})$ with $\mu = n/\lambda$.

Now we add a number of single-minded traders who trade between different groups. Recall that we say a trader is a $(r, G_1 : G_2)$-trader, if her endowment consists of $r$ units of $G_1$ and she is only interested in $G_2$; We say a trader is a $(r, G_1, G_2 : G_3)$-trader, if her endowment consists of $r$ units of $G_1$ and $G_2$ each, and is only interested in $G_3$. At the same time, we define a weighted directed graph $\mathcal{G} = (V, E)$ which will be used in the proof of correctness later. The vertices of $\mathcal{G}$ correspond to the $n(m+1)$ groups $\mathcal{R}_{i,j}$. The meaning of an edge and its weight in $\mathcal{G}$ is the same as the graph $\mathcal{G}$ defined in Section 3. Here is the construction:

1. For each $i \in [2n]$, we use $G_i$ to denote $G_{i,0}$ and $H_i$ to denote $G_{i,m}$ for convenience. For each pair $i, j \in [n]$, we add to $M_\textbf{P}$ the following four traders who trade from group $\mathcal{R}_{i,m}$ to group $\mathcal{R}_{j,0}$: one $(P_{2i-1,2j-1}, H_{2i-1} : G_{2j-1})$-trader, one $(P_{2i-1,2j}, H_{2i-1} : G_{2j})$-trader, one $(P_{2i,2j-1}, H_{2i} : G_{2j-1})$ trader, and one $(P_{2i,2j}, H_{2i} : G_{2j})$-trader. Since $\textbf{P}$ is normalized, we have

   $$P_{2i-1,2j-1} + P_{2i-1,2j} = P_{2i,2j-1} + P_{2i,2j} = 1$$

   Thus, the total endowment of these four traders consists of one unit of $H_{2i-1}$ and $H_{2i}$ each, so we add an edge in $\mathcal{G}$ from $\mathcal{R}_{i,m}$ to $\mathcal{R}_{j,0}$ with weight 1. At this moment, the total out-weight of each $\mathcal{R}_{i,m}$ in $\mathcal{G}$ (a complete bipartite graph) is $n$, and the total in-weight of each $\mathcal{R}_{j,0}$ in $\mathcal{G}$ is $n$.

2. Next for each $i \in [n]$ and $j \in [m]$, we add two traders who trade from group $\mathcal{R}_{i,j-1}$ to group $\mathcal{R}_{i,j}$: one $(n, G_{2i-1,j-1} : G_{2i-1,j})$-trader and one $(n, G_{2i,j-1} : G_{2i,j})$-trader. As their total endowment consists of $n$ units of $G_{2i-1,j-1}$ and $G_{2i,j-1}$ each, we add an edge from $\mathcal{R}_{i,j-1}$ to $\mathcal{R}_{i,j}$ of weight $n$.

This finishes the construction of $\mathcal{G}$. It is also easy to verify that $\mathcal{G}$ is strongly connected and each vertex has both its total in-weight and out-weight equal to $n$.

Finally, for each $j \in [n]$ we add traders between $\text{AUX}_j$ and $\mathcal{R}_{j,0}$. Let $r_{2j-1}$ and $r_{2j}$ be the two numbers defined in (6). Let $\theta^*$ denote a $\gamma$-rational approximation of $\theta$, the positive root of $f$, where $\gamma = 1/n^7$. Then we add the following three traders: one $((1-\theta^*)r_{2j-1}, \text{AUX}_j : G_{2j-1})$-trader, one $((1-\theta^*)r_{2j}, \text{AUX}_j : G_{2j})$ trader, and one $((1-\theta^*)n, G_{2j-1}, G_{2j} : \text{AUX}_j)$-trader. Note that $r_{2j-1} + r_{2j} = 2n$ as $\mathbf{P}$ is normalized.

This finishes the construction of $M_\mathbf{P}$. It follows immediately from the strong connectivity of $\mathcal{G}$ that the economy graph of $M_\mathbf{P}$ is strongly connected as well. Thus, $M_\mathbf{P}$ is a valid input of problem $\rho$-**CES-APPROX** and can be constructed from $\mathbf{P}$ in polynomial time. We also record the following properties of $M_\mathbf{P}$:

**Lemma 17.** *For each $i \in [n]$, the total supply of good $\text{AUX}_i$ is $2n(1-\theta^*)$;*
*For each $i \in [2n]$, the total supply of good $G_{i,0}$ is $\tau + (2-\theta^*)n$; and*
*For each $i \in [2n]$ and $j \in [m]$, the total supply of good $G_{i,j}$ is $\mu + n = \Theta(n)$.*

### 4.3 Proof of Correctness

Now let $\boldsymbol{\pi}$ denote an $\epsilon$-*additively* approximate market equilibrium of $M_\mathbf{P}$ where $\epsilon = 1/n^{14}$. We show in the rest of this section that given $\boldsymbol{\pi}$, one can compute a $(1/n)$-well-supported Nash equilibrium of $\mathbf{P}$ efficiently in polynomial time. Theorem 6 then follows. Below for each $\mathcal{R}_{i,j}$, we let $\pi_{i,j} = \pi(G_{2i-1,j}) + \pi(G_{2i,j})$.

First note that only one trader is interested in $\text{AUX}_j$ and thus,

**Lemma 18.** *Let $\boldsymbol{\pi}$ denote an $\epsilon$-additively approximate equilibrium of $M_\mathbf{P}$, where $\epsilon = 1/n^{14}$. If we scale $\boldsymbol{\pi}$ so that $\pi_{j,0} = \pi(G_{2j-1}) + \pi(G_{2j}) = 2$ for some $j \in [n]$, then we have $\pi(\text{AUX}_j) \geq 1 - O(\epsilon/n)$.*

Second, by using the strong connectivity of $\mathcal{G}$ and the property that every vertex in $\mathcal{G}$ has the same total in-weight and out-weight, we can follow the proof of Lemma 6 (replacing $4t$ with $m$) to prove

**Lemma 19.** *Let $\boldsymbol{\pi}$ denote an $\epsilon$-additively approximate equilibrium of $M_\mathbf{P}$. Let $\pi_{\max}$ and $\pi_{\min}$ denote*

$$\pi_{\max} = \max_{i,j} \pi_{i,j} \quad \text{and} \quad \pi_{\min} = \min_{i,j} \pi_{i,j}$$

*both taken over $i \in [n]$ and $j \in [0:m]$. If we scale $\boldsymbol{\pi}$ so that $\pi_{\min} = 2$, then $\pi_{\max} = 2 + O(m\epsilon)$.*

Then we can follow the proof of Lemma 7 to prove the following upper bound on $\pi(\text{AUX}_j)$:

**Lemma 20.** *Let $\boldsymbol{\pi}$ denote an $\epsilon$-additively approximate equilibrium of $M_\mathbf{P}$ with $\epsilon = 1/n^{14}$. If we scale $\boldsymbol{\pi}$ so that $\pi_{j,0} = 2$ for some $j \in [n]$, then we have $\pi(\text{AUX}_j) \leq 1 + O(m\epsilon)$.*

From now on, for each $i \in [n]$ and $j \in [0:m]$ we use $x_{i,j}$ to denote the unique number that satisfies

$$\frac{1+x_{i,j}}{1-x_{i,j}} = \frac{\pi(G_{2i-1,j})}{\pi(G_{2i,j})}$$

Note that the $x_{i,j}$'s are invariant under scaling of $\boldsymbol{\pi}$. If we scale $\boldsymbol{\pi}$ so that $\pi_{i,j} = 2$ for some $i$ and $j$, then we must have $\pi(G_{2i-1,j}) = 1 + x_{i,j}$ and $\pi(G_{2i,j}) = 1 - x_{i,j}$. Moreover, even if we scale $\boldsymbol{\pi}$ so that the sum of prices of another group becomes 2, we still have the following estimations by Lemma 18, 19 and 20:

$$\pi(G_{2i-1,j}) = 1 + x_{i,j} \pm O(m\epsilon), \quad \pi(G_{2i,j}) = 1 - x_{i,j} \pm O(m\epsilon) \quad \text{and} \quad \pi(\text{AUX}_j) = 1 \pm O(m\epsilon) \quad (25)$$

It would be great if we can prove a lemma similar to Lemma 8. However, right now we have no bound on the ratio of $\pi(G_{2i-1,j})$ and $\pi(G_{2i,j})$. Next we show that $x_{i,0}$ must be very close to 0 for all $i \in [n]$.

**Lemma 21.** *If $\boldsymbol{\pi}$ is an $\epsilon$-additively approximate equilibrium, then $|x_{i,0}| = O(1/n^3)$ for all $i \in [n]$.*

*Proof.* Fix an $i \in [n]$. We first scale $\boldsymbol{\pi}$ so that $\pi_{i,0} = 2$, and use $x$ to denote $x_{i,0}$.

We let $T$ denote the trader with $\tau$ units of $G_{2i-1}$ and $G_{2i}$ each. We let $y_1$ denote the demand of $G_{2i-1}$, and $y_2$ to denote the demand of $G_{2i}$ from $T$. Then $y_1(1+x) + y_2(1-x) = 2\tau$ and by (18) we have

$$\frac{y_1}{y_2} = \left(\frac{1-x}{1+x}\right)^{1/(1+r)}$$

Now assume without loss of generality that $x > 0$, we will show that $x = O(1/n^3)$. To this end, we have

$$y_2 = \frac{2\tau}{(1-x) + (1-x)^{1/(1+r)}(1+x)^{r/(1+r)}} \leq \tau + O(n)$$

which follows from the assumption of $\boldsymbol{\pi}$ being an additively approximate equilibrium. It then follows that

$$(1-x)^{1/(1+r)} \geq (1+x)^{1/(1+r)} - O(1/n^3) > 1 - O(1/n^3)$$

Since $r$ is a positive constant, we have $x = O(1/n^3)$ and the lemma follows. □

From now on we set $N = n^6$. Recall the definition of $A_N, B_N, C_N, B'_N, C'_N$, and $S_N$ from (24). Using Lemma 21, we have $x_{i,0} \in S_N$. Next we show that $x_{i,j} \in S_N$ for all $i$ and $j$.

**Lemma 22.** *If $\boldsymbol{\pi}$ is an $\epsilon$-additively approximate equilibrium, then $x_{i,j} \in S_N$ for all $i \in [n]$ and $j \in [m]$.*

Lemma 22 follows directly from the following three lemmas by induction:

**Lemma 23.** *For any $i \in [n]$ and $j \in [m]$, if $x_{i,j-1} \in A_N$, then we have $x_{i,j} \in A_N \cup B_N \cup B'_N$.*

**Lemma 24.** *For any $i \in [n]$ and $j \in [m]$, if $x_{i,j-1} \in B_N$, then we have $x_{i,j} \in B_N \cup C_N$; and if $x_{i,j-1} \in B'_N$ then we have $x_{i,j} \in B'_N \cup C'_N$*

**Lemma 25.** *For any $i \in [n]$ and $j \in [m]$, if $x_{i,j-1} \in C_N$, then we have $x_{i,j} \in C_N$; and if $x_{i,j-1} \in C'_N$, then we have $x_{i,j} \in C'_N$.*

*Proof of Lemma 23.* First we scale $\boldsymbol{\pi}$ so that $\pi_{i,j} = 2$. We let $x$ denote $x_{i,j}$, so that the prices of $\pi(G_{2i-1,j})$ and $\pi(G_{2i,j})$ are $1+x$ and $1-x$, respectively. We also let

$$y = x_{i,j-1}, \quad \pi(G_{2i-1,j-1}) = 1 + y_1 \quad \text{and} \quad \pi(G_{2i,j-1}) = 1 - y_2$$

By Lemma 19, $y_1$ and $y_2$ are both $y \pm O(m\epsilon)$. The excess spending of $G_{2i-1,j}$ of the whole market is

$$\mu \cdot f(x) + n(1+y_1) - n(1+x) = n(1/\lambda)(f(x) - \lambda x + \lambda y_1) = n(1/\lambda)(g(x) + \lambda y_1) \quad (26)$$

while the excess spending of $G_{2i,j}$ of the whole market is

$$-\mu \cdot f(x) + n(1-y_2) - n(1-x) = -n(1/\lambda)(g(x) + \lambda y_2) \quad (27)$$

As $\pi_{i,j} = 2$ and $\boldsymbol{\pi}$ is an $\epsilon$-additively approximate equilibrium, both (26) and (27) are at most $O(\epsilon)$. Thus

$$\big|n(1/\lambda)(g(x) + \lambda y)\big| = O(nm\epsilon) \ \Rightarrow\ |g(x) + \lambda y| = O(m\epsilon) \quad (28)$$

as $\lambda$ is a positive constant. As $|y| = |x_{i,j-1}| \leq 1/N = 1/n^6$ and $m\epsilon = 1/n^7$, we have $|g(x)| = O(1/N)$.

The lemma now follows from Corollary 3 and Corollary 5. □

*Proof of Lemma 24.* Suppose that $x_{i,j-1} \in B_N$; the proof for the case of $x_{i,j-1} \in B'_N$ is similar. Using the same notation and argument of Lemma 23, we start with (28) and get

$$g(x) \geq -\lambda y - O(m\epsilon) \geq -\lambda(\theta - 1/N + O(m\epsilon)) > -\lambda\theta$$

where the second inequality used $y = x_{i,j-1} \in B_N$ and thus, $y \leq \theta - 1/N$. We also have

$$g(x) \leq -\lambda y + O(m\epsilon) \leq -\lambda/N + O(m\epsilon) < 0$$

where the second inequality used $y = x_{i,j-1} \geq 1/N$. By Corollary 3, we have $x \in A_N \cup B_N \cup C_N$.

Assume for contradiction that $x \in A_N$, then by Corollary 5 we have

$$-\lambda/N + O(m\epsilon) \geq g(x) \geq -\lambda x/2$$

and thus, $x \geq 2/N - O(m\epsilon) \notin A_N$ and we get a contradiction. The lemma now follows. □

*Proof of Lemma 25.* Suppose that $x_{i,j-1} \in C_N$; the proof for the case of $x_{i,j-1} \in C'_N$ is similar. Using the same notation and argument of Lemma 23, we start with (28) and get

$$O(m\epsilon) = |g(x) + \lambda y| = |g(x) + \lambda\theta \pm \lambda/N|$$

and thus, we have the following upper bound

$$|g(x) + \lambda\theta| \leq \lambda/N + O(m\epsilon)$$

The right side is smaller than $(\lambda - \sigma/2)/N$ as $N = n^6$, $\epsilon = 1/n^{14}$ and $m = n^7$. It follows from Lemma 16 that $x \in C_N$. The lemma follows directly. □

We now construct a $2n$-dimensional vector $\mathbf{y}$ from $\boldsymbol{\pi}$ as follows. Recall $\theta^*$ is a $\gamma$-rational approximation of $\theta$ with $\gamma = 1/n^7$. Let $\delta = 1/N$. For each $i \in [n]$, if $x_{i,m} \geq \theta^* - 2\delta$, then we set $y_{2i-1} = 1$ and $y_{2i} = 0$; if $x_{i,m} \leq -(\theta^* - 2\delta)$ then we set $y_{2i-1} = 0$ and $y_{2i} = 1$; otherwise, we set $y_{2i-1}$ and $y_{2i}$ to be

$$y_{2i-1} = \frac{\theta^* + x_{i,m}}{2\theta^*} \quad \text{and} \quad y_{2i} = \frac{\theta^* - x_{i,m}}{2\theta^*}$$

By definition, $\mathbf{y}$ is a nonnegative vector and $y_{2i-1} + y_{2i} = 1$ for all $i \in [n]$. Note that when $x_{i,m} \in C_N$, we must have $x_{i,m} \geq \theta^* - 2\delta$ since $\gamma < \delta$, and hence $y_{2i-1} = 1$ and $y_{2i} = 0$. Similarly, if $x_{i,m} \in C'_N$ then $y_{2i-1} = 0$ and $y_{2i} = 1$. By Lemma 19, for every $i \in [2n]$ we have

$$y_i = \frac{\theta^* + \pi(G_i) - 1}{2\theta^*} \pm \big(O(\gamma + m\epsilon + 1/N)\big) \Rightarrow \pi(G_i) = 2\theta^* y_i + (1 - \theta^*) \pm O(1/N)$$

To finish the proof of Theorem 6, we prove the following theorem:

**Theorem 10.** *When $n$ is sufficiently large, $\mathbf{y}$ built above is a $(1/n)$-well-supported Nash equilibrium of $\mathbf{P}$.*

To prove Theorem 10, we need the following key lemma:

**Lemma 26.** *For every $i \in [n]$, if $x_{i,0} \in B_N \cup C_N$ then we have $x_{i,m} \in C_N$, and $y_{2i-1} = 1$, $y_{2i} = 0$. Similarly, if $x_{i,0} \in B'_N \cup C'_N$ then we have $x_{i,m} \in C'_N$, and $y_{2i-1} = 0$, $y_{2i} = 1$.*

*Proof.* By Lemma 25, we assume that $x_{i,0} \in B_N$ without loss of generality.

Now assume for contradiction that $x_{i,m} \notin C_N$. By Lemma 25 again we have $x_{i,j} \in B_N$ for all $j \in [m]$. This contradicts with the following lemma:

**Lemma 27.** *For any $j \in [m]$, if $x_{i,j-1}$ and $x_{i,j}$ are both in $B_N$, then we have $x_{i,j} = x_{i,j-1} + \Omega(1/N)$.*

*Proof.* Using the same notation and argument of Lemma 23, we start with (28) and get

$$g(x_{i,j}) = -\lambda x_{i,j-1} \pm O(m\epsilon)$$

Using Lemma 15, we have $g(x_{i,j}) + \lambda x_{i,j} = f(x_{i,j}) = \Omega(1/N)$ since $x_{i,j} \in B_N$. As a result, we get

$$-\lambda x_{i,j} + \Omega(1/N) = g(x_{i,j}) = -\lambda x_{i,j-1} \pm O(m\epsilon)$$

and thus, $x_{i,j} = x_{i,j-1} + \Omega(1/N)$ using $m = n^7$ and $\epsilon = 1/n^{14}$. The lemma then follows. □

We get a contradiction from Lemma 27 because $m = n^7$ and $N = n^6$. Lemma 26 follows. □

Finally we prove Theorem 10:

*Proof of Theorem 10.* We assume for contradiction that $\mathbf{y}$ constructed above is not a $(1/n)$-well-supported Nash equilibrium of $\mathbf{P}$. Without loss of generality, we assume that

$$\mathbf{y}^T \cdot \mathbf{P}_1 > \mathbf{y}^T \cdot \mathbf{P}_2 + 1/n \qquad (29)$$

where $\mathbf{P}_1$ and $\mathbf{P}_2$ denote the first and second columns of $\mathbf{P}$, respectively, but $y_2 > 0$. To reach a contradiction, by Lemma 26, it suffices to show that (29) implies that $x_{1,0} \in B_N \cup C_N$.

To this end, we first scale $\boldsymbol{\pi}$ so that $\pi(G_1) + \pi(G_2) = 2$, and use $x$ to denote $x_{1,0}$. By Lemma 21, we have $\pi(G_1), \pi(G_2) = 1 \pm O(1/n^3)$ are very close to 1. By applying Walras' law over the whole market $M_\mathbf{P}$ and using the assumption that $\boldsymbol{\pi}$ is an $\epsilon$-additively approximate equilibrium, we have

$$\epsilon \geq \text{the excess demand of } G_1 \text{ (or } G_2) \geq -O(mn\epsilon) \qquad (30)$$

Now we compare the total money spent on $G_1$ and $G_2$, by all traders in $M_\mathbf{P}$ except the one, denoted by $T$, who owns $\tau$ units of $G_1$ and $G_2$ each. Here is the list of such traders:

1. For each $i \in [2n]$, there is a $(P_{i,1}, H_i : G_1)$-trader. The total money these traders spend on $G_1$ is

$$\sum_{i \in [2n]} P_{i,1} \cdot \pi(H_i) = \sum_{i \in [2n]} P_{i,1} \cdot \left(2\theta^* y_i + (1 - \theta^*) \pm O(1/N)\right)$$

2. For each $i \in [2n]$, there is a $(P_{i,2}, H_i : G_2)$-trader. The total money these traders spend on $G_2$ is

$$\sum_{i \in [2n]} P_{i,2} \cdot \pi(H_i) = \sum_{i \in [2n]} P_{i,2} \cdot \left(2\theta^* y_i + (1 - \theta^*) \pm O(1/N)\right)$$

3. There are one $((1 - \theta^*)r_1, \text{AUX}_1 : G_1)$-trader and one $((1 - \theta^*)r_2, \text{AUX}_1 : G_2)$ trader.

Recall the definition of $r_1$ and $r_2$ in (6). As a result, the total money these traders spend on $G_1$ is

$$M_1 = 2\theta^* \cdot \mathbf{y}^T \cdot \mathbf{P}_1 + 2n(1-\theta^*) \pm O(n/N)$$

using $N = n^6$ and $m\epsilon = 1/n^7$, and the total money these traders spend on $G_2$ is

$$M_2 = 2\theta^* \cdot \mathbf{y}^T \cdot \mathbf{P}_2 + 2n(1-\theta^*) \pm O(n/N)$$

Thus $M_1 - M_2 = \Omega(1/n)$ and the demand for $G_1$ is larger than the demand for $G_2$, from these traders, by

$$\frac{M_1}{\pi(G_1)} - \frac{M_2}{\pi(G_2)} \geq \frac{M_2 + \Omega(1/n)}{\pi(G_1)} - \frac{M_2}{(1-O(1/n^3)) \cdot \pi(G_1)} = \Omega(1/n)$$

where both inequalities used $\pi(G_1), \pi(G_2) = 1 \pm O(1/n^3)$ and $M_1, M_2 = O(n)$.

Let $d_1$ (or $d_2$) denote the demand of $G_1$ (or $G_2$, respectively) from $T$. Using (30) and $mn\epsilon \ll 1/n$ we must have $d_2 - d_1 = \Omega(1/n)$. On the other hand, we have from (18):

$$\frac{d_1}{d_2} = \left(\frac{1-x}{1+x}\right)^{1/(1+r)}$$

As $\pi(G_1), \pi(G_2)$ are close to 1, $d_1$ and $d_2$ are $O(\tau)$ even if $T$ spends all the budget on one of them. Thus,

$$\left(\frac{1+x}{1-x}\right)^{1/(1+r)} = \frac{d_2}{d_1} = 1 + \frac{d_2 - d_1}{d_1} = 1 + \Omega(1/n^5)$$

and thus, $x = \Omega(1/n^5)$. It follows from Lemma 21 and $N = n^6$ that $x \in B_N$. The theorem follows. $\square$

## 5 Membership in FIXP

We are given a market $M$ with $n$ traders and $m$ goods. Each trader $T_i$, $i \in [n]$, has an endowment $w_{i,j} \geq 0$ of each good $j \in [m]$, and has a CES utility with coefficients $\alpha_{i,j} \geq 0$ and parameter $\rho_i < 1$. The endowments $w_{i,j}$ and coefficients $\alpha_{i,j}$ are rationals given in binary and the parameters $\rho_i < 1$ are rationals given in unary; the parameter $\rho_i$'s for different traders may be the same or different, and there may be a mixture of positive and negative $\rho_i$'s. We also assume that the economy graph is strongly connected.

We prove the following theorem in this section:

**Theorem 11.** CES *is in* FIXP.

We first introduce some notation:

- Let $w_{\min} = \min_{i,j}\{w_{i,j} : w_{i,j} > 0\}$ and $w_{\max} = \max_{i,j}\{w_{ij}\}$ be respectively the minimum non-zero and the maximum endowment of a good, i.e., if a trader owns a good then he owns at least $w_{\min}$ and at most $w_{\max}$ units of this good.

- Let $\alpha_{\min} = \min_{i,j}\{\alpha_{i,j} : \alpha_{i,j} > 0\}$ and $\alpha_{\max} = \max_{i,j}\{\alpha_{ij}\}$ be respectively the minimum non-zero and the maximum CES coefficient over utilities of all traders.

- Finally, let $\mu = (h^m/m)^{t^m}$, where

$$t = \max\left(\{\lceil 1-\rho_i \rceil : \rho_i < 0\} \cup \{\lceil 1/(1-\rho_i)\rceil : \rho_i > 0\}\right) \quad \text{and} \quad h = \frac{\alpha_{\min}}{\alpha_{\max}} \cdot \frac{w_{\min}}{w_{\max}} \cdot \frac{1}{2nm^2}$$

Without loss of generality we focus on price vectors $\boldsymbol{\pi}$ that are normalized, i.e., the entries sum to 1. Let

$$S = \left\{ \boldsymbol{\pi} \in \mathbb{R}_+^m : \sum_{i=1}^m \pi_i = 1 \right\}$$

be the unit simplex in the $m$-dimensional space. To prove membership in FIXP, given a market $M$ we will construct in polynomial time an algebraic circuit $C$ with operations from

$$\{+, -, *, /, \max, \min, \sqrt[k]{\ }\}$$

with $m$ inputs and $m$ outputs, which defines a continuous function $F : S \to S$ such that the fixed points of $F$ coincide with the market equilibria of $M$. As $F$ is continuous on $S$ which is convex and compact, a fixed point always exists because of Brouwer's fixed point theorem. We also point out that in the construction, the $k$'s in the $\sqrt[k]{\ }$ operations are encoded in unary.

Given an input vector $\boldsymbol{\pi} \in S$, the circuit $C$ first computes $\mu$, and then computes a new vector $\hat{\boldsymbol{\pi}}$, where $\hat{\pi}_j = \max(\pi_j, \mu)$ for all $j \in [m]$. Then it computes and outputs for each $j \in [m]$:

$$F_j(\boldsymbol{\pi}) = \frac{\hat{\pi}_j + \max\{0, Z_j(\hat{\boldsymbol{\pi}})\}}{\sum_{k=1}^m (\hat{\pi}_k + \max\{0, Z_k(\hat{\boldsymbol{\pi}})\})}$$

where we use $Z_j(\hat{\boldsymbol{\pi}})$ to denote the excess demand of the $j$th good at $\hat{\boldsymbol{\pi}}$ in $M$.

Note first that $\mu$ can be generated by the circuit $C$ with only a polynomial number of operations. This is because for any number $c$, we can generate $c^\ell$ with $O(\log(\ell))$ multiplications by using successive squaring. Hence, we can generate $\mu$ from $h^m/m$ using $O(m \log t)$ multiplications, from which we then compute the new vector $\hat{\boldsymbol{\pi}}$. From $\hat{\boldsymbol{\pi}}$, the excess demand $Z_k(\hat{\boldsymbol{\pi}})$ can be computed using (1) with a polynomial number of operations (it is important here that FIXP allows roots and hence fractional powers), and so can $F(\boldsymbol{\pi})$.

Note that all the operations of $C$ are well-defined. In particular, all the fractional powers are applied to positive numbers and all the denominators are positive. The map $F$ is clearly continuous. Furthermore, we have $\sum_j F_j(\boldsymbol{\pi}) = 1$ and $F_j(\boldsymbol{\pi}) \geq 0$ for all $j \in [m]$. Thus, $F$ is a map from $S$ to itself.

Next we show that the fixed points of $F$ are precisely the market equilibria of $M$ in the unit simplex $S$. We need the following key lemma, which we will prove later.

**Lemma 28.** *Let $\boldsymbol{\pi}$ be a price vector with $\sum_{j=1}^m \pi_j \geq 1$ and suppose that $\pi_i \leq \mu$ for some good $i$. Then there must be a good $\ell \in [m]$ for which $Z_\ell(\boldsymbol{\pi}) > nmw_{\max}$.*

Using this lemma we prove the following theorem. Theorem 11 then follows.

**Theorem 12.** *The fixed points of $F$ are precisely the market equilibria of $M$ in $S$.*

*Proof.* Assume $\boldsymbol{\pi} \in S$ is an equilibrium of $M$. Then $\pi_j > \mu$ for all $j$ by Lemma 28, and hence $\hat{\boldsymbol{\pi}} = \boldsymbol{\pi}$ and $Z_j(\hat{\boldsymbol{\pi}}) = Z_j(\boldsymbol{\pi}) = 0$ for all $j$. Therefore, $F_j(\boldsymbol{\pi}) = \pi_j$ for all $j$ and $\boldsymbol{\pi}$ is a fixed point of $F$.

Now let $\boldsymbol{\pi}$ be a fixed point of $F$. We first show that $\pi_j > \mu$ for all $j \in [m]$.

Suppose that there is a good $i$ with $\pi_i \leq \mu$. From $\sum_{j=1}^m \pi_j = 1$ and $\hat{\pi}_j \geq \pi_j$, we have $\sum_{j=1}^m \hat{\pi}_j \geq 1$. Then because of Lemma 28 there is a good $\ell$ with $Z_\ell(\hat{\boldsymbol{\pi}}) > nmw_{\max}$. We partition the goods into two sets $H = \{j : \pi_j > \mu\}$ and $L = \{j : \pi_j \leq \mu\}$. Since $\boldsymbol{\pi}$ is a fixed point, from the definition of $F$ we get

$$\max\{0, Z_j(\hat{\boldsymbol{\pi}})\} \geq \pi_j \sum_{k=1}^m \max\{0, Z_k(\hat{\boldsymbol{\pi}})\}, \quad \text{for every good } j \text{ in } H.$$

From $Z_\ell(\hat{\boldsymbol{\pi}}) > nmw_{\max}$ we get $\sum_{k=1}^{m} \max\{0, Z_k(\hat{\boldsymbol{\pi}})\} > 0$. Therefore, we have $Z_j(\hat{\boldsymbol{\pi}}) > 0$ for all $j \in H$. Moreover, note that the excess demand of any good cannot be less than $-nw_{\max}$ because each trader owns at most $w_{\max}$ units of any good. Combining these we get:

$$\sum_{j=1}^{m} \hat{\pi}_j \cdot Z_j(\hat{\boldsymbol{\pi}}) = \sum_{j \in H} \hat{\pi}_j \cdot Z_j(\hat{\boldsymbol{\pi}}) + \sum_{j \in L} \mu \cdot Z_j(\hat{\boldsymbol{\pi}}) \geq \hat{\pi}_\ell \cdot Z_\ell(\hat{\boldsymbol{\pi}}) - \sum_{j \in L} \mu \cdot nw_{\max} > 0$$

However, by Warlas' law, $\sum_{j=1}^{m} \hat{\pi}_j \cdot Z_j(\hat{\boldsymbol{\pi}}) = 0$, a contradiction. Therefore, $\pi_j > \mu$ for all $j$ and $\hat{\boldsymbol{\pi}} = \boldsymbol{\pi}$.

Finally, assume $\boldsymbol{\pi}$ is not an equilibrium, i.e., $Z_\ell(\boldsymbol{\pi}) > 0$ for some $\ell$. Then $\sum_{k=1}^{m} \max\{0, Z_k(\boldsymbol{\pi})\} > 0$. For any good $j$ we have $F_j(\hat{\boldsymbol{\pi}}) = F_j(\boldsymbol{\pi}) = \pi_j > \mu$ and from the definition of $F$ we get

$$\max\{0, Z_j(\boldsymbol{\pi})\} = \pi_j \sum_{k=1}^{m} \max\{0, Z_k(\boldsymbol{\pi})\} > 0$$

and thus, $Z_j(\boldsymbol{\pi}) > 0$ for all $j$. Therefore, $\sum_{j=1}^{m} \pi_j \cdot Z_j(\boldsymbol{\pi}) > 0$ which again violates Walras' law. It follows that $\boldsymbol{\pi}$ must be a market equilibrium of $M$, and the theorem is proven. $\square$

It remains to prove Lemma 28. We show first the following lemma.

**Lemma 29.** *If the economy graph contains an edge $q \to j$ such that $\pi_j \leq h^t \cdot \pi_q^t$, then there is a good $\ell$ with excess demand $Z_\ell(\boldsymbol{\pi}) > nmw_{\max}$.*

*Proof.* Suppose that the excess demand $Z_j(\boldsymbol{\pi}) \leq nmw_{\max}$, for all goods $j$. Since the total supply of every good is at most $nw_{\max}$, this implies that the total demand for every good is less than $2nmw_{\max}$.

Suppose that the economy graph has an edge $q \to j$ such that $\pi_j \leq h^t \cdot \pi_q^t$, and let $T_i$ denote a trader with $w_{i,q} > 0$ and $\alpha_{i,j} > 0$. We may assume, without loss of generality, that good $j$ has the lowest price among those goods that $T_i$ is interested in, i.e., that $\pi_j \leq \pi_k$ for all $k$ such that $\alpha_{i,k} > 0$. (If not, then let

$$\ell = \arg\min_{k} \{\pi_k : \alpha_{i,k} > 0\}$$

and consider the edge $q \to \ell$ instead of $q \to j$; clearly $\pi_\ell$ also satisfies $\pi_\ell \leq h^t \cdot \pi_q^t$.) We distinguish two cases depending on the sign of $\rho_i$.

- *Case 1*: $\rho_i < 0$. Using (1), and since $\rho_i < 0$, we have

$$x_{i,j} = \frac{\alpha_{i,j}^{\frac{1}{1-\rho_i}} \cdot \sum_{k=1}^{m} w_{i,k} \cdot \pi_k}{\pi_j^{\frac{1}{1-\rho_i}} \cdot \sum_{k=1}^{m} \alpha_{i,k}^{\frac{1}{1-\rho_i}} \cdot \pi_k^{\frac{-\rho_i}{1-\rho_i}}} \geq \frac{\alpha_{\min}^{\frac{1}{1-\rho_i}} \cdot w_{\min} \cdot \pi_q}{\pi_j^{\frac{1}{1-\rho_i}} \cdot m \cdot \alpha_{\max}^{\frac{1}{1-\rho_i}}}$$

We must have $x_{i,j} < 2nmw_{\max}$. Thus, solving for $\pi_j$ and using the fact that $\rho_i < 0$ and $t \geq 1 - \rho_i$:

$$\pi_j > \frac{\alpha_{\min}}{\alpha_{\max}} \cdot \left(\frac{w_{\min}}{w_{\max}} \cdot \frac{1}{2nm^2}\right)^{1-\rho_i} \cdot \pi_q^{1-\rho_i} \geq h^t \cdot \pi_q^t$$

- *Case 2*: $\rho_i > 0$. Using (1), and since $\rho_i > 0$ and $\pi_j \leq \pi_k$ for all $k$ such that $\alpha_{i,k} > 0$, we have:

$$x_{i,j} = \frac{\alpha_{i,j}^{\frac{1}{1-\rho_i}} \cdot \sum_{k=1}^{m} w_{i,k} \cdot \pi_k}{\pi_j^{\frac{1}{1-\rho_i}} \cdot \sum_{k=1}^{m} \alpha_{i,k}^{\frac{1}{1-\rho_i}} \cdot \pi_k^{\frac{-\rho_i}{1-\rho_i}}} \geq \frac{\alpha_{\min}^{\frac{1}{1-\rho_i}} \cdot w_{\min} \cdot \pi_q}{\pi_j^{\frac{1}{1-\rho_i}} \cdot m \cdot \alpha_{\max}^{\frac{1}{1-\rho_i}} \cdot \pi_j^{\frac{-\rho_i}{1-\rho_i}}} \geq \left(\frac{\alpha_{\min}}{\alpha_{\max}}\right)^{\frac{1}{1-\rho_i}} \cdot \frac{w_{\min}}{m} \cdot \frac{\pi_q}{\pi_j}$$

From $x_{i,j} < 2nmw_{\max}$, solving for $\pi_j$ and using the fact that $\rho_i > 0$ and $t \geq 1/(1-\rho_i) > 1$:

$$\pi_j > \left(\frac{\alpha_{\min}}{\alpha_{\max}}\right)^{\frac{1}{1-\rho_i}} \cdot \frac{w_{\min}}{w_{\max}} \cdot \frac{1}{2nm^2} \cdot \pi_q > h^t \cdot \pi_q^t$$

Combining the two cases, the lemma is now proven. □

We can prove now Lemma 28:

*Proof of Lemma 28.* Let $\boldsymbol{\pi}$ be a price vector with $\sum_{j=1}^m \pi_j \geq 1$. If a good $j$ has $\pi_j = 0$, then it has infinite demand. So assume without loss of generality that $\pi_j > 0$ for all $j$. Suppose that all goods $j$ have excess demand $Z_j(\boldsymbol{\pi}) \leq nmw_{\max}$. Then $\pi_j > h^t \cdot \pi_q^t$, for all edges $q \to j$ by Lemma 29.

We use $G_{\max}$ and $G_{\min}$ to denote a good with the maximum and minimum price, respectively. We also use $\pi_{\max}$ and $\pi_{\min}$ to denote their prices, respectively. Because the economy graph is strongly connected, it contains a simple path from $G_{\max}$ to $G_{\min}$. Let

$$G_{\max} = j_0 \to j_1 \to j_2 \to \ldots \to j_\ell = G_{\min}$$

be one such simple path. By Lemma 29 we have $\pi_{j_k} > h^t \cdot \pi_{j_{k-1}}^t$ for $k = 1, \ldots, \ell$. Therefore, we have

$$\pi_{\min} = \pi_{j_\ell} > h^{\ell t^\ell} \cdot \pi_{j_0}^{t^\ell} = h^{\ell t^\ell} \cdot \pi_{\max}^{t^\ell}$$

by induction on $k$. As the path is simple, $\ell \leq m$. Also, since $\pi_{\max}$ is the maximum price and $\sum_j \pi_j \geq 1$, we have $\pi_{\max} \geq 1/m$. It then follows from the definition of $\mu$ that

$$\pi_{\min} > (h^m \cdot \pi_{\max})^{t^m} \geq (h^m/m)^{t^m} = \mu$$

and the lemma follows. □

## 6 Membership in PPAD

We focus on CES markets with $n$ traders, $m$ goods, and strongly connected economy graphs. Each trader $T_i$ uses a CES utility function, with its coefficients $\alpha_{i,j}$'s encoded in binary and its parameter $\rho_i < 1$ encoded in unary. Again we allow traders to use different $\rho_i$'s. We prove the following theorem in this section.

**Theorem 13.** CES-APPROX *is in* PPAD.

We use the definition of $w_{\min}$, $w_{\max}$, $\alpha_{\min}$ and $\alpha_{\max}$ in Section 5. For convenience, we assume without loss of generality that $w_{\min} \leq w_{\max} = 1$ and $1 = \alpha_{\min} \leq \alpha_{\max}$, since scaling the $\alpha_{i,j}$'s and $w_{i,j}$'s does not change the set of $\epsilon$-approximate market equilibria (as the approximation here is multiplicative).

The following lemma implies that we may assume without loss of generality that there is a trader in the market who owns a positive amount of all $m$ goods and is interested in all of them.

**Lemma 30.** *Let $M$ be a market with $n$ traders and $m$ goods. For any $\epsilon : 0 < \epsilon < 1$, we construct from $M$ a market $M'$ by adding a new trader $T^*$ who initially owns $\epsilon w_{\min}/4$ units of each good; and equally likes all the $m$ goods with a CES function of parameter $\rho^* = -1$. Then any $(\epsilon/4)$-approximate equilibrium of $M'$ is also an $\epsilon$-approximate equilibrium of $M$.*

*Proof.* Let $\boldsymbol{\pi}$ denote an $(\epsilon/4)$-approximate equilibrium of $M'$. We let $Z_j(\boldsymbol{\pi})$ and $Z'_j(\boldsymbol{\pi})$ denote the excess demand of good $j$ under pricing $\boldsymbol{\pi}$ in $M$ and $M'$, respectively. By the definition of approximate equilibria,

$$Z'_j(\boldsymbol{\pi}) \leq \frac{\epsilon}{4} \cdot \left( \sum_{i=1}^{n} w_{i,j} + \frac{\epsilon w_{\min}}{4} \right) = \frac{\epsilon}{4} \cdot \sum_{i=1}^{n} w_{i,j} + \frac{\epsilon^2 w_{\min}}{16}$$

for all $j$. At the same price vector $\boldsymbol{\pi}$, all traders in $M$ have the same demands as in $M'$ but trader $T^*$ is not present anymore to provide $\epsilon w_{\min}/4$ units of supply of each good $j$. This implies that for each good $j$:

$$Z_j(\boldsymbol{\pi}) = Z'_j(\boldsymbol{\pi}) - \big(\text{demand of } j \text{ from } T^* \text{ in } M'\big) + \epsilon w_{\min}/4 \leq Z'_j(\boldsymbol{\pi}) + \epsilon w_{\min}/4$$

Combining the two inequalities, we have for each good $j$:

$$Z_j(\boldsymbol{\pi}) \leq \frac{\epsilon}{4} \cdot \sum_{i=1}^{n} w_{i,j} + \frac{\epsilon^2 w_{\min}}{16} + \frac{\epsilon w_{\min}}{4} < \epsilon \sum_{i=1}^{n} w_{i,j}$$

where the last inequality follows from $\sum_{i=1}^{n} w_{i,j} \geq w_{\min}$, as the economy graph is strongly connected and thus, at least one trader owns good $j$. It follows that $\boldsymbol{\pi}$ is an $\epsilon$-approximate equilibrium of $M$. □

As a result, from now on, we always assume that there is a trader $T^*$ in the input market $M$, who owns $w_{\min}$ units of each good (notice that after adding $T^*$ to $M$ as described in Lemma 30, one needs to update $w_{\min}$) and equally likes all the goods (i.e., all coefficients are 1) with a CES function of $\rho^* = -1$. Let

$$\xi = \left( \frac{w_{\min}}{4nm} \right)^2$$

It is clear that $\xi : 0 < \xi < 1$ can be computed efficiently. Now we prove the following useful lemma:

**Lemma 31.** *Given a vector $\boldsymbol{\pi}$ with $1 \leq \sum_{j=1}^{m} \pi_j \leq 2$, if $\pi_\ell \leq \xi$ for some $\ell$ then $Z_\ell(\boldsymbol{\pi}) > n$.*

*Proof.* As $T^*$ is equally interested in all the goods and $\rho^* = -1$, by (1) his demand for good $\ell$ is

$$\frac{1}{\sqrt{\pi_\ell}} \cdot \frac{\sum_{k=1}^{m} w_{\min} \cdot \pi_k}{\sum_{k=1}^{m} \sqrt{\pi_k}} > \frac{1}{\xi^{1/2}} \cdot \frac{w_{\min}}{2m} = 2n$$

where we used $1 \leq \sum_{j=1}^{m} \pi_j \leq 2$. The lemma follows as the supply of good $\ell$ is at most $nw_{\max} = n$. □

Let $S$ denote the unit simplex in the $m$-dimensional space and $\hat{\pi}_j = \max(\pi_j, \xi)$, for all $j \in [m]$. We are going to use the following continuous map $F : S \to S$ with

$$F_j(\boldsymbol{\pi}) = \frac{\hat{\pi}_j + \max\{0, Z_j(\hat{\boldsymbol{\pi}})\}}{\sum_{k=1}^{m} (\hat{\pi}_k + \max\{0, Z_k(\hat{\boldsymbol{\pi}})\})} \tag{31}$$

We need the following definition of *approximate fixed points*:

**Definition 11** (Approximate Fixed Points). *We say $\boldsymbol{\pi} \in S$ is a $c$-approximate fixed point of $F : S \to S$ for some $c \geq 0$, if $\|F(\boldsymbol{\pi}) - \boldsymbol{\pi}\| \leq c$, where we use $\|\cdot\|$ to denote the $L_\infty$ norm of a vector.*

We prove that every $c$-approximate fixed point of $F$, where $c = \xi \epsilon w_{\min}/2$, is an $\epsilon$-approximate equilibrium of market $M$.

**Lemma 32.** *When $0 < \epsilon < 1$, any $c$-approximate fixed point $\boldsymbol{\pi}$ of $F$ is an $\epsilon$-approximate equilibrium of $M$ where $c = \xi \epsilon w_{\min}/2$.*

*Proof.* First we show that $\pi_j \geq \xi$ for all $j \in [m]$. Assume for contradiction that $\pi_\ell < \xi$ for some $\ell$.

We divide the traders into two groups: $H = \{j : \pi_j > \xi\}$ and $L = \{j : \pi_j \leq \xi\}$. Then $L$ is nonempty by our assumption since $\ell \in L$. By Lemma 31, $Z_j(\hat{\boldsymbol{\pi}}) > n$ for all $j \in L$, since $\sum_{j \in [m]} \hat{\pi}_j$ is between $1$ and $1 + n\xi < 2$. On the other hand, because $\boldsymbol{\pi}$ is a $c$-approximate fixed point, for each good $j$ we have

$$\hat{\pi}_j + \max\{0, Z_j(\hat{\boldsymbol{\pi}})\} \geq (\pi_j - c) \cdot \sum_{k=1}^m \left(\hat{\pi}_k + \max\{0, Z_k(\hat{\boldsymbol{\pi}})\}\right) \tag{32}$$

For each good $j \in H$, we have $\hat{\pi}_j = \pi_j > \xi$. Using $\sum_{k=1}^m \hat{\pi}_k \geq 1$, we have for each good $j \in H$:

$$\max\{0, Z_j(\hat{\boldsymbol{\pi}})\} \geq -c + (\pi_j - c)\sum_{k=1}^m \max\{0, Z_k(\hat{\boldsymbol{\pi}})\} > -c + (\xi - c)n > 0$$

by the definition of $c$ and the assumption that $w_{\min} \leq w_{\max} = 1$. This contradicts with Walras' law.

Now $\pi_j \geq \xi$ for all $j$. Assume for contradiction $\boldsymbol{\pi} = \hat{\boldsymbol{\pi}}$ is not an $\epsilon$-approximate equilibrium. Then

$$\sum_{k=1}^m \max\{0, Z_k(\hat{\boldsymbol{\pi}})\} = \sum_{k=1}^m \max\{0, Z_k(\boldsymbol{\pi})\} > \epsilon w_{\min}$$

Since $\boldsymbol{\pi}$ is a $c$-approximate fixed point of $F$, using (32) we have for each good $j \in [m]$:

$$\max\{0, Z_j(\boldsymbol{\pi})\} \geq -c + (\pi_j - c)\epsilon w_{\min} \geq -c + (\xi - c)\epsilon w_{\min} = c(1 - \epsilon w_{\min}) > 0$$

since $\epsilon < 1$ and $w_{\min} \leq 1$. This again contradicts with Walras' law. The lemma then follows. $\square$

Thus, the approximate market equilibrium problem reduces to the approximate fixed point computation problem for the functions (31) that arise from CES markets.

Let $\mathcal{F}$ be a family of functions, where each function $F$ in $\mathcal{F}$ is represented (encoded) by a binary string. Denote by $F_I$ the function represented by string $I$, and assume that every $F_I$ in the family $\mathcal{F}$ is a continuous function whose domain and range is a convex polytope $D_I$, described by a set of linear inequalities all with rational coefficients that can be computed from the string $I$ in polynomial time.

An example of such a family is the family $\mathcal{F}_{\text{CES}}$ of functions in equation (31) which correspond to CES markets; every CES market (represented by its string encoding) induces a corresponding function from the unit simplex $S$ to itself given in (31).

Given a function $F_I$ from $\mathcal{F}$ we define its size, denoted $\mathsf{size}(F_I)$, to be the length $|I|$ of the string $I$. As usual, we define also the size of a rational number $r$, denoted $\mathsf{size}(r)$, to be the number of bits in the binary representation of $r$. We will show that the family $\mathcal{F}_{\text{CES}}$ has the following two crucial properties, and that for every family of functions that has these properties, the problem of computing an approximate fixed point is in PPAD. Theorem 13 will then follow.

**Definition 12** (Polynomially Continuous Families). *We say that a family of functions $\mathcal{F}$ is polynomially continuous if there is a polynomial $q$ such that for every $F_I \in \mathcal{F}$ and every rational $c > 0$, there is a rational $\delta$ such that $\log(1/\delta) \leq q(|I| + \mathsf{size}(c))$ and such that $\|\mathbf{x} - \mathbf{y}\| \leq \delta$ implies $\|F_I(\mathbf{x}) - F_I(\mathbf{y})\| \leq c$ for any $\mathbf{x}, \mathbf{y} \in D_I$.*

**Definition 13** (Approximately Polynomially Computable Families). *We say that a family of functions $\mathcal{F}$ is* approximately polynomially computable *if there is a polynomial $q$ and an algorithm which, given (the string encoding $I$ of) a function $F_I \in \mathcal{F}$, a rational vector $\mathbf{x} \in D_I$ and a rational number $c > 0$, computes a vector $\mathbf{y}$ that satisfies $\|F_I(\mathbf{x}) - \mathbf{y}\| \leq c$ in time $q(|I| + \mathsf{size}(\mathbf{x}) + \mathsf{size}(c))$.*

It was shown in [EY10] that, if a family of functions is both polynomially continuous and polynomially computable, then the following problem, called the *Weak Approximation problem*, is in PPAD: Given (the string encoding $I$ of) a function $F_I \in \mathcal{F}$ and a rational number $c > 0$ (in binary), compute a $c$-approximate fixed point of $F$. We show that the same is true if $\mathcal{F}$ is approximately polynomially computable.

**Theorem 14.** *If a family of functions $\mathcal{F}$ is polynomially continuous and approximately polynomially computable then the weak approximation problem for $\mathcal{F}$ (i.e. given rational $c > 0$ and function $F \in \mathcal{F}$ compute a $c$-approximate fixed point of $F$) is in* PPAD.

*Proof.* The proof is similar to that of Proposition 2.2, Part 2, in [EY10].

First, from Lemma 2.1 of [EY10], the problem can be reduced to the case where all the functions in $\mathcal{F}$ have a unit simplex as their domain and range, so we assume this is the case from now on.

Let $F \in \mathcal{F}$ be a given function (represented by its string encoding) that maps the unit simplex $S$ in $\mathbb{R}_+^m$ to itself, and $c > 0$ a given rational number, where $c < 1$ without loss of generality. By Definition 12, we can pick an integer $N$ with polynomially many bits in $\mathsf{size}(F)$ and $\mathsf{size}(c/3m)$ such that $\delta = 1/N$ satisfies both $\delta < c/(3m)$ and the following condition:

$$\|F(\mathbf{x}) - F(\mathbf{y})\| < c/(3m), \quad \text{for all } \mathbf{x}, \mathbf{y} \in S \text{ that satisfy } \|\mathbf{x} - \mathbf{y}\| < \delta.$$

We discretize $S$ into a regular simplicial decomposition [Kuh68] with the following set of vertices $T$:

$$T = \{\mathbf{p} \in S : \text{each } p_i \text{ is a multiple of } 1/N\}$$

For each $\mathbf{p} \in T$, we define $g(\mathbf{p})$ to be the output of the algorithm from Definition 13, given (the encoding $I$ of) $F$, vector $\mathbf{p}$ and $c/(9m^2)$. We can assume, without loss of generality, that $g(\mathbf{p}) \geq \mathbf{0}$. As $g(\mathbf{p})$ may not lie on the unit simplex $S$, we scale it to a vector $f(\mathbf{p}) = g(\mathbf{p})/\sum_k g_k(\mathbf{p})$ that lies on $S$. We have:

$$|F_i(\mathbf{p}) - f_i(\mathbf{p})| = \left|\frac{F_i(\mathbf{p})(\sum_k g_k(\mathbf{p})) - g_i(\mathbf{p})}{\sum_k g_k(\mathbf{p})}\right| \leq \frac{(m+1)c/(9m^2)}{1 - mc/(9m^2)} \leq \frac{c}{3m}$$

for every $i \in [m]$. Therefore, $\|F(\mathbf{p}) - f(\mathbf{p})\| \leq c/(3m)$ and $f(\mathbf{p}) \in S$ for all $\mathbf{p} \in T$.

We consider the following $m$-coloring on $T$:

> Vertex $\mathbf{p}$ is colored $i \in [m]$ if $f(\mathbf{p}) \neq \mathbf{p}$ and $i$ is the smallest coordinate such that $f_i(\mathbf{p}) < p_i$, or $f(\mathbf{p}) = \mathbf{p}$ and $i$ is the smallest coordinate such that $p_i = \max_j p_j$.

Note that for any $\mathbf{p} \in T$, if $f(\mathbf{p}) \neq \mathbf{p}$ then at least one of the coordinates satisfies $f_i(\mathbf{p}) < p_i$, since $\mathbf{p}$ and $f(\mathbf{p})$ are both in $S$ and their coordinates sum to 1. Hence, the coloring rule above is well defined. Note also that the $m$ unit vectors $e_i$, $i \in [m]$, at the corners of the unit simplex $S$ are labeled $i$, and all the vertices of $T$ on the facet $p_i = 0$ are labeled with a color $\neq i$. Hence, this $m$-coloring of $T$ satisfies the conditions of Sperner's Lemma and therefore a panchromatic simplex of diameter $1/N$ must exist, i.e., a simplex whose $m$ vertices have different colors.

It is known that finding a panchromatic simplex of the regular simplicial decomposition of $S$ in such an $m$-coloring that satisfies Sperner's Lemma is in PPAD, e.g., using the method described in [EY10]. Now it suffices to prove that one of the vertices of a panchromatic simplex is a $c$-approximate fixed point of $F$.

For this purpose, consider a panchromatic simplex with the following $m$ vertices $\mathbf{p}^1, \ldots, \mathbf{p}^m$. Assume, without loss of generality, that $\mathbf{p}^i$ is colored $i$ for all $i$. We next prove that any point $\mathbf{p} \in \{\mathbf{p}^1, \ldots, \mathbf{p}^m\}$ is a $c$-approximate fixed point of $F$. First notice that $f_i(\mathbf{p}^i) \leq p_i^i$ for all $i \in [m]$ since $\mathbf{p}^i$ is colored $i$. Since $\|\mathbf{p}^i - \mathbf{p}\| \leq \delta$, we have

$$F_i(\mathbf{p}) - p_i = F_i(\mathbf{p}) - F_i(\mathbf{p}^i) + F_i(\mathbf{p}^i) - f_i(\mathbf{p}^i) + f_i(\mathbf{p}^i) - p_i^i + p_i^i - p_i \leq \frac{c}{m}$$

and hence, $F_i(\mathbf{p}) \leq p_i + c/m$ for every $i \in [m]$.

On the other hand, as $\sum_i F_i(\mathbf{p}) = \sum_i p_i = 1$, if we sum the previous inequalities for all $i \neq j$, we get

$$1 - F_j(\mathbf{p}) \leq 1 - p_j + c$$

and thus, $F_j(\mathbf{p}) \geq p_j - c$ for all $j \in [m]$. It follows that $\mathbf{p}$ is a $c$-approximate fixed point of $F$. □

We show now that the family $\mathcal{F}_{\text{CES}}$ for the CES markets satisfies the two conditions of Theorem 14.

**Lemma 33.** *The family $\mathcal{F}_{\text{CES}}$ is polynomially continuous.*

*Proof.* Let $M$ be a given CES market and $F \in \mathcal{F}_{\text{CES}}$ the corresponding function given in (31). First, we let $r$ be the largest positive $\rho_i$ in the market, with $r = 0$ when all $\rho_i$'s are negative. Since each trader can only use a nonzero $\rho_i < 1$, we have $r < 1$.

Let $\boldsymbol{\pi}, \boldsymbol{\pi}' \in S$ denote two vectors with $\|\boldsymbol{\pi} - \boldsymbol{\pi}'\| \leq \delta$ for some parameter $\delta > 0$ satisfying

$$h = \frac{\delta}{\xi(1-r)} < \frac{1}{4} \tag{33}$$

but to be specified later. Let $\mathbf{y} = \hat{\boldsymbol{\pi}}$ and $\mathbf{z} = \hat{\boldsymbol{\pi}}'$ then we have $y_j, z_j \geq \xi$ for all $j$ and $\|\mathbf{y} - \mathbf{z}\| \leq \delta$. Let $x_{i,j}$ and $x'_{i,j}$ denote the demand of good $j$ from trader $i$ at $\mathbf{y}$ and $\mathbf{z}$, respectively.

For any $\delta > 0$ satisfying (33), we will prove the following inequality:

$$|x_{i,j} - x'_{i,j}| \leq q, \quad \text{where } q = 5m^2 h \cdot \xi^{\frac{r-4}{1-r}} \cdot (\alpha_{\max})^{\frac{2}{1-r}} \tag{34}$$

Assume this inequality holds. Notice that

$$|Z_j(\mathbf{y}) - Z_j(\mathbf{z})| = \left|\sum_{i=1}^n (x_{i,j} - x'_{i,j})\right| \leq \sum_{i=1}^n |x_{i,j} - x'_{i,j}| \leq nq$$

Recall the definition of the function $F$. For each $j$, we have

$$F_j(\boldsymbol{\pi}) = \frac{y_j + \max\{0, Z_j(\mathbf{y})\}}{\sum_{k=1}^m (y_k + \max\{0, Z_k(\mathbf{y})\})}$$

Replacing $\boldsymbol{\pi}, \mathbf{y}$ with $\boldsymbol{\pi}', \mathbf{z}$, the change in the numerator is at most $\delta + nq$. The change in the denominator is at most $m\delta + nmq$. Now we show that if $\delta$ is small enough so that (33) and the following hold:

$$\delta + nq \leq c\xi/3 \quad \text{and} \quad m\delta + nmq \leq c/3 \tag{35}$$

then $\|F(\boldsymbol{\pi}) - F(\boldsymbol{\pi}')\| \leq c$. To see this, assume without loss of generality that $F_j(\boldsymbol{\pi}') > F_j(\boldsymbol{\pi})$, for some $j$. Since the numerator of $F_j(\boldsymbol{\pi})$ is at least $\xi$ and the denominator of $F_j(\boldsymbol{\pi})$ is at least 1, we have

$$F_j(\boldsymbol{\pi}') - F_j(\boldsymbol{\pi}) < F_j(\boldsymbol{\pi}) \cdot \frac{1 + c/3}{1 - c/3} - F_j(\boldsymbol{\pi}) \leq F_j(\boldsymbol{\pi}) \cdot c \leq c$$

Because all $\rho_i$'s are given in unary, $1/(1-r)$ can be bounded from above by $\mathsf{size}(M)$. It is now clear that $\delta$ satisfies (33) and (35) when $\log(1/\delta)$ is polynomially large in $\mathsf{size}(M) + \mathsf{size}(c)$, for some large enough polynomial. We now prove inequality (34).

Using (1) we have the following explicit form of $|x_{i,j} - x'_{i,j}|$ (we let $\rho$ denote $\rho_i$ for convenience):

$$\alpha_{i,j}^{\frac{1}{1-\rho}} \cdot \frac{\left|\left(\sum_{k=1}^m w_{i,k} \cdot y_k\right)\left(z_j^{\frac{1}{1-\rho}} \sum_{k=1}^m \alpha_{i,k}^{\frac{1}{1-\rho}} \cdot z_k^{\frac{-\rho}{1-\rho}}\right) - \left(\sum_{k=1}^m w_{i,k} \cdot z_k\right)\left(y_j^{\frac{1}{1-\rho}} \sum_{k=1}^m \alpha_{i,k}^{\frac{1}{1-\rho}} \cdot y_k^{\frac{-\rho}{1-\rho}}\right)\right|}{\left(y_j^{\frac{1}{1-\rho}} \sum_{k=1}^m \alpha_{i,k}^{\frac{1}{1-\rho}} \cdot y_k^{\frac{-\rho}{1-\rho}}\right)\left(z_j^{\frac{1}{1-\rho}} \sum_{k=1}^m \alpha_{i,k}^{\frac{1}{1-\rho}} \cdot z_k^{\frac{-\rho}{1-\rho}}\right)}$$

We start with a few easy bounds that work for both positive and negative $\rho$. First, using $1/(1-\rho) > 0$:

$$\alpha_{i,j}^{1/(1-\rho)} \geq \alpha_{\min}^{1/(1-\rho)} = 1$$

since $\alpha_{\min} = 1$; and $\alpha_{i,j}^{1/(1-\rho)} \leq (\alpha_{\max})^{1/(1-r)}$. Let $\beta$ denote a number in $[\xi, 1]$, then $\beta^{1/(1-\rho)} \leq 1$ and

$$\beta^{1/(1-\rho)} \geq \xi^{1/(1-\rho)} \geq \xi^{1/(1-r)}$$

Note that this holds even when $r = 0$. Finally, we have

$$\beta^{-\rho/(1-\rho)} \leq \xi^{-r/(1-r)} \quad \text{and} \quad \beta^{-\rho/(1-\rho)} \geq \xi$$

The first one follows from $\beta \leq 1$ and $\beta^{-\rho/(1-\rho)} \leq 1$ when $\rho < 0$. The second follows similarly.

We will now bound the numerator. For this purpose we use the following two inequalities

$$\left|\sum_{k=1}^m w_{i,k} \cdot y_k - \sum_{k=1}^m w_{i,k} \cdot z_k\right| \leq \delta m \tag{36}$$

and

$$\left|z_j^{\frac{1}{1-\rho}} \sum_{k=1}^m \alpha_{i,k}^{\frac{1}{1-\rho}} \cdot z_k^{\frac{-\rho}{1-\rho}} - y_j^{\frac{1}{1-\rho}} \sum_{k=1}^m \alpha_{i,k}^{\frac{1}{1-\rho}} \cdot y_k^{\frac{-\rho}{1-\rho}}\right| \leq 4mh \cdot (\alpha_{\max})^{\frac{1}{1-r}} \cdot \xi^{\frac{-r}{1-r}} \tag{37}$$

which we prove later. Using (36), (37) and $|ab - cd| \leq |(b-d)a| + |(a-c)d|$, the numerator is at most

$$(\alpha_{\max})^{\frac{1}{1-r}} \cdot \left(4mh \cdot (\alpha_{\max})^{\frac{1}{1-r}} \cdot \xi^{\frac{-r}{1-r}} \cdot \sum_{k=1}^m w_{i,k} \cdot y_k + \delta m \cdot z_j^{\frac{1}{1-\rho}} \cdot \sum_{k=1}^m \alpha_{i,k}^{\frac{1}{1-\rho}} \cdot z_k^{\frac{-\rho}{1-\rho}}\right) \tag{38}$$

Using $\xi \leq y_k, z_k \leq 1$, $w_{i,k} \leq w_{\max} = 1$ and $\alpha_{i,k} \leq \alpha_{\max}$, the last expression is at most

$$(\alpha_{\max})^{\frac{1}{1-r}} \cdot \left(4m^2 h \cdot (\alpha_{\max})^{\frac{1}{1-r}} \cdot \xi^{\frac{-r}{1-r}} + \delta m^2 \cdot (\alpha_{\max})^{\frac{1}{1-r}} \cdot \xi^{\frac{-r}{1-r}}\right) \leq 5m^2 h \cdot (\alpha_{\max})^{\frac{2}{1-r}} \cdot \xi^{\frac{-r}{1-r}}$$

using (33). The inequality (34) follows since the denominator is at least

$$\left(\xi^{\frac{1}{1-r}} \cdot \xi\right)^2 = \xi^{\frac{4-2r}{1-r}}$$

It remains to prove the two inequalities (36) and (37). For (36) we have:

$$\left|\sum_{k=1}^{m} w_{i,k} \cdot y_k - \sum_{k=1}^{m} w_{i,k} \cdot z_k\right| \leq \sum_{k=1}^{m} w_{i,k} \cdot |y_k - z_k| \leq \delta m$$

For (37) we first give an upper bound for $|\gamma^{1/(1-\rho)} - \beta^{1/(1-\rho)}|$, where $\beta + \delta \geq \gamma \geq \beta \geq \xi$ and $\gamma \leq 1$:

$$\gamma^{\frac{1}{1-\rho}} - \beta^{\frac{1}{1-\rho}} \leq \beta^{\frac{1}{1-\rho}} \left(\left(1+\frac{\delta}{\beta}\right)^{\frac{1}{1-\rho}} - 1\right) \leq \left(1+\frac{\delta}{\beta}\right)^{\frac{1}{1-\rho}} - 1$$

since $\beta \leq \gamma \leq 1$ and $1/(1-\rho) > 0$. By (33), we have $\delta/(\xi(1-\rho)) \leq \delta/(\xi(1-r)) \leq 1/4$ and thus,

$$\left(1+\frac{\delta}{\beta}\right)^{\frac{1}{1-\rho}} - 1 \leq \left(1+\frac{\delta}{\xi}\right)^{\frac{1}{1-\rho}} - 1 \leq e^{\frac{\delta}{\xi(1-\rho)}} - 1 \leq \frac{2\delta}{\xi(1-\rho)} \leq \frac{2\delta}{\xi(1-r)} = 2h$$

where we used $e^x \geq 1 + x \geq e^{x/2}$ for all $0 \leq x \leq 1$; and $1 - \rho \geq 1 - r > 0$.

We also need to give an upper bound for $|\gamma^{-\rho/(1-\rho)} - \beta^{-\rho/(1-\rho)}|$. When $\rho > 0$, we have

$$\left|\gamma^{\frac{-\rho}{1-\rho}} - \beta^{\frac{-\rho}{1-\rho}}\right| = \gamma^{\frac{-\rho}{1-\rho}} \left|1 - \left(\frac{\gamma}{\beta}\right)^{\frac{\rho}{1-\rho}}\right| \leq \xi^{\frac{-r}{1-r}} \cdot \left(\left(1+\frac{\delta}{\beta}\right)^{\frac{\rho}{1-\rho}} - 1\right)$$

By (33), we have $\delta\rho/(\beta(1-\rho)) \leq \delta r/(\xi(1-r)) < 1/4$ and thus, by the same argument

$$\left|\gamma^{\frac{-\rho}{1-\rho}} - \beta^{\frac{-\rho}{1-\rho}}\right| \leq \xi^{\frac{-r}{1-r}} \cdot \frac{2\delta r}{\xi(1-r)} < 2h \cdot \xi^{\frac{-r}{1-r}}$$

On the other hand, when $\rho < 0$, we have

$$\left|\gamma^{\frac{-\rho}{1-\rho}} - \beta^{\frac{-\rho}{1-\rho}}\right| = \beta^{\frac{-\rho}{1-\rho}} \left(\left(\frac{\gamma}{\beta}\right)^{\frac{-\rho}{1-\rho}} - 1\right) \leq \left(1+\frac{\delta}{\beta}\right)^{\frac{-\rho}{1-\rho}} - 1 \leq \frac{2\delta}{\beta} \leq \frac{2\delta}{\xi} \leq 2h$$

since $0 < -\rho/(1-\rho) < 1$. Thus for both cases, $2h \cdot \xi^{-r/(1-r)}$ is a valid upper bound.

Using the second bound, we immediately have

$$\left|\sum_{k=1}^{m} \alpha_{i,k}^{\frac{1}{1-\rho}} \cdot z_k^{\frac{-\rho}{1-\rho}} - \sum_{k=1}^{m} \alpha_{i,k}^{\frac{1}{1-\rho}} \cdot y_k^{\frac{-\rho}{1-\rho}}\right| \leq \sum_{k=1}^{m} \alpha_{i,k}^{\frac{1}{1-\rho}} \cdot \left|z_k^{\frac{-\rho}{1-\rho}} - y_k^{\frac{-\rho}{1-\rho}}\right| \leq 2mh \cdot (\alpha_{\max})^{\frac{1}{1-r}} \cdot \xi^{\frac{-r}{1-r}} \quad (39)$$

Using the inequality on $|ab - cd| \leq |(b-d)a| + |(a-c)d|$ again, the left side of (37) is at most

$$z_j^{\frac{1}{1-\rho}} \cdot 2mh \cdot (\alpha_{\max})^{\frac{1}{1-r}} \cdot \xi^{\frac{-r}{1-r}} + 2h \cdot \sum_{k=1}^{m} \alpha_{i,k}^{\frac{1}{1-\rho}} \cdot y_k^{\frac{-\rho}{1-\rho}}$$

Plugging $z_j^{\frac{1}{1-\rho}} \le 1$ and $y_k^{-\rho/(1-\rho)} \le \xi^{-r/(1-r)}$, we can upperbound it by

$$2mh \cdot (\alpha_{\max})^{\frac{1}{1-r}} \cdot \xi^{\frac{-r}{1-r}} + 2mh \cdot (\alpha_{\max})^{\frac{1}{1-r}} \cdot \xi^{\frac{-r}{1-r}} = 4mh \cdot (\alpha_{\max})^{\frac{1}{1-r}} \cdot \xi^{\frac{-r}{1-r}}$$

and (37) follows. The lemma is now proven. $\square$

We now show that $\mathcal{F}_{\text{CES}}$ satisfies the second condition of Theorem 14.

**Lemma 34.** *The family $\mathcal{F}_{\text{CES}}$ is approximately polynomially computable.*

*Proof.* Let $M$ be a CES market and $F$ be the corresponding function in (31). Let $\pi \in S$ be a given rational vector in the unit simplex, and $c > 0$ be a given rational number, where $c < 1$ without loss of generality.

First we can clearly compute $\xi$ and $\hat{\pi}$ in polynomial time. From the definition of $F$ and the fact that the denominator of $F_j$ is at least 1, it suffices to approximate the demand $x_{i,j}(\hat{\pi})$ of each trader within (additive) precision of, e.g., $c/(2nm^2)$. An easy calculation shows then that the approximate values for the $F_j$ that we derive will have error at most $c$.

For each $x_{i,j}$, we use the explicit form in (1), applied to $\hat{\pi}$. Because all the prices in $\hat{\pi}$ are at least $\xi$, we have lower bounds for both the denominator and numerator of $x_{i,j}$ (e.g., see the proof of Lemma 33). With these lower bounds it then suffices to approximate the rational powers of $\hat{\pi}_k$'s and $\alpha_{i,k}$'s in (1) to sufficient precision. This can be done efficiently because the exponents are either $1/(1-\rho_i)$ or $-\rho_i/(1-\rho_i)$ and $\rho_i$ is encoded in unary. $\square$

Theorem 13 now follows from Lemma 32, Theorem 14, Lemma 12, and Lemma 13. Using Lemma 32, to compute an $\epsilon$-approximate equilibrium of a market $M$, it suffices to compute a $c$-approximate fixed point of the corresponding function $F$ in (31), where $c = \xi \epsilon w_{\min}/2$. By Lemmas 12 and 13, the family $\mathcal{F}_{\text{CES}}$ of functions (31) corresponding to CES markets is polynomially continuous and approximately polynomially computable. Theorem 14 then implies that the problem of computing a $c$-approximate fixed point of $F$ is in PPAD, and hence so is the problem of computing an $\epsilon$-approximate equilibrium of a CES market $M$.

## 7 PPAD-Completeness of Two-Strategy Polymatrix Games

In this section we prove Theorem 8. Membership in PPAD for the exact equilibrium problem (and thus the approximation as well) was shown in [EY10], Corollary 5.3. The proof of its PPAD-hardness below follows the techniques developed in [DGP09, CDT09].

### 7.1 Generalized Circuits and Their Assignment Problem

Syntactically, a generalized circuit $\mathcal{S}$ is a pair $(V, \mathcal{T})$, in which $V$ is a set of nodes; and $\mathcal{T}$ is a set of gates. Every gate $T \in \mathcal{T}$ is a 5-tuple $T = (G, v_1, v_2, v, \alpha)$ in which

1. $G \in \{G_\zeta, G_{\times\zeta}, G_=, G_+, G_-, G_<, G_\wedge, G_\vee, G_\neg\}$ is the type of the gate. Among the nine types of gates, $G_\zeta, G_{\times\zeta}, G_=, G_+$ and $G_-$ are arithmetic gates implementing arithmetic constraints. $G_<$ is called a brittle comparator: it only distinguishes two values that are properly separated. Finally, $G_\wedge$, $G_\vee$ and $G_\neg$ are logic gates.

2. $v_1, v_2 \in V \cup \{\text{nil}\}$ are the first and second input nodes of the gate.

3. $v \in V$ is the output node, and $\alpha \in \mathbb{R}_{\ge 0} \cup \{\text{nil}\}$.

$$
\begin{aligned}
G = G_\zeta : \quad & \mathcal{P}[T, \epsilon] = \Big[\ \mathbf{x}[v] = \alpha \pm \epsilon\ \Big] \\
G = G_{\times\zeta} : \quad & \mathcal{P}[T, \epsilon] = \Big[\ \mathbf{x}[v] = \min\big(\alpha \mathbf{x}[v_1], 1\big) \pm \epsilon\ \Big] \\
G = G_= : \quad & \mathcal{P}[T, \epsilon] = \Big[\ \mathbf{x}[v] = \min\big(\mathbf{x}[v_1], 1\big) \pm \epsilon\ \Big] \\
G = G_+ : \quad & \mathcal{P}[T, \epsilon] = \Big[\ \mathbf{x}[v] = \min\big(\mathbf{x}[v_1] + \mathbf{x}[v_2], 1\big) \pm \epsilon\ \Big] \\
G = G_- : \quad & \mathcal{P}[T, \epsilon] = \Big[\ \min\big(\mathbf{x}[v_1] - \mathbf{x}[v_2], 1\big) - \epsilon \le \mathbf{x}[v] \le \max\big(\mathbf{x}[v_1] - \mathbf{x}[v_2], 0\big) + \epsilon\ \Big] \\
G = G_< : \quad & \mathcal{P}[T, \epsilon] = \Big[\ \mathbf{x}[v] =^\epsilon_B 1 \text{ if } \mathbf{x}[v_1] < \mathbf{x}[v_2] - \epsilon;\ \mathbf{x}[v] =^\epsilon_B 0 \text{ if } \mathbf{x}[v_1] > \mathbf{x}[v_2] + \epsilon\ \Big] \\
G = G_\vee : \quad & \mathcal{P}[T, \epsilon] = \begin{bmatrix} \mathbf{x}[v] =^\epsilon_B 1 \text{ if } \mathbf{x}[v_1] =^\epsilon_B 1 \text{ or } \mathbf{x}[v_2] =^\epsilon_B 1 \\ \mathbf{x}[v] =^\epsilon_B 0 \text{ if } \mathbf{x}[v_1] =^\epsilon_B 0 \text{ and } \mathbf{x}[v_2] =^\epsilon_B 0 \end{bmatrix} \\
G = G_\wedge : \quad & \mathcal{P}[T, \epsilon] = \begin{bmatrix} \mathbf{x}[v] =^\epsilon_B 0 \text{ if } \mathbf{x}[v_1] =^\epsilon_B 0 \text{ or } \mathbf{x}[v_2] =^\epsilon_B 0 \\ \mathbf{x}[v] =^\epsilon_B 1 \text{ if } \mathbf{x}[v_1] =^\epsilon_B 1 \text{ and } \mathbf{x}[v_2] =^\epsilon_B 1 \end{bmatrix} \\
G = G_\neg : \quad & \mathcal{P}[T, \epsilon] = \Big[\ \mathbf{x}[v] =^\epsilon_B 0 \text{ if } \mathbf{x}[v_1] =^\epsilon_B 1;\ \mathbf{x}[v] =^\epsilon_B 1 \text{ if } \mathbf{x}[v_1] =^\epsilon_B 0\ \Big]
\end{aligned}
$$

Table 1: Constraints $\mathcal{P}[T, \epsilon]$, where $T = (G, v_1, v_2, v, \alpha)$ and $K = |V|$

The set $\mathcal{T}$ of gates must satisfy the following important property:

No Conflict: For any gates $T = (G, v_1, v_2, v, \alpha) \ne T' = (G', v'_1, v'_2, v', \alpha')$ in $\mathcal{T}$, we have $v \ne v'$.

Suppose $T = (G, v_1, v_2, v, \alpha)$ in $\mathcal{T}$, then

1. If $G = G_\zeta$, then the gate has no input node and $v_1 = v_2 = $ nil.

2. If $G \in \{G_{\times\zeta}, G_=, G_\neg\}$, then the gate has one input node: $v_1 \in V$ and $v_2 = $ nil.

3. If $G \in \{G_+, G_-, G_<, G_\wedge, G_\vee\}$, then the gate has two input nodes: $v_1, v_2 \in V$ and $v_1 \ne v_2$.

Moreover, the parameter $\alpha$ is only used in $G_\zeta$ and $G_{\times\zeta}$ gates. If $G = G_\zeta$ or $G_{\times\zeta}$, then $\alpha \in [0, 1]$.

Semantically we associate each node $v \in V$ with a real variable $\mathbf{x}[v]$. Each gate $T$ requires the variables of its input and output nodes to satisfy a certain constraint, logical or arithmetic, depending on the type of $T$ (see Table 1 for the details of the constraints). Here the notation $b = a \pm \epsilon$ means $b \in [a - \epsilon, a + \epsilon]$ and the notation $=^\epsilon_B$ is defined as follows. Given an assignment $(\mathbf{x}[v] : v \in V)$ to the variables, we say the value of $\mathbf{x}[v]$ represents Boolean 1 with precision $\epsilon$, denoted by $\mathbf{x}[v] =^\epsilon_B 1$ if

$$1 - \epsilon \le \mathbf{x}[v] \le 1 + \epsilon;$$

it represents Boolean 0 with precision $\epsilon$, denoted by $\mathbf{x}[v] =^\epsilon_B 0$ if $0 \le \mathbf{x}[v] \le \epsilon$. One can see that the logic constraints required by the three logic gates $G_\wedge, G_\vee$ and $G_\neg$ are defined similarly to the classical ones.

**Definition 14.** *Suppose $\mathcal{S} = (V, \mathcal{T})$ is a generalized circuit, where $K = |V|$. For $\epsilon \geq 0$, an $\epsilon$-approximate solution to $\mathcal{S}$ is an assignment $(\mathbf{x}[v] : v \in V)$ to the variables such that*

$$0 \leq \mathbf{x}[v] \leq 1 + \epsilon, \quad \text{for all } v \in V;$$

*and for each gate $T = (G, v_1, v_2, v, \alpha) \in \mathcal{T}$, the values of $\mathbf{x}[v_1]$, $\mathbf{x}[v_2]$ and $\mathbf{x}[v]$ must satisfy the constraint $\mathcal{P}[T, \epsilon]$ defined in* Table 1.

We use **POLY-GCIRCUIT** to denote the following problem: given a generalized circuit $\mathcal{S}$ with $K$ nodes, find an $\epsilon$-approximate solution, where $\epsilon = 1/K$. It is known that:

**Theorem 15.** **POLY-GCIRCUIT** *is* PPAD-hard.

## 7.2 PPAD-Hardness of Two-Strategy Polymatrix Games

We present a polynomial-time reduction from **POLY-GCIRCUIT** to **POLYMATRIX**.

Let $\mathcal{S} = (V, \mathcal{T})$ be a generalized circuit with $K = |V|$. Let $\mathcal{C}$ be an arbitrary bijection from $V$ to

$$\{1, 3, \ldots, 2K - 3, 2K - 1\}$$

Let $n = 2K$. We construct from $\mathcal{S}$ a $2n \times 2n$ matrix

$$\mathbf{P} = \begin{pmatrix} \mathbf{0} & \mathbf{B} \\ \mathbf{A} & \mathbf{0} \end{pmatrix}$$

where $\mathbf{A}, \mathbf{B} \in [0, 1]^{n \times n}$, as follows

$$\mathbf{A} = \sum_{T \in \mathcal{T}} \mathbf{L}[T] \quad \text{and} \quad \mathbf{B} = \sum_{T \in \mathcal{T}} \mathbf{R}[T]$$

The construction of $\mathbf{L}[T]$ and $\mathbf{R}[T]$ can be found in Table 2. For each $T \in \mathcal{T}$, it is easy to check that $\mathbf{L}[T]$ and $\mathbf{R}[T]$ defined in Table 2 satisfy the following property.

**Property 1.** *Let $T = (G, v_1, v_2, v, \alpha)$, $\mathbf{L}[T] = (L_{i,j})$ and $\mathbf{R}[T] = (R_{i,j})$. Suppose $\mathcal{C}(v) = 2k - 1$, then*

$$j \notin \{2k, 2k - 1\} \Rightarrow L_{i,j} = R_{i,j} = 0, \quad \forall i \in [2K];$$
$$j \in \{2k, 2k - 1\} \Rightarrow 0 \leq L_{i,j}, R_{i,j} \leq 1, \quad \forall i \in [2K].$$

**Corollary 6.** $\mathbf{A}, \mathbf{B} \in [0, 1]^{n \times n}$ *and* $\mathbf{P} \in [0, 1]^{2n \times 2n}$.

We denote an $\epsilon$-well-supported Nash equilibrium of $\mathbf{P}$, where $\epsilon = 1/n < 1/K$, by a pair of $n$-dimensional vectors $(\mathbf{x}, \mathbf{y})$, instead of a single $2n$-dimensional vector. For each node $v \in V$, we let $\mathbf{x}[v] = x_{2k-1}$ where $2k - 1 = \mathcal{C}(v)$. As $\epsilon < 1/K$, PPAD-hardness of **POLYMATRIX** follows from the following lemma:

**Lemma 35.** $(\mathbf{x}[v] : v \in V)$ *is an $\epsilon$-approximate solution to the generalized circuit $\mathcal{S}$.*

It is clear that $0 \leq \mathbf{x}[v] \leq 1$ for all $v \in V$ simply because $x_{2k-1} + x_{2k} = 1$ for all $k \in [K]$.
So to prove Lemma 35, it suffices to show that $(\mathbf{x}[v] : v \in V)$ satisfies all the constraints $\mathcal{P}[T, \epsilon]$.

**Lemma 36** (Constraints $P[T, \epsilon]$)**.** *Let $(\mathbf{x}, \mathbf{y})$ be an $\epsilon$-well-supported Nash equilibrium of $\mathbf{P}$, then for each gate $T \in \mathcal{T}$, $(\mathbf{x}[v] : v \in V)$ satisfies the constraint $\mathcal{P}[T, \epsilon]$ defined in* Table 1.

Matrices $\mathbf{L}[T]$ and $\mathbf{R}[T]$, where $T = (G, v_1, v_2, v, \alpha)$ is a gate in $\mathcal{T}$

Set $\mathbf{L}[T]$ and $\mathbf{R}[T]$ to be the zero matrix. Let $2k-1 = \mathcal{C}(v)$, $2k_1 - 1 = \mathcal{C}(v_1)$ and $2k_2 - 1 = \mathcal{C}(v_2)$:

$G_\zeta$: $L_{2k-1,2k} = L_{2k,2k-1} = R_{2k-1,2k-1} = 1$, $R_{i,2k} = \alpha/K$, $\forall i: 1 \le i \le 2K$

$G_{\times\zeta}$: $L_{2k-1,2k-1} = L_{2k,2k} = R_{2k-1,2k} = 1$, $R_{2k_1-1,2k-1} = \alpha$

$G_=$: $L_{2k-1,2k-1} = L_{2k,2k} = R_{2k_1-1,2k-1} = R_{2k-1,2k} = 1$

$G_+$: $L_{2k-1,2k-1} = L_{2k,2k} = R_{2k_1-1,2k-1} = R_{2k_2-1,2k-1} = R_{2k-1,2k} = 1$

$G_-$: $L_{2k-1,2k-1} = L_{2k,2k} = R_{2k_1-1,2k-1} = R_{2k_2-1,2k} = R_{2k-1,2k} = 1$

$G_<$: $L_{2k-1,2k} = L_{2k,2k-1} = R_{2k_1-1,2k-1} = R_{2k_2-1,2k} = 1$

$G_\vee$: $L_{2k-1,2k-1} = L_{2k,2k} = R_{2k_1-1,2k-1} = R_{2k_2-1,2k-1} = 1$, $R_{i,2k} = 1/(2K)$, $\forall i: 1 \le i \le 2K$

$G_\wedge$: $L_{2k-1,2k-1} = L_{2k,2k} = R_{2k_1-1,2k-1} = R_{2k_2-1,2k-1} = 1$, $R_{i,2k} = 3/(2K)$, $\forall i: 1 \le i \le 2K$

$G_\neg$: $L_{2k-1,2k} = L_{2k,2k-1} = R_{2k_1-1,2k-1} = R_{2k_1,2k} = 1$

Table 2: Matrices $\mathbf{L}[T]$ and $\mathbf{R}[T]$

*Proof.* Let $T = (G, v_1, v_2, v, \alpha)$ be a gate in $\mathcal{T}$, with $\mathcal{C}(v) = 2k - 1$. Let $\mathbf{A}_i$, $\mathbf{L}_i$, $\mathbf{B}_i$, and $\mathbf{R}_i$ denote $i$th column vector of $\mathbf{A}, \mathbf{B}, \mathbf{L}[T]$ and $\mathbf{R}[T]$, respectively. By Property 1, $\mathbf{L}[T]$ and $\mathbf{R}[T]$ are the only two gadget matrices that modify the entries in columns $\mathbf{A}_{2k-1}, \mathbf{A}_{2k}$ or columns $\mathbf{B}_{2k-1}, \mathbf{B}_{2k}$. Thus, we have

$$\mathbf{A}_{2k-1} = \mathbf{L}_{2k-1}, \quad \mathbf{A}_{2k} = \mathbf{L}_{2k}, \quad \mathbf{B}_{2k-1} = \mathbf{R}_{2k-1} \quad \text{and} \quad \mathbf{B}_{2k} = \mathbf{R}_{2k} \tag{40}$$

We start with the addition gate $G = G_+$. Let $\mathcal{C}(v_1) = 2k_1 - 1$ and $\mathcal{C}(v_2) = 2k_2 - 1$. We need to show

$$\mathbf{x}[v] = \min(\mathbf{x}[v_1] + \mathbf{x}[v_2], 1) \pm \epsilon$$

From (40) and Table 2, we have

$$\mathbf{x} \cdot \mathbf{B}_{2k-1} - \mathbf{x} \cdot \mathbf{B}_{2k} = \mathbf{x} \cdot \mathbf{R}_{2k-1} - \mathbf{x} \cdot \mathbf{R}_{2k} = \mathbf{x}[v_1] + \mathbf{x}[v_2] - \mathbf{x}[v] \tag{41}$$

$$\mathbf{y} \cdot \mathbf{A}_{2k-1} - \mathbf{y} \cdot \mathbf{A}_{2k} = \mathbf{y} \cdot \mathbf{L}_{2k-1} - \mathbf{y} \cdot \mathbf{L}_{2k} = y_{2k-1} - y_{2k} \tag{42}$$

In a proof by contradiction, we consider two cases. First, assume $\mathbf{x}[v] > \min(\mathbf{x}[v_1] + \mathbf{x}[v_2], 1) + \epsilon$. Since $\mathbf{x}[v] \le 1$, the assumption would imply $\mathbf{x}[v] > \mathbf{x}[v_1] + \mathbf{x}[v_2] + \epsilon$. From Equation (41) and the definition of $\epsilon$-well-supported Nash equilibria, we have $y_{2k-1} = 0$ and $y_{2k} = 1$. By combining this with (42), we finally get $\mathbf{x}[v] = x_{2k-1} = 0$, contradicting our assumption that $\mathbf{x}[v] > \mathbf{x}[v_1] + \mathbf{x}[v_2] + \epsilon > 0$.

Next, we assume $\mathbf{x}[v] < \min(\mathbf{x}[v_1] + \mathbf{x}[v_2], 1) - \epsilon \le \mathbf{x}[v_1] + \mathbf{x}[v_2] - \epsilon$. Then, Equation (41) implies that $y_{2k} = 0$ and $y_{2k-1} = 1$. By Equation (42), we have $x_{2k} = 0$ and $\mathbf{x}[v] = x_{2k-1} = 1$, contradicting our assumption that $\mathbf{x}[v] < \min(\mathbf{x}[v_1] + \mathbf{x}[v_2], 1) - \epsilon \le 1 - \epsilon$. □

*Proof for $G_\zeta$ Gates.* From (40) and Table 2, we have

$$\mathbf{x} \cdot \mathbf{B}_{2k-1} - \mathbf{x} \cdot \mathbf{B}_{2k} = \mathbf{x} \cdot \mathbf{R}_{2k-1} - \mathbf{x} \cdot \mathbf{R}_{2k} = \mathbf{x}[v] - \alpha$$

$$\mathbf{y} \cdot \mathbf{A}_{2k-1} - \mathbf{y} \cdot \mathbf{A}_{2k} = \mathbf{y} \cdot \mathbf{L}_{2k-1} - \mathbf{y} \cdot \mathbf{L}_{2k} = y_{2k} - y_{2k-1}$$

If $\mathbf{x}[v] > \alpha + \epsilon$, then from the first equation, we have $y_{2k} = 0$ and $y_{2k-1} = 1$. But by the second equation we have $\mathbf{x}[v] = x_{2k-1} = 0$, which contradicts our assumption that $\mathbf{x}[v] > \alpha + \epsilon > 0$.

If $\mathbf{x}[v] < \alpha - \epsilon$, then from the first equation, we have $y_{2k-1} = 0$ and $y_{2k} = 1$. But the second equation implies $x_{2k} = 0$ and $\mathbf{x}[v] = x_{2k-1} = 1$, contradicting the assumption that $\mathbf{x}[v] < \alpha - \epsilon$ and $\alpha \leq 1$.  □

*Proof for $G_{\times\zeta}$ Gates.* From (40) and Table 2, we have

$$\mathbf{x} \cdot \mathbf{B}_{2k-1} - \mathbf{x} \cdot \mathbf{B}_{2k} = \mathbf{x} \cdot \mathbf{R}_{2k-1} - \mathbf{x} \cdot \mathbf{R}_{2k} = \alpha \cdot \mathbf{x}[v_1] - \mathbf{x}[v]$$

$$\mathbf{y} \cdot \mathbf{A}_{2k-1} - \mathbf{y} \cdot \mathbf{A}_{2k} = \mathbf{y} \cdot \mathbf{L}_{2k-1} - \mathbf{y} \cdot \mathbf{L}_{2k} = y_{2k-1} - y_{2k}$$

If $\mathbf{x}[v] > \min(\alpha\mathbf{x}[v_1], 1) + \epsilon$, then we have $\mathbf{x}[v] > \alpha\mathbf{x}[v_1] + \epsilon$ as $\mathbf{x}[v] = x_{2k-1} \leq 1$. By the first equation, we have $y_{2k-1} = 0$ and $y_{2k} = 1$. The second equation implies that $\mathbf{x}[v] = x_{2k-1} = 0$, which contradicts the assumption that $\mathbf{x}[v] > \min(\alpha\mathbf{x}[v_1], 1) + \epsilon > 0$.

If $\mathbf{x}[v] < \min(\alpha\mathbf{x}[v_1], 1) - \epsilon \leq \alpha\mathbf{x}[v_1] - \epsilon$, then the first equation implies that $y_{2k} = 0$ and $y_{2k-1} = 1$. From the second equation we have $x_{2k} = 0$ and $\mathbf{x}[v] = x_{2k-1} = 1$, which contradicts the assumption that $\mathbf{x}[v] < \min(\alpha\mathbf{x}[v_1], 1) - \epsilon \leq 1 - \epsilon$.  □

*Proof for $G_=$ Gates.* $G_=$ is a special case of $G_{\times\zeta}$ with parameter $\alpha = 1$.  □

*Proof for $G_-$ Gates.* From (40) and Table 2, we have

$$\mathbf{x} \cdot \mathbf{B}_{2k-1} - \mathbf{x} \cdot \mathbf{B}_{2k} = \mathbf{x} \cdot \mathbf{R}_{2k-1} - \mathbf{x} \cdot \mathbf{R}_{2k} = \mathbf{x}[v_1] - \mathbf{x}[v_2] - \mathbf{x}[v]$$

$$\mathbf{y} \cdot \mathbf{A}_{2k-1} - \mathbf{y} \cdot \mathbf{A}_{2k} = \mathbf{y} \cdot \mathbf{L}_{2k-1} - \mathbf{y} \cdot \mathbf{L}_{2k} = y_{2k-1} - y_{2k}$$

If $\mathbf{x}[v] > \max(\mathbf{x}[v_1] - \mathbf{x}[v_2], 0) + \epsilon \geq \mathbf{x}[v_1] - \mathbf{x}[v_2] + \epsilon$, then the first equation shows that $y_{2k-1} = 0$ and $y_{2k} = 1$. But from the second equation, we have $\mathbf{x}[v] = x_{2k-1} = 0$ which contradicts with the assumption that $\mathbf{x}[v] > \max(\mathbf{x}[v_1] - \mathbf{x}[v_2], 0) + \epsilon > 0$.

If $\mathbf{x}[v] < \min(\mathbf{x}[v_1] - \mathbf{x}[v_2], 1) - \epsilon \leq \mathbf{x}[v_1] - \mathbf{x}[v_2] - \epsilon$, then by the first equation we have $y_{2k} = 0$ and $y_{2k-1} = 1$. By the second equation, we have $x_{2k} = 0$ and $\mathbf{x}[v] = x_{2k-1} = 1$, contradicting the assumption that $\mathbf{x}[v] < \min(\mathbf{x}[v_1] - \mathbf{x}[v_2], 1) - \epsilon \leq 1 - \epsilon < 1$.  □

*Proof for $G_<$ Gates.* From (40) and Table 2, we have

$$\mathbf{x} \cdot \mathbf{B}_{2k-1} - \mathbf{x} \cdot \mathbf{B}_{2k} = \mathbf{x} \cdot \mathbf{R}_{2k-1} - \mathbf{x} \cdot \mathbf{R}_{2k} = \mathbf{x}[v_1] - \mathbf{x}[v_2]$$

$$\mathbf{y} \cdot \mathbf{A}_{2k-1} - \mathbf{y} \cdot \mathbf{A}_{2k} = \mathbf{y} \cdot \mathbf{L}_{2k-1} - \mathbf{y} \cdot \mathbf{L}_{2k} = y_{2k} - y_{2k-1}$$

If $\mathbf{x}[v_1] < \mathbf{x}[v_2] - \epsilon$, then we have $y_{2k-1} = 0$ and $y_{2k} = 1$, from the first equation. But the second equation implies that $x_{2k} = 0$ and $\mathbf{x}[v] = x_{2k-1} = 1$ and thus, $\mathbf{x}[v] =_B^\epsilon 1$.

If $\mathbf{x}[v_1] > \mathbf{x}[v_2] + \epsilon$, then $y_{2k} = 0$ and $y_{2k-1} = 1$ according to the first equation. By the second one we have $\mathbf{x}[v] = x_{2k-1} = 0$ and thus, $\mathbf{x}[v] =_B^\epsilon 0$.  □

*Proof for $G_\vee$ Gates.* From (40) and Table 2, we have

$$\mathbf{x} \cdot \mathbf{B}_{2k-1} - \mathbf{x} \cdot \mathbf{B}_{2k} = \mathbf{x} \cdot \mathbf{R}_{2k-1} - \mathbf{x} \cdot \mathbf{R}_{2k} = \mathbf{x}[v_1] + \mathbf{x}[v_2] - (1/2)$$

$$\mathbf{y} \cdot \mathbf{A}_{2k-1} - \mathbf{y} \cdot \mathbf{A}_{2k} = \mathbf{y} \cdot \mathbf{L}_{2k-1} - \mathbf{y} \cdot \mathbf{L}_{2k} = y_{2k-1} - y_{2k}$$

If $\mathbf{x}[v_1] =_B^\epsilon 1$ or $\mathbf{x}[v_2] =_B^\epsilon 1$, then $\mathbf{x}[v_1] + \mathbf{x}[v_2] \geq 1 - \epsilon$. By the first equation $y_{2k} = 0$ and $y_{2k-1} = 1$. By the second equation, we have $x_{2k} = 0$ and $\mathbf{x}[v] = x_{2k-1} = 1$ and thus, $\mathbf{x}[v] =_B^\epsilon 1$.

If $\mathbf{x}[v_1] =_B^\epsilon 0$ and $\mathbf{x}[v_2] =_B^\epsilon 0$, then we have $\mathbf{x}[v_1] + \mathbf{x}[v_2] \leq 2\epsilon$. From the first equation we must have $y_{2k-1} = 0$ and $y_{2k} = 1$. Then, the second equation implies $\mathbf{x}[v] = x_{2k-1} = 0$ and thus, $\mathbf{x}[v] =_B^\epsilon 0$. □

*Proof for $G_\wedge$ Gates.* From (40) and Table 2, we have

$$\mathbf{x} \cdot \mathbf{B}_{2k-1} - \mathbf{x} \cdot \mathbf{B}_{2k} = \mathbf{x} \cdot \mathbf{R}_{2k-1} - \mathbf{x} \cdot \mathbf{R}_{2k} = \mathbf{x}[v_1] + \mathbf{x}[v_2] - (3/2)$$

$$\mathbf{y} \cdot \mathbf{A}_{2k-1} - \mathbf{y} \cdot \mathbf{A}_{2k} = \mathbf{y} \cdot \mathbf{L}_{2k-1} - \mathbf{y} \cdot \mathbf{L}_{2k} = y_{2k-1} - y_{2k}$$

If $\mathbf{x}[v_1] =_B^\epsilon 0$ or $\mathbf{x}[v_2] =_B^\epsilon 0$, then $\mathbf{x}[v_1] + \mathbf{x}[v_2] \leq 1 + \epsilon$. From the first equation, we have $y_{2k-1} = 0$ and $y_{2k} = 1$. By the second equation, we have $\mathbf{x}[v] = x_{2k-1} = 0$ and thus, $\mathbf{x}[v] =_B^\epsilon 0$.

If $\mathbf{x}[v_1] =_B^\epsilon 1$ and $\mathbf{x}[v_2] =_B^\epsilon 1$, then $\mathbf{x}[v_1] + \mathbf{x}[v_2] \geq 2 - 2\epsilon$. The first equation shows that $y_{2k} = 0$ and $y_{2k-1} = 1$. By the second equation, $x_{2k} = 0$ and $\mathbf{x}[v] = x_{2k-1} = 1$ and thus, $\mathbf{x}[v] =_B^\epsilon 1$. □

*Proof for $G_\neg$ Gates.* From (40) and Table 2, we have

$$\mathbf{x} \cdot \mathbf{B}_{2k-1} - \mathbf{x} \cdot \mathbf{B}_{2k} = \mathbf{x} \cdot \mathbf{R}_{2k-1} - \mathbf{x} \cdot \mathbf{R}_{2k} = \mathbf{x}[v_1] - (1 - \mathbf{x}[v_1]) = 2\mathbf{x}[v_1] - 1$$

$$\mathbf{y} \cdot \mathbf{A}_{2k-1} - \mathbf{y} \cdot \mathbf{A}_{2k} = \mathbf{y} \cdot \mathbf{L}_{2k-1} - \mathbf{y} \cdot \mathbf{L}_{2k} = y_{2k} - y_{2k-1}$$

If $\mathbf{x}[v_1] =_B^\epsilon 1$, then by the first equation, $y_{2k} = 0$ and $y_{2k-1} = 1$. The second one implies $\mathbf{x}[v] = x_{2k-1} = 0$ and thus, $\mathbf{x}[v] =_B^\epsilon 0$. If $\mathbf{x}[v_1] =_B^\epsilon 0$, then the first equation implies $y_{2k-1} = 0$ and $y_{2k} = 1$. By the second equation, we have $x_{2k} = 0$ and $\mathbf{x}[v] = x_{2k-1} = 1$ and thus, $\mathbf{x}[v] =_B^\epsilon 1$. □

## 8 Conclusions

This paper is a first step towards a systematic understanding of what features make the equilibrium analysis of markets computationally hard. We introduced the notion of non-monotone utilities, which covers a wide variety of important utility functions. We then showed that for any family $\mathcal{U}$ of non-monotone utilities, it is PPAD-hard to compute an approximate equilibrium for a market with utilities that are drawn from $\mathcal{U}$ or are linear. Using our general approach, and a further, customized analysis, we resolved the long-standing open problem on the complexity of CES markets when the parameter $\rho$ is less than $-1$, showing that for any fixed value of $\rho < -1$, the problem of computing an approximate equilibrium is PPAD-hard.

This work raises clearly many questions. First, can we dispense with the linear functions in the general theorem, i.e., is it true that for any family $\mathcal{U}$ of non-monotone utilities, the approximate equilibrium problem is PPAD-hard for markets that use utilities from $\mathcal{U}$ only? For the important class of CES utilities with (any) $\rho < -1$, we were able to show this, using a deeper analysis of the class of CES utilities, and appropriate adaptations of the construction. Can a similar approach work in general for all non-monotone utilities?

Second, what other general features of utilities (if any) are there that make the market equilibrium problem hard? Non-monotonicity is connected with markets that can have disconnected sets of market equilibria for which currently we do not have any efficient algorithmic methods to deal with. Convexity has been critical essentially in all tractable cases so far, whether the set of market equilibria itself is convex or whether a convex formulation can be obtained after a change of variables.

Most ambitiously, can we obtain a complexity dichotomy theorem that allows us to classify any family of utility functions (under standard, generally acceptable, mild assumptions for utilities) into those that can be solved efficiently and those that are apparently intractable (PPAD-hard and/or FIXP-hard)? The present paper takes a first step towards this goal.

## Acknowledgment

The authors would like to thank Costis Daskalakis and Shang-Hua Teng for very useful discussions.